\documentclass[modern,twocolumn]{aastex631}

\usepackage{bm,amsmath}

\shorttitle{Generation of solar-like differential rotation}
\shortauthors{Hotta, Kusano, \& Shimada}

\graphicspath{{./}{figures/}}
\begin{document}
  \title{Generation of solar-like differential rotation}

  \author[0000-0002-6312-7944]{H. Hotta}
  \affiliation{Department of Physics, Graduate School of Science,
  Chiba University \\
  1-33 Yayoi-cho, Inage-ku, Chiba 263-8522 Japan
  }
  \author[0000-0002-6814-6810]{K. Kusano}
  \affiliation{Institute for Space-Earth Environmental Research, Nagoya University \\
  Chikusa-ku, Nagoya, Aichi 464-8601, Japan}
  \author[0000-0002-8507-3633]{R. Shimada}
  \affiliation{Department of Earth and Planetary Science, The University of Tokyo \\
  7-3-1, Hongo, Bunkyo-ku, Tokyo, 113-0033, Japan}

  \begin{abstract}
    We analyze the simulation result shown in \cite{hotta_2021NatAs...5.1100H} in which the solar-like differential rotation is reproduced. The Sun is rotating differentially with the fast equator and the slow pole. It is widely thought that the thermal convection maintains the differential rotation, but recent high-resolution simulations tend to fail to reproduce the fast equator. This fact is an aspect of one of the biggest problems in solar physics called the convective conundrum. \cite{hotta_2021NatAs...5.1100H} succeed in reproducing the solar-like differential rotation without using any manipulation with unprecedentedly high-resolution simulation. In this study, we analyze the simulation data to understand the maintenance mechanism of the fast equator. Our analyses lead to conclusions that are summarized as follows. 1. Superequipatition magnetic field is generated by the compression, which can indirectly convert the massive internal energy to magnetic energy. 
    2. The efficient small-scale energy transport suppresses large-scale convection energy.
    3. Non-Taylor--Proudman differential rotation is maintained by the entropy gradient caused by the anisotropic latitudinal energy transport enhanced by the magnetic field. 4. The fast equator is maintained by the meridional flow mainly caused by the Maxwell stress. The Maxwell stress itself also has a role in the angular momentum transport for fast near-surface equator (we call it the {\it Punching ball} effect).
    The fast equator in the simulation is reproduced not due to the low Rossby number regime but due to the strong magnetic field.
    This study newly finds the role of the magnetic field in the maintenance of differential rotation.
  \end{abstract}

  \keywords{Solar convection zone(1998) --- Solar differential rotation(1996) --- Solar dynamo(2001) --- Solar magnetic fields(1503)}


\section{Introduction}
\begin{figure}[htbp]
  \centering
  \includegraphics[width=0.4\textwidth]{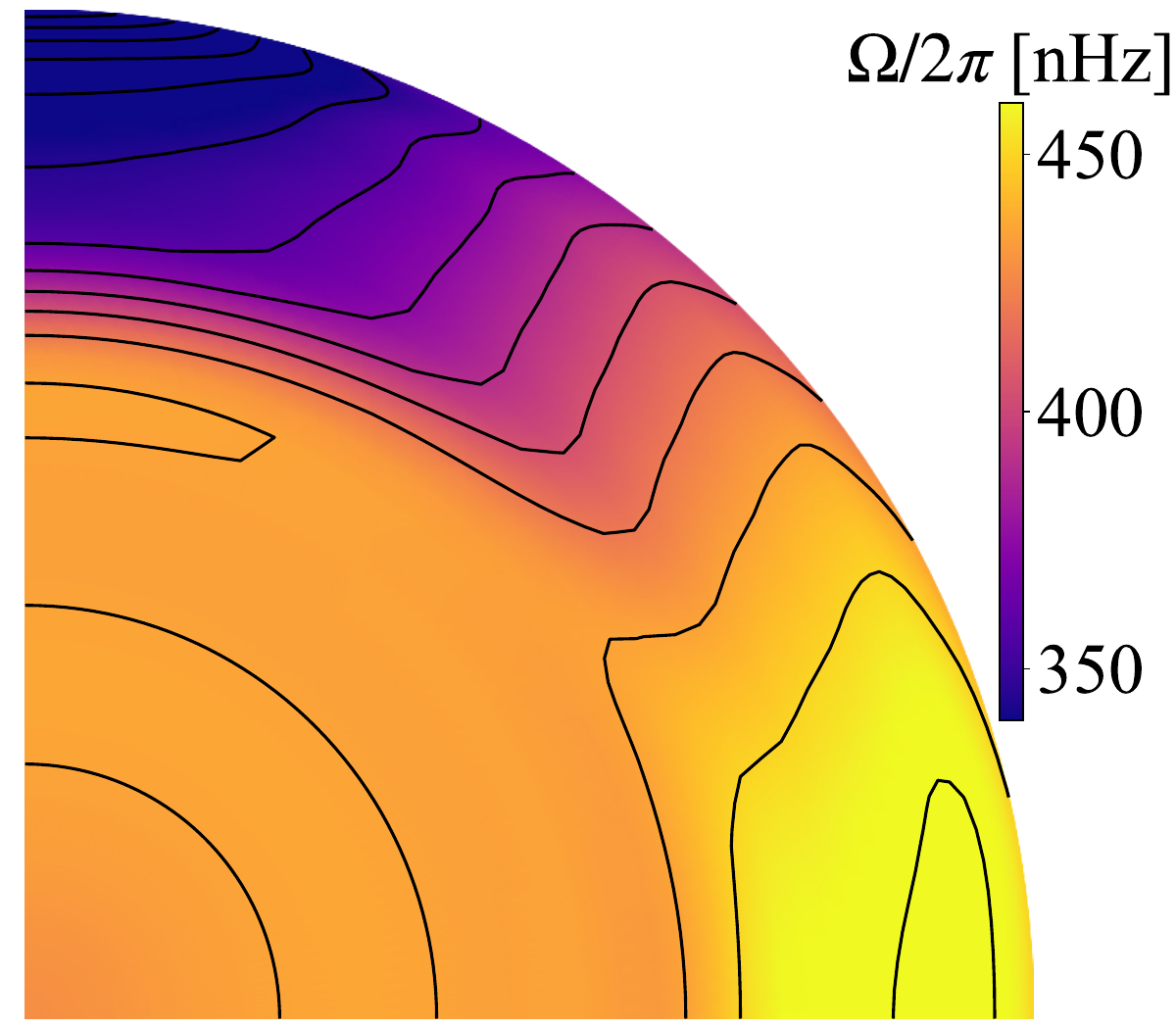}
  \caption{Inversion of the helioseismic data from Helioseismic and Magnetic Imager (HMI)
  of SDO (Solar Dynamics Observatory) satellite for the angular velocity ($\Omega/2\pi$) in the unit of nHz \citep{howe_2011JPhCS.271a2061H}. The solid lines show the values from 340 to 460 nHz in 10 nHz increments.  \label{difrot_observe}}
\end{figure}

The Sun is rotating differentially, i.e., different latitudes have different rotation rates, which is called differential rotation. The solar rotation has a long observational history. In 1630, Christoph Scheiner found the different rotation periods between latitudes using the trajectory of the sunspots \citep{Paterno_2010Ap&SS.328..269P}. In modern-day observations, the Doppler effect is used to measure the rotation rate \citep[e.g.,][]{Howard_1970SoPh...12...23H}. After the appearance of helioseismology that uses acoustic waves to detect the internal structure of the Sun, the internal profile of the differential rotation has been measured \citep{Schou_1998ApJ...505..390S}. Fig.~\ref{difrot_observe} shows one of the helioseismic results of the differential rotation $\Omega/2\pi$ from \cite{howe_2011JPhCS.271a2061H}, where $\Omega$ is the angular velocity. While we observe interesting features of the shear layers, i.e., tachocline at the base of the convection zone and the near-surface shear layer, one of the most prominent features of the solar differential rotation is the fast equator and slow pole. The equator and the polar region rotate in 25 and 30 days, respectively.\par

It has been thought that thermal convection is a key to understanding the generation mechanism of the solar differential rotation. Around the solar center, nuclear fusion generates thermal energy. The radiation transports the energy outward in the radiation zone in the inner part of the solar interior (radiation zone: $<0.71R_\odot$, where $R_\odot$ is the solar radius). In the outer part ($>0.71R_\odot$, convection zone), opacity increases, and the radiation energy transport becomes inefficient. Then, the thermal convection transports the energy. Because of large Reynolds numbers, the thermal convection is turbulent. The turbulence is influenced by the Coriolis force and becomes anisotropic. Angular momentum is transported by the anisotropic turbulence, and the large-scale flow is constructed. Because the turbulence in the convection zone is highly non-linear and chaotic, scientific research with numerical simulation is an essential approach to understanding the differential rotation.\par
By using the numerical simulations, the generation mechanism of the solar differential rotation was thought to be understood at the beginning of the 2000s, but recent high-resolution simulations have crucial problems in reproducing the rotation observed. As a pioneering work, \cite{Gilman_1977GApFD...8...93G} performed solar global convection simulations while ignoring the stratification using the Boussinesq approximation. After the standard model of the solar stratification is established \citep{Christensen-Dalsgaard_1996Sci...272.1286C}, global solar calculations with realistic stratification and other solar parameters are widely performed \citep{Miesch_2000ApJ...532..593M,Brun_2002ApJ...570..865B,Miesch_2006ApJ...641..618M,brun_2011ApJ...742...79B,kapyla_2014A&A...570A..43K,Hotta_2015ApJ...798...51H}. In general, convection with a faster (slower) rotation rate tends to show a fast equator (pole) \citep{Gastine_2013Icar..225..156G}. The essential control parameter for the differential rotation is the Rossby number Ro = $v/(2\Omega_0 L)$ \citep{Miesch_2005LRSP....2....1M,Featherstone_2015ApJ...804...67F}, where $v$, $\Omega_0$, and $L$ are the typical convection velocity, angular velocity of the system, and typical spatial scale of the convection, respectively. The Rossby number measures the effect of the rotation on the convection. A system with a low Rossby number has rotationally constrained convection, which is essential to reproduce a fast equator. Low-resolution calculations in the early years of the global solar convection studies were able to reproduce the solar-like differential rotation (fast equator) because only the large-scale convection is included in their system. Higher-resolution calculations, in other words high Rayleigh and Reynolds numbers, however, have difficulties in reproducing it because small-scale turbulence is introduced and the effective convection scale, $L$, becomes small \citep[see parameter survey by][]{hindman_2020ApJ...898..120H}. This fact is problematic because the real Sun must have much smaller turbulence down to a centimeter scale. Currently, there are three numerical manipulation methods to produce the solar-like differential rotation: 
\begin{enumerate}
  \item To increase the rotation rate \citep{Brown_2008ApJ...689.1354B,Nelson_2013ApJ...762...73N,Hotta_2018ApJ...860L..24H}.
  \item To decrease the luminosity \citep{Hotta_2015ApJ...798...51H}.
  \item To adopt large viscosity and/or thermal conductivity \citep{Miesch_2000ApJ...532..593M,Miesch_2008ApJ...673..557M,Fan_2014ApJ...789...35F,Hotta_2016Sci....351..1427}.
\end{enumerate}
These manipulations aim to reduce the Rossby number. Manipulation 1 increases $\Omega_0$ and directly reduces the Rossby number. In manipulation 2, the convection velocity $v$ is reduced with smaller luminosity and energy flux. Manipulation 3 decreases convection velocity $v$ and increases the effective spatial scale $L$ with the large diffusivities. Early calculations implicitly adopt manipulation 3 because of their low resolution. 
\cite{Fan_2014ApJ...789...35F} find that the magnetic field may contribute to suppressing the convection velocity and decreasing the Rossby number. This effect is extensively investigated by the following researchers.
\cite{gastine_2014MNRAS.438L..76G} carry out a comprehensive parameter study for the fast equator and poles. They find that the existence of the magnetic field relaxes the required rotation rate to the smaller value for the fast equator. 
\cite{mabuchi_2015ApJ...806...10M} pointed out that the Rossby number evaluated with the root-mean-square (RMS) velocity in simulations is a good measure of this issue since the relaxation found in \cite{gastine_2014MNRAS.438L..76G} is caused by the reduction of the convection velocity by the magnetic field \citep[see also][]{karak_2015A&A...576A..26K}.
While the magnetic field certainly has a role(s) in the construction of the differential rotation, large thermal conductivity ($\sim3\times10^{13}~\mathrm{cm^2~s^{-1}}$) is still required to maintain solar-like differential rotation \citep{Fan_2014ApJ...789...35F}.
Because the solar angular velocity $\Omega_0$ and the solar luminosity $L_\odot$ are well-determined values, we should not change these. The viscosity and the thermal conductivity are extremely small and cannot be reproduced in modern computers, and we should keep these as small as possible. We note that there should be larger turbulent diffusivities, but these should be automatically reproduced in three-dimensional simulations. In summary, we have not reproduced the solar-like differential rotation without introducing artificial effects and do not know the valid reason why the fast equator is produced in the Sun. The problem is that low diffusivities accomplished with high resolution hinder the reproduction of the solar-like differential rotation.\par
This problem is one of the most critical and difficult problems in solar physics, called the convective conundrum \citep{OMara_2016AdSpR..58.1475O}. An observational estimate using helioseismology suggests that the convective flow in numerical simulations is much faster than in reality. \cite{Hanasoge_2012PNAS..10911928H} show convective energy spectra in large-scale ($\ell<60$, where $\ell$ is the spherical harmonic degree). The observational estimate is more than two orders of magnitude smaller than a simulation \citep{Miesch_2008ApJ...673..557M}. We note that the helioseismic result is still controversial, and another study shows a consistent result with the simulation \citep{Greer_2015ApJ...803L..17G}. \
The overpowering of the large-scale convection also causes a problem with the supergranulation. The supergranulation is a 30 Mm-scale flow pattern observed at the solar surface, which has a prominent peak in the energy spectrum \citep[e.g.][]{hathaway_2015ApJ...811..105H}. On the solar surface, kinetic energy larger than supergranulation decreases with increasing the scale. 
\cite{Lord_2014ApJ...793...24L} carry out realistic convection simulations for the photosphere and show that a larger calculation box tends to show a kinetic energy peak at a larger scale.  Thus, the large-scale motion excited in the deep layer should be suppressed to obtain the supergranulation peak. \cite{featherstone_2016ApJ...830L..15F} suggest that provided the convective amplitude is suppressed; the rotational influence can construct the supergranulation peak \citep[see also][]{vasil_2021PNAS..11822518V}.
\par
The problem of the previously presented differential rotation is one aspect of the convective conundrum because fast convection flow leads to a large Rossby number and a resulting fast pole. Regarding differential rotation, the observational results confirm the existence of a fast equator; thus, we are confident in the results obtained, but numerical simulations fail in reproducing the real solar differential rotation. Consequently, the numerical simulation has problems.\par
\cite{hotta_2021NatAs...5.1100H} (hereafter, HK21) have
suggested promising possible solution to the problem in the differential rotation aspect of the convective conundrum. We carried out unprecedented high-resolution simulations, and the solar-like differential rotation, i.e., the fast equator, is reproduced without using any manipulation. In this study, we analyze the simulation result to understand the physical mechanism to maintain the solar-like differential rotation, i.e., the fast equator.

\section{Model}
The simulations analyzed in this study are introduced in HK21\footnote{Statistical data are available at \url{https://doi.org/10.5281/zenodo.5919257}}.
We solve three-dimensional magnetohydrodynamic equations in the spherical geometry $(r,\theta,\phi)$ using the Yin-Yang grid \citep{Kageyama_2004GGG.....5.9005K}. The radial computational domain extends $0.71R_\odot<r<0.96R_\odot$. The magnetohydrodynamic equations are 
\begin{align}
  \frac{\partial \rho_1}{\partial t} &= -\frac{1}{\xi^2}\nabla\cdot\left(\rho \bm{v}\right), \\
  \frac{\partial}{\partial t}\left(\rho \bm{v}\right) &= -\nabla\cdot\left(\rho\bm{vv}\right) - \nabla p_1 - \rho_1 g \bm{e}_r \nonumber \\
  &+2 \rho \bm{v}\times\bm{\Omega}_0 + \frac{1}{4\pi}\left(\nabla\times\bm{B}\right)\times\bm{B}, \\
  \frac{\partial \bm{B}}{\partial t} &= \nabla\times\left(\bm{v}\times\bm{B}\right), \\
  \rho T \frac{\partial s_1}{\partial t} &= -\rho T \left(\bm{v}\cdot\nabla\right) s + Q_s, \label{eq:entropy}\\
  p_1 &= \left(\frac{\partial p}{\partial \rho}\right)_s \rho_1 + \left(\frac{\partial p}{\partial s}\right)_\rho s_1,
\end{align}
where $\rho$, $\bm{v}$, $\bm{B}$, $s$, and $p$ are the density, velocity, magnetic field, specific entropy, and gas pressure, respectively. $\bm{e}_r$ is the radial unit vector. To deal with the small perturbation $\rho_1/\rho_0\sim p_1/p_0\sim T_1/T_0\sim10^{-6}$, we separate the quantities to the zeroth order spherically symmetric values (subscript 0) and the perturbation from the background (subscript 1). The zeroth order quantities and the gravitational acceleration $g$ are adopted from Model S \citep{Christensen-Dalsgaard_1996Sci...272.1286C}. The linearized equation of state is used for the pressure to deal with the small perturbation. The coefficient $\left(\partial p/\partial \rho\right)_s$ and $\left(\partial p/\partial s\right)_\rho$ are calculated with the OPAL repository \citep{Rogers_1996ApJ...456..902R}. We use the system rotation rate $\bm{\Omega_0}$ of the solar value, i.e., $\Omega_0=2.6\times10^{-6}~\mathrm{s^{-1}}$ with $\bm{\Omega}_0=\Omega_0\left(\cos\theta \bm{e}_r - \sin\theta \bm{e}_\theta\right)$, where $\bm{e}_\theta$ is the colatitudinal unit vector.\par
We use the reduced speed of sound technique \cite[RSST:][]{Hotta_2012A&A...539A..30H,Hotta_2015ApJ...798...51H}. The effective speed of sound is reduced by a factor of $\xi$. We keep the adiabatic reduced speed of sound to 3 $\mathrm{km~s^{-1}}$ throughout the convection zone.\par
The heating term $Q_s$ at the entropy equation (eq.~(\ref{eq:entropy})) is expressed with two radial flux densities as
\begin{align}
  Q_s &= -\frac{1}{r^2}\frac{\partial }{\partial r}\left[r^2 \left(F_\mathrm{rad} + F_\mathrm{art}\right)\right], \\
  F_\mathrm{rad} &= -\kappa_r \frac{dT_0}{dr}, \label{eq:frad}\\
  F_\mathrm{art} &= \frac{L_\odot}{4\pi r_\mathrm{max}^2}\left(\frac{r}{r_\mathrm{max}}\right)^2\exp\left[-\left(\frac{r-r_\mathrm{max}}{d_\mathrm{art}}\right)^2\right],\label{eq:fart}
\end{align}
where $F_\mathrm{rad}$ and $F_\mathrm{art}$ are the radiative flux and the artificial energy flux. For the radiative energy flux $F_\mathrm{rad}$, we use the diffusion approximation, and the radiative diffusion coefficient is adopted from Model S. Because we do not include the photosphere where the radiation extracts the energy, in this calculation, we need an artificial energy flux around the top boundary. We extract the solar luminosity $L_\odot$ from the top boundary $r=r_\mathrm{max}=0.96R_\odot$. The depth of the cooling layer is defined as $d_\mathrm{art}=2H_p(r_\mathrm{max})$, where $H_p(r_\mathrm{max})=9.46~\mathrm{Mm}$ is the pressure scale height at $r=0.96R_\odot$.
\par
The magnetohydrodynamic equations are solved with R2D2 (Radiation and RSST for Deep Dynamics) code \citep[][HK21]{Hotta_2019SciA....eaau2307,Hotta_2020MNRAS.494.2523H} with the fourth-order space centered difference and the four-step Runge--Kutta time integration. To maintain the numerical stability, we use the slope-limited artificial diffusivity suggested by \cite{Rempel_2014ApJ...789..132R} for all variables. We use $h=2$ for the parameter for the artificial diffusivity shown in eq.~(10) of \cite{Rempel_2014ApJ...789..132R}.
\par
Because the whole sphere is covered with the Yin-Yang grid, we only need the radial boundary condition. We adopt the stress-free and impenetrable boundary condition both at the top and bottom boundaries for the flow. The magnetic field is radial and horizontal at the top and the bottom boundaries, respectively. The density and entropy perturbations are symmetric about the radial boundaries.
\par
We perform four cases, Low, Middle, High, and High-HD. The basic parameters are summarized in Table \ref{summary}.
Information on non-dimensional parameters is provided in Subsection \ref{sec:non-dimensional}.
The High and High-HD cases have the same number of grid points. The High-HD case does not include the magnetic field. The magnetic field is included in the other cases. The calculations continue for 4000 days.
The convection time scale in the deep convection zone is 20 days and 4000 days corresponds to 200 turnover time.
The diffusion time scale for the High case is around 500 years. Our calculation is much shorter than that.
Around 2700 days, however, the flow and the differential rotation reach a statistically steady-state (see Supplementary Figure 1 of HK21). We cannot rule out the further evolution in the longer study, but as shown in the result section, the large-scale flows (differential rotation and meridional flow) are mainly determined by the convection and magnetic field. The contributions from the diffusivities are tiny. It seems that we do not have to extend our calculation to the diffusion time scale to discuss the maintenance mechanism of the large-scale flows.
The period between 3600--4000 days is used in the following analysis unless otherwise noted. The typical time spacing $\Delta t=100~\mathrm{s}$ and 3 million time steps are integrated for the High case.

\begin{table*}[htbp]
  \begin{center}
    \caption{Summary of calculations.\label{summary}}
    \begin{tabular}{ccccc}
      \hline
      \hline
      Case    & Low  & Middle & High & High-HD \\
      \hline
      No. of Grids \\
      $N_r\times N_\theta \times N_\phi\times2$
      & $96\times384\times1152$ & $192\times768\times2304$ & $384\times1536\times4608$& $384\times1536\times4608$\\
      Magnetic field & Yes & Yes & Yes & No \\
      $\nu_\mathrm{eff},~\eta_\mathrm{eff},\kappa_\mathrm{eff}$$~\mathrm{[cm^2~s^{-1}]}$  & 
      $2.41\times10^{11}$ & $8.44\times10^{10}$ & $2.64\times10^{10}$ & $2.55\times10^{10}$\\
      $\overline{v}_\mathrm{RMS}~\mathrm{[m~s^{-1}]}$ & 142 & 127 & 108 & 163\\
      Reynolds number $\mathrm{(Re)}$ &$1.02\times10^3$ & $2.62\times10^3$ & $7.11\times10^3$ & $1.11\times10^4$\\
      Rayleigh number $\mathrm{(Ra)}$ & $1.68\times10^7$ & $1.35\times10^8$ & $1.38\times10^9$ & $1.7\times10^9$\\
      flux Rayleigh number $\mathrm{(Ra_F)}$ & $3.09\times10^{10}$ & $7.24\times10^{11}$ & $2.36\times10^{11}$ & $2.62\times10^{13}$  \\
      Ekman number $\mathrm{(Ek)}$ & $3.1\times10^{-4}$ & $1.1\times10^{-4}$ & $3.3\times10^{-5}$ & $3.2\times10^{-5}$\\
      Convective Rossby number \\
      $\mathrm{(Ro_c)}$ & 1.26 & 1.25 & 1.25 & 1.32 \\
      Rossby number $\mathrm{(Ro)}$ & 0.157 & 0.141 & 0.120 & 0.181\\
      Mean spherical harmonic degree\\
      $\tilde{\ell}_\mathrm{u}$ & 62.5 & 104.6 & 201.8 & 177.0\\
      Local Rossby number $\mathrm{(Ro_\ell)}$ & 6.26 & 9.36 & 15.36 & 20.34\\
      $E_\mathrm{k,turb}~[\mathrm{erg}]$ & $2.78\times10^{39}$ & $2.11\times10^{39}$ & $1.50\times10^{39}$ & $4.30\times10^{39}$\\
      $E_\mathrm{k,mean}~[\mathrm{erg}]$ & $4.71\times10^{38}$ & $5.25\times10^{38}$ & $1.01\times10^{39}$ & $2.09\times10^{40}$\\
      $E_\mathrm{m,turb}~[\mathrm{erg}]$ & $1.23\times10^{39}$ & $2.03\times10^{39}$ & $2.68\times10^{39}$ & N/A\\
      $E_\mathrm{m,mean}~[\mathrm{erg}]$ & $4.55\times10^{36}$ & $1.83\times10^{36}$ & $2.56\times10^{36}$ & N/A\\
      \hline
    \end{tabular}\\
    We convert the Yin-Yang grid to the spherical geometry for analyses. In the spherical geometry, the number of grid is $N_r\times 2N_\theta \times 4N_\phi/3$
  \end{center}
\end{table*}

\section{Result}
\subsection{Overall structure}
\begin{figure*}[htpb]
  \begin{center}
    \includegraphics[width=0.95\textwidth]{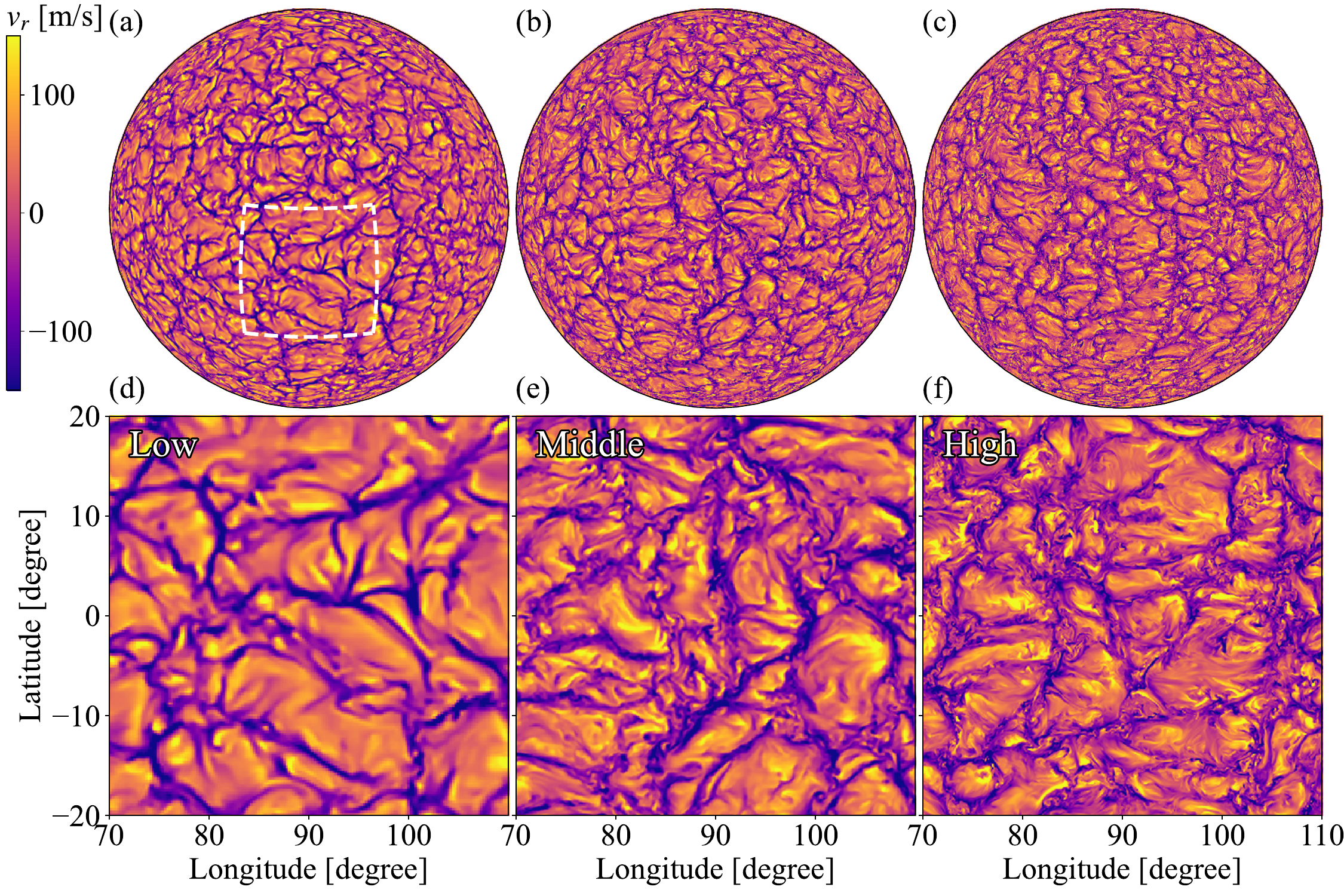}
    \caption{
      Radial velocity $v_r$ at $r=0.95R_\odot$ in Low (panels a, d), Middle (panels b, e), and High (panels c, f) cases are shown. The lower panels (d, e, f) show the subset of the calculation domain indicated by the white dashed box in panel a. 
      Movie is available at \url{https://youtu.be/GXwnIIOJxvY}.
      Movie continues 4min27sec and covers whole evolution of the calculation period, i.e., 4000 days.
    \label{overall_095_vx}}
  \end{center}
\end{figure*}

\begin{figure*}[htpb]
  \begin{center}
    \includegraphics[width=0.95\textwidth]{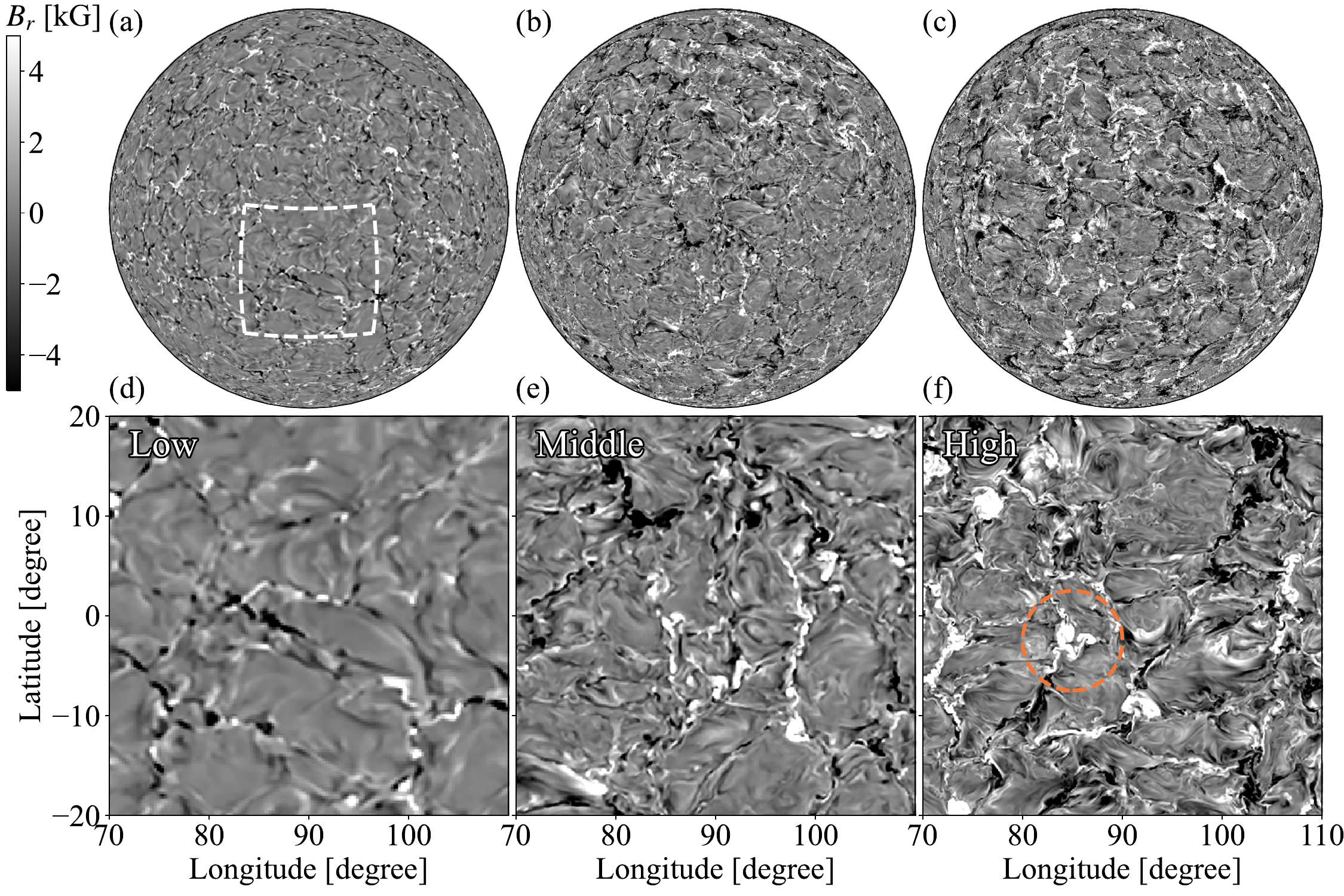}
    \caption{
      The radial magnetic field $B_r$ at $r=0.95R_\odot$ is shown. The format is the same as Fig.~\ref{overall_095_vx}.
      Movie is available at \url{https://youtu.be/ULPPKKGwJNw}.
    \label{overall_095_bx}}
  \end{center}
\end{figure*}

\begin{figure*}[htpb]
  \begin{center}
    \includegraphics[width=0.95\textwidth]{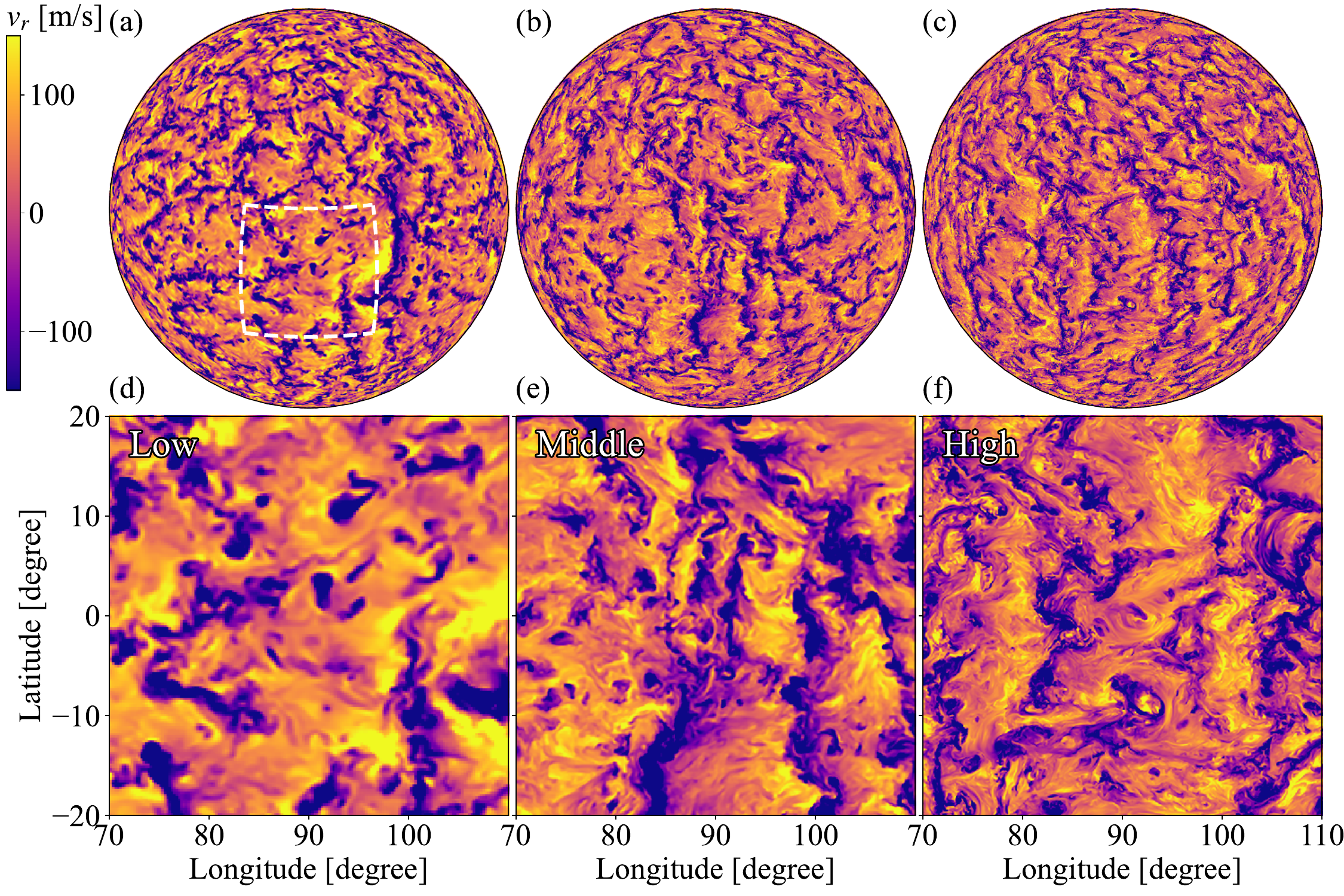}
    \caption{The radial velocity $v_r$ at $r=0.9R_\odot$ is shown. The format is the same as Fig.~\ref{overall_095_vx}.
    Movie is available at \url{https://youtu.be/Ne0jsSCTXX4}.
    \label{overall_090_vx}}
  \end{center}
\end{figure*}

\begin{figure*}[htpb]
  \begin{center}
    \includegraphics[width=\textwidth]{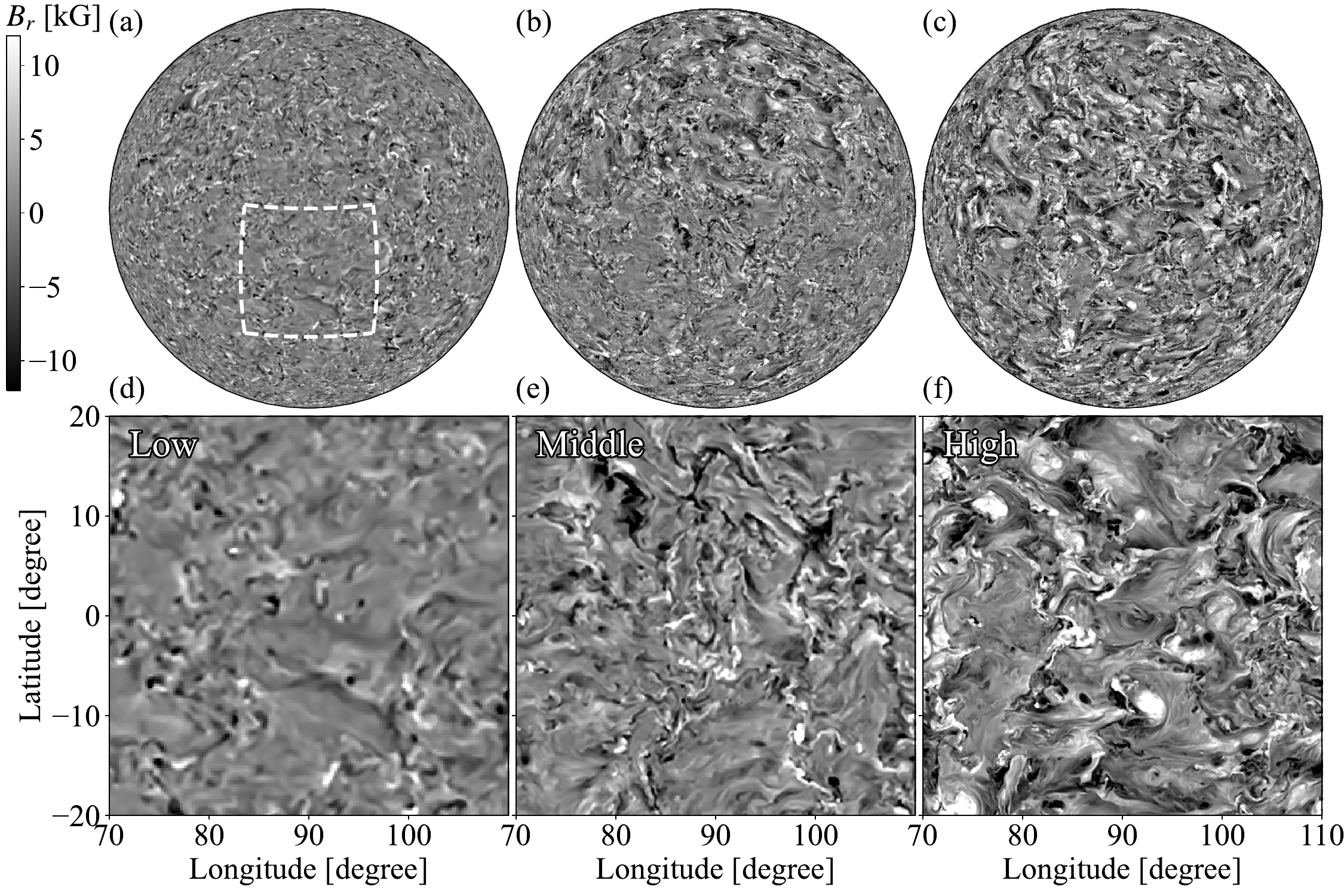}
    \caption{
      The radial magnetic field $B_r$ at $r=0.9R_\odot$ is shown. The format is the same as Fig.~\ref{overall_095_vx}. Movie is available at \url{https://youtu.be/cYZqLUHNMt4}.
      \label{overall_090_bx}}
  \end{center}
\end{figure*}

\begin{figure*}[htpb]
  \begin{center}
    \includegraphics[width=0.95\textwidth]{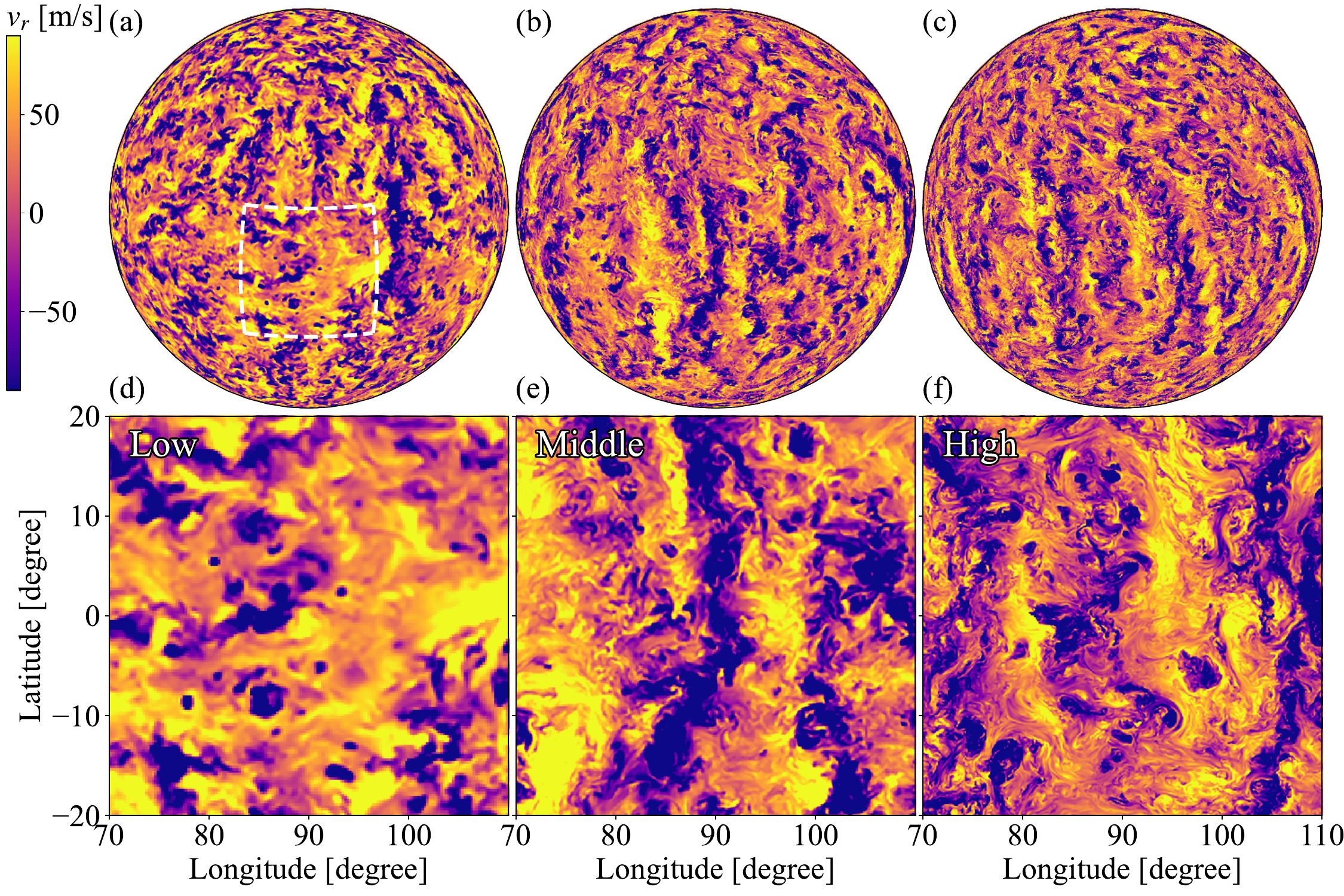}
    \caption{
      The radial velocity $v_r$ at $r=0.85R_\odot$ is shown. The format is the same as Fig.~\ref{overall_095_vx}. Movie is available at \url{https://youtu.be/8zZW8OP9i7Y}.
    \label{overall_085_vx}}
  \end{center}
\end{figure*}

\begin{figure*}[htpb]
  \begin{center}
    \includegraphics[width=\textwidth]{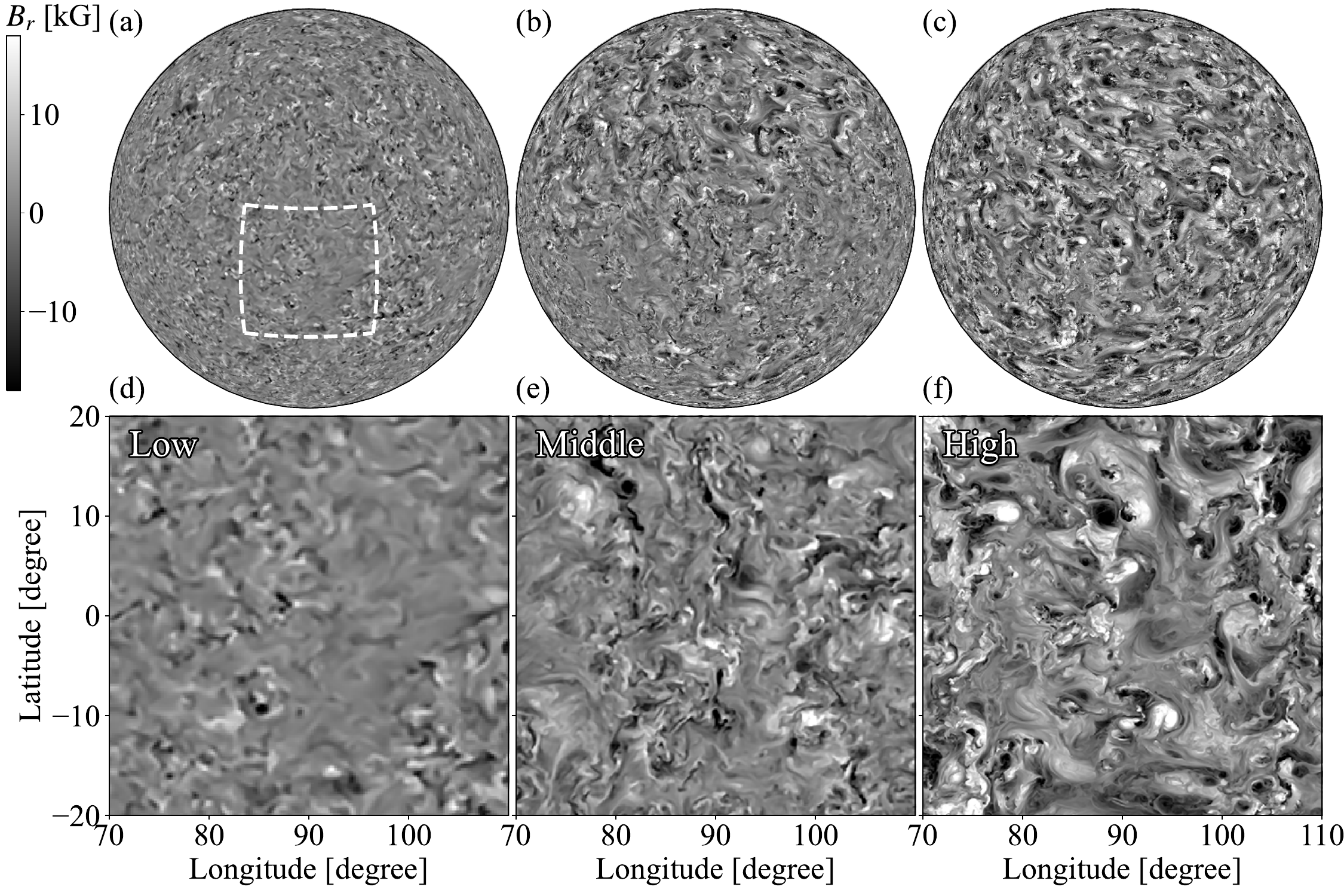}
    \caption{
      The radial magnetic field $B_r$ at $r=0.85R_\odot$ is shown. The format is the same as Fig.~\ref{overall_095_vx}. Movie is available at \url{https://youtu.be/0C6XFdYDkKk}.
      \label{overall_085_bx}}
  \end{center}
\end{figure*}

In this subsection, the overall convection and magnetic field are discussed. Figs.~\ref{overall_095_vx} -- \ref{overall_085_bx} show the overall structure of the radial velocity and the radial magnetic field. Figs.~\ref{overall_095_vx}, \ref{overall_090_vx}, and \ref{overall_085_vx} show the radial velocity $v_r$ at $r=0.95R_\odot$, $0.9R_\odot$, and $0.85R_\odot$, respectively. Figs.~\ref{overall_095_bx}, \ref{overall_090_bx}, and \ref{overall_085_bx} show the radial magnetic field $B_r$ at $r=0.95R_\odot$, $0.9R_\odot$, and $0.85R_\odot$, respectively. The results from Low (panels a, d), Middle (panels b, e), and High (panels c, f) cases are shown in these figures. The panels d, e, and f show zoomed views indicated by a white dashed box in panel a. The radial velocity at $r=0.95R_\odot$ (Fig.~\ref{overall_095_vx}) shows a typical convection pattern, i.e., thin concentrated downflows surrounded by broad upflows. Two effects cause this pattern. The first effect is stratification. Because the solar convection zone is gravitationally stratified, the upper layer has a lower gas pressure. A rising fluid parcel expands because of the stratification, while the descending parcel contracts. This asymmetry of the upflows and downflows cause the typical convection pattern. In addition, we should see a boundary effect at this depth. A wall exists at $r=0.96R_\odot$ where the radial motion stops. This process leads to diverging and converging motions in the upflows and downflows, respectively. The convection patterns in the Low case (Figs. \ref{overall_095_vx}a and d) are similar to previous calculations \citep{Miesch_2008ApJ...673..557M} i.e., the smallest scale is the downflow lane. In the High case, we can see smaller-scale structures even in the downflow lanes (Fig.~\ref{overall_095_vx}f). The banana-cell, the north-south aligned convection cell, cannot be seen in all the cases at this depth.\par
Fig.~\ref{overall_095_bx} shows the radial magnetic field $B_r$ at $r=0.95R_\odot$. The magnetic field strength increases from the Low case to the High case. This tendency is also seen in the other depth (see Figs.\ \ref{overall_090_bx} and \ref{overall_085_bx}). In all the cases, the radial magnetic field is swept up to the downflow region. This concentration is also seen in the previous calculation in the deep interior \citep{Brun_2004ApJ...614.1073B} and the photosphere \citep{Vogler_2005A&A...429..335V}. While the previous simulations and the Low case in this study typically show sheet-like magnetic flux aligned to the downflow lane, we can occasionally observe blob-shaped magnetic flux (a notable one is indicated by the dashed orange circle in Fig.~\ref{overall_095_bx}f). This structure shows significantly superequipartition magnetic field strength and low gas pressure. The convection is suppressed in this region. This structure is important for magnetic field generation (see discussion at Subsection \ref{sec:magnetic_field_generation}). At $r=0.9R_\odot$, the convection shows larger-scale pattern. The small-scale convection around the top boundary is merged to construct the larger-scale while increasing the pressure/density scale height in the deep region 
\cite[see][]{Stein_1998ApJ...499..914S,Lord_2014ApJ...793...24L}. The banana-cell-like feature begins to appear in this depth. In the deeper layer ($r=0.85R_\odot$, middle of the convection zone), the flow pattern shows the banana-cell-like features more clearly than the upper layers (Fig.~\ref{overall_085_vx}). In the mixing length theory, the convection velocity $v_\mathrm{c}$ scales as $\rho_0 v_\mathrm{c}^3\sim L_\odot/4\pi r^2$ \citep{Biermann_1948ZA.....25..135B}, where $L_\odot$ is the solar luminosity. This dependence indicates that the convection velocity decreases in the deeper layers with the larger density $\rho_0$. The convection time scale is $\tau\sim H_p/v_\mathrm{c}$. These relations mean that the convection time scale increases in the deeper layers by increasing the pressure scale height and decreasing the convection velocity. As a result, the convection tends to obey the rotation influence (Coriolis force) and show the banana-cell in the deep layers. The magnetic field distribution is chaotic at this depth. The strong magnetic field tends to locate at the downflow plume, but the coincidence between the downflow and the strong magnetic field is worse than the upper layers.

\subsection{Statistical properties of convection and magnetic field}

\begin{figure}
  \begin{center}
    \includegraphics[width=0.5\textwidth]{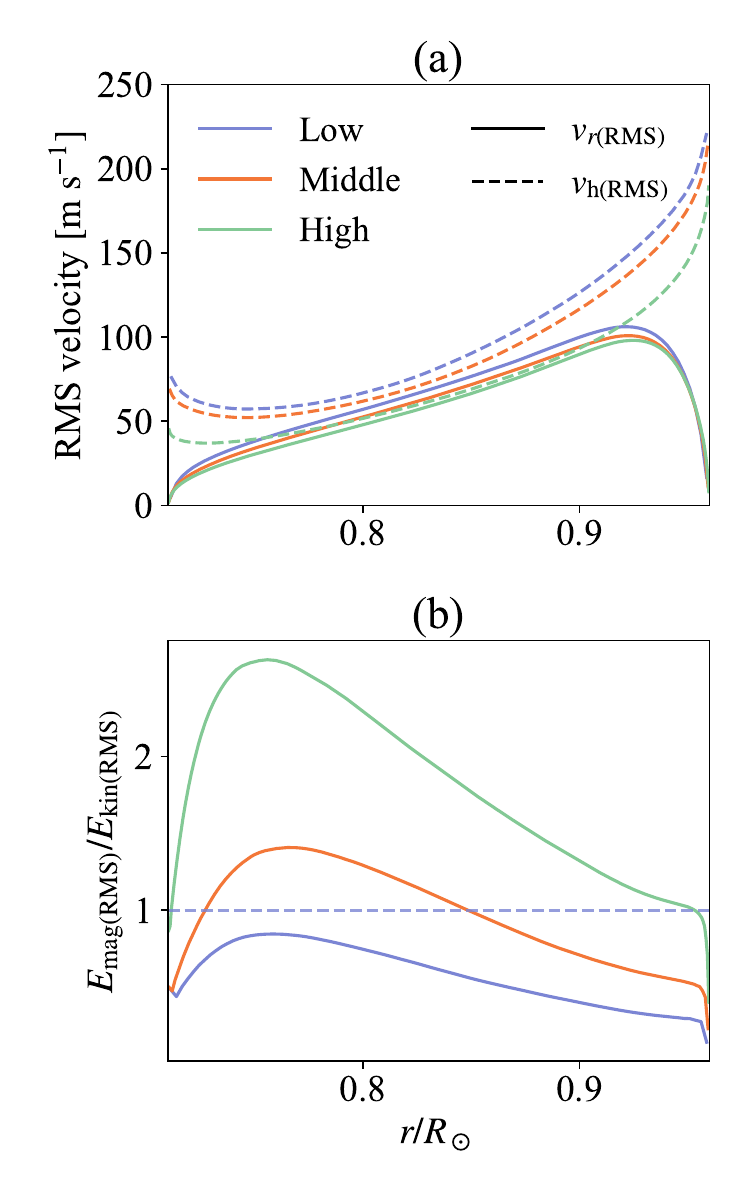}
    \caption{(a) Longitudinal RMS velocity (see eq.~(\ref{eq:longitudinal_rms})) and (b) the ratios of the magnetic energy to the kinetic energy  are shown. The blue, orange, and green colors show the results in Low, Middle, High cases. The same color format is used in the following figures unless otherwise noted. The solid and dashed lines in panel a are radial $v_{r\mathrm{(RMS)}}$, and horizontal $v_\mathrm{h(RMS)}$ longitudinal RMS velocities, where the horizontal velocity is defined as $v_\mathrm{h}=\sqrt{v^2_\theta + v^2_\phi}$. The black dashed line in panel b indicates the equiparition level, i.e., $E_\mathrm{mag(RMS)}/E_\mathrm{kin(RMS)}=1$. The results show that increasing the resolution decreases the convection velocity while increasing the magnetic field strength.
    \label{rms_velocity}}
  \end{center}
\end{figure}

In this subsection, we discuss statistical properties of the convection and magnetic fields. Here, we define statistical values of a quantity $Q$, the longitudinal average $\langle Q\rangle$, the longitudinal RMS (root-mean-square) $Q'_\mathrm{(RMS)}$, and the latitudinally averaged longitudinal RMS $Q_\mathrm{(RMS)}$ as follows.

\begin{align}
  \langle Q \rangle (r,\theta)&= \frac{1}{2\pi}\int_0^{2\pi} Q d\phi \label{eq:longitudinal_average}\\
  Q'_\mathrm{(RMS)} (r,\theta) &= \sqrt{\frac{1}{2\pi}\int_0^{2\pi} \left(Q-\langle Q\rangle\right)^2d\phi} \\
  Q_\mathrm{(RMS)} (r) &=\sqrt{ \frac{1}{2}\int_0^\pi {Q'}_\mathrm{(RMS)}^2 \sin\theta d\theta } \label{eq:longitudinal_rms}
\end{align}

We note that we define spherical average $\widetilde{Q}$ and spherical RMS $Q_\mathrm{(rms)}$ in Subsection \ref{sec:convection_driving} differently from the current definition.\par
Fig.~\ref{rms_velocity} shows the longitudinal RMS velocity (panel a) and ratio of the magnetic energy $E_\mathrm{mag(RMS)}$ to the kinetic energy $E_\mathrm{kin(RMS)}$, where the energies are defined as:

\begin{align}
  E_\mathrm{kin(RMS)} &= \frac{1}{2}\rho_0 v^2_\mathrm{(RMS)} \\
  E_\mathrm{mag(RMS)} &= \frac{B^2_\mathrm{(RMS)}}{8\pi}
\end{align}

Fig.~\ref{rms_velocity}a shows that the convection velocity is suppressed in the higher resolution. This is a general tendency of the high-resolution simulations \citep[e.g.,][]{Hotta_2015ApJ...803...42H}. The horizontal velocity (dashed lines) is more suppressed than the radial velocity (solid lines). A key to understanding convection suppression is the magnetic field. Fig.~\ref{rms_velocity}b shows the ratio of the magnetic energy to the kinetic energy. While in the Low case (blue line), the magnetic energy is smaller than the kinetic energy throughout the convection zone, and we observe a superequipartition magnetic field in the High case (green line). The ratio exceeds 2.5 at maximum. This result indicates the strong influence of the magnetic field on the convective flow. The reason why the High case has such a strong magnetic field is discussed in Subsection \ref{sec:magnetic_field_generation}, and the suppression mechanism of the convection velocity by the magnetic field is shown in Subsection \ref{sec:convection_driving}.

\subsection{Energy spectra}

\begin{figure}[htpb]
  \begin{center}
    \includegraphics[width=0.45\textwidth]{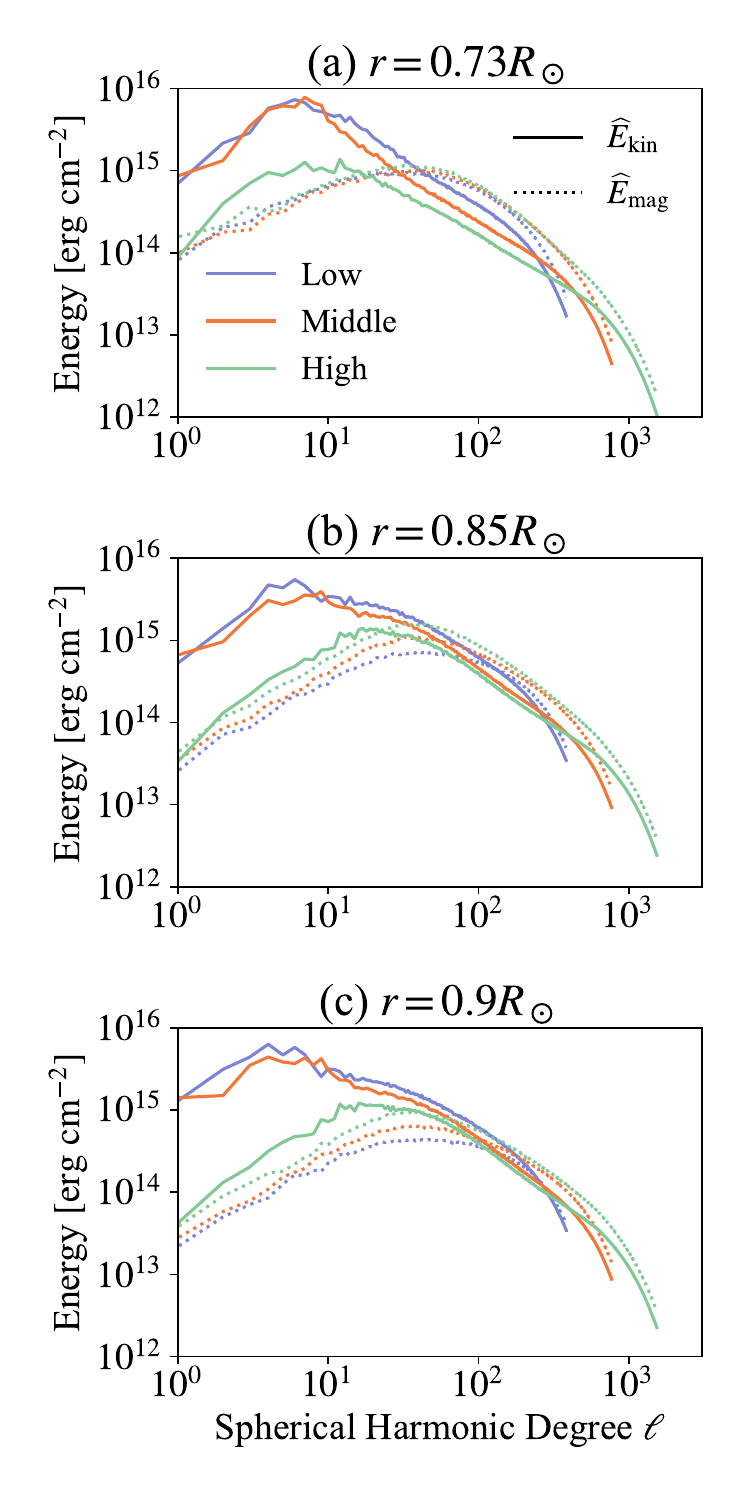}
    \caption{The energy spectra at $r=0.73R_\odot$ (panel a), $r=0.85R_\odot$ (panel b), and $r=0.9R_\odot$ (panel c) are shown. The solid and dotted lines show the kinetic $E_\mathrm{k}$ and magnetic $E_\mathrm{m}$ energies, respectively. In this plot, only the $m\neq0$ mode is shown to exclude contributions from the differential rotation. The large-scale kinetic energy is significantly suppressed in the High case (green line) compared with the other cases.
    \label{spectra}}
  \end{center}
\end{figure}

Fig.~\ref{spectra} shows the kinetic and magnetic energy spectra. We show the results at the layers at $r=0.73R_\odot$ (panel a), $0.85R_\odot$ (panel b), and $0.9R_\odot$ (panel c). 
The definition of the spherical harmonic expansion is for an arbitrary quantity $Q(\theta,\phi)$ is
\begin{align}
  \widehat{Q}_m^\ell = \int_0^{\pi} \int_{0}^{2\pi} Q(\theta,\phi) Y_m^\ell(\theta,\phi) d\phi d\theta,
\end{align}
where $Y_m^\ell$, $\ell$, $m$ are the spherical harmonics, the spherical harmonic degree, and the spherical harmonic order, respectively. We use a normalization which satisfies
\begin{align}
  Q^2_\mathrm{(RMS)} = \sum_\ell \sum_{m,m\neq0} \frac{\widehat{Q}_m^\ell \widehat{Q}_m^{\ell *}}{r},
\end{align}
where $*$ denotes the complex conjugate.
The kinetic $\widehat{E}_\mathrm{kin}(\ell)$ and magnetic $\widehat{E}_\mathrm{mag}(\ell)$ energy spectra are calculated as
\begin{align}
  \widehat{E}_\mathrm{kin}(\ell) =& \frac{1}{2} \rho_0 \sum_m \widehat{\bm{v}}\cdot\widehat{\bm{v}}^* \\
  \widehat{E}_\mathrm{mag}(\ell) =& \frac{1}{8\pi} \sum_m \widehat{\bm{B}}\cdot\widehat{\bm{B}}^*.
\end{align}
The tendency of the kinetic energy spectra is almost the same among the different layers. While the large-scale ($\ell<10$) energy does not change from the Low to Middle cases (blue and green lines, respectively), the energy is significantly reduced in the High case (green line). This reduction is one of the main topics in this paper. The relation between the kinetic and magnetic energies depends on the resolution. When the magnetic energy surpasses the kinetic energy, we expect an efficient small-scale dynamo 
\citep[e.g.][]{Rempel_2014ApJ...789..132R}. In the Low case, while the magnetic energy exceeds the kinetic energy in the deep layer ($r=0.73R_\odot$, panel a) in the small scale ($\ell\sim40$), this clear excess cannot be seen at the shallower layer ($r=0.9R_\odot$, panel c). The inefficient small-scale dynamo in a shallower layer is a common feature in the global dynamo calculation \cite[e.g.,][]{Hotta_2014ApJ...786...24H}. Because the shallower layer has a smaller energy injection scale of the convection because of a small pressure/density scale height and a short time scale for downward magnetic energy transport, we need a high resolution to resolve the small-scale dynamo \citep[see discussion by][]{Stein_2002ESASP.505...83S,Vogler_2007A&A...465L..43V}. This difficulty of the small-scale dynamo is solved in the Middle case (orange line). While the turnover scale depends on the layer depth, the excess of the magnetic field in the small scales is achieved in all the layers in the Middle case. The situation drastically changes in the High case (green line). The kinetic energy is reduced in all the scales but especially in the large-scale ($\ell<30$). This significant suppression is seen at all depths. As a result, the magnetic energy exceeds or is comparable to the kinetic energy in all scales.

\subsection{Mean flows}

\begin{figure*}[htpb]
  \begin{center}
    \includegraphics[width=\textwidth]{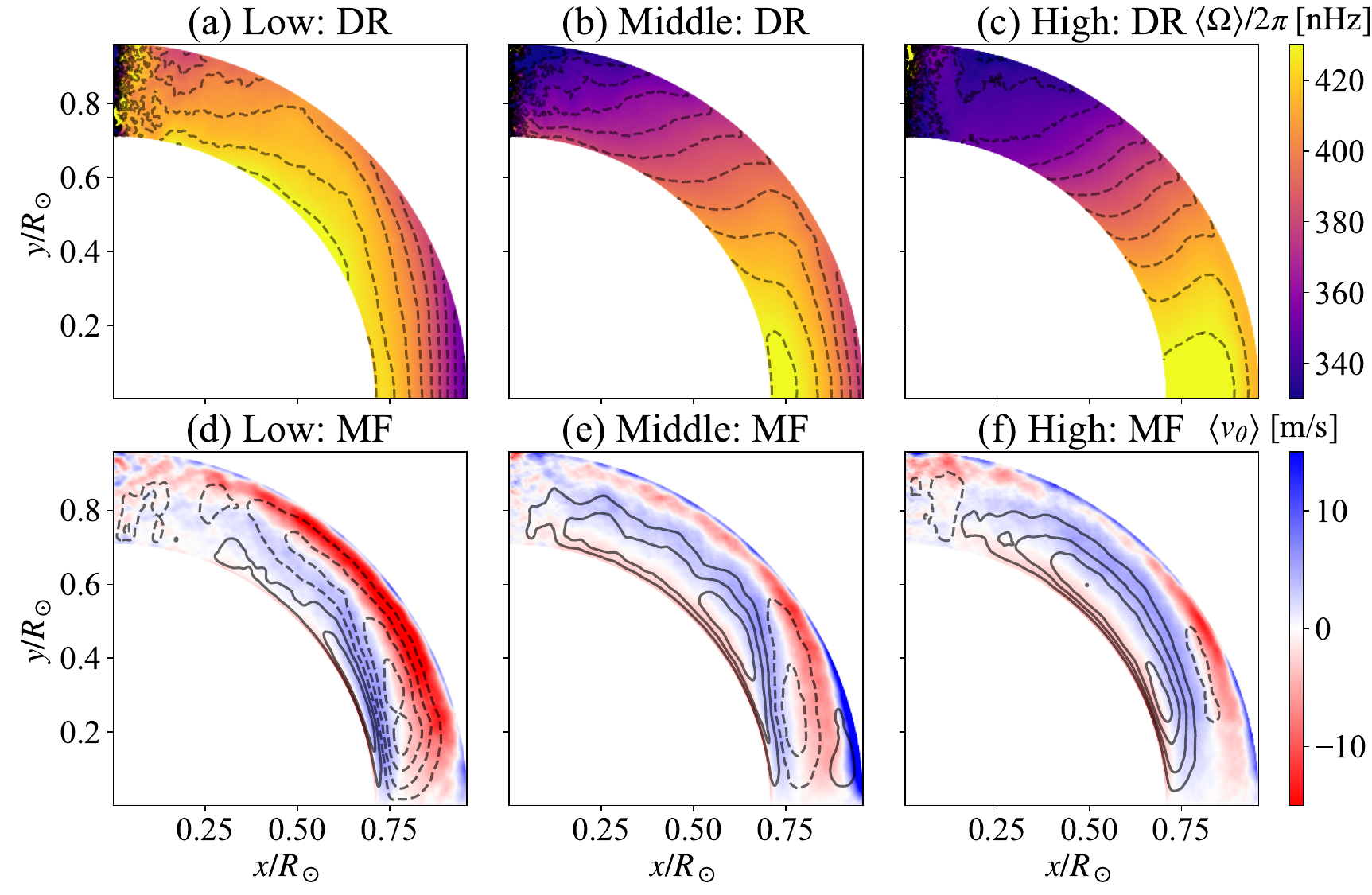}
    \caption{Differential rotation $\langle\Omega\rangle/2\pi$ (panels a, b, c) and meridional flow $\langle v_\theta \rangle$ (panels d, e, and f) in Low (panels a, d), Middle (panels b, e), and High (panels c, f) are shown. The black lines in the lower panels are stream lines of the mass flux $\rho_0\bm{v}_\mathrm{m}$, see Appendix \ref{sec:stream_function}). The solid and dashed lines indicate the clockwise and the counter clockwise flows, respectively.
    The solar-like differential rotation, i.e., the fast equator is reproduced in the High case (panel c).
    \label{mean_flow}}
  \end{center}
\end{figure*}

Fig.~\ref{mean_flow} shows the differential rotation $\langle\Omega\rangle/2\pi$ and the meridional flow $\langle\bm{v}_\mathrm{m}\rangle=\langle v_r\rangle \bm{e}_r+\langle v_\theta\rangle \bm{e}_\theta$. The angular velocity is defined as $\Omega=\Omega_0 + \Omega_1$ and $\Omega_1=v_\phi/(r\sin\theta)$. While the Low case shows the fast pole (panel a), we reproduce the fast equator in the High case as shown in HK21. The reason why we have the fast equator in the high-resolution calculation is discussed in Subsection \ref{sec:angular_momentum_transport}. Also, the differential rotation succeeds in avoiding the Taylor--Proudman constraint, i.e., $\partial \Omega/\partial z\neq 0 $, where $z$ is the direction of the rotational axis. This topic is discussed in Subsection \ref{sec:meridional_force_balance}. The meridional flow structure also depends on the resolution (Figs.~\ref{mean_flow}d, e, f). In the Low case, an anti-clockwise flow is dominant, and we can see a clear poleward flow around the surface. We can observe a tiny clockwise cell around the base of the convection zone. In the Middle case, the meridional flow is separated around the tangential cylinder of the base of the convection zone. Anti-clockwise and clockwise flow cells are seen in low and high latitudes, respectively. In the High case, a clockwise meridional flow is dominant throughout the convection zone. The poleward flow around the base of the convection zone is an essential feature for the fast equator (see Subsection \ref{sec:angular_momentum_transport}). The poleward meridional flow around the surface becomes weak in Middle and High cases. Note that we can recover clear poleward meridional flow when the top boundary is closer to the real solar surface \citep{Hotta_2015ApJ...798...51H}. In a high-resolution calculation, we have already checked this tendency and will introduce it in a future publication (Hotta, Kusano \& Sekii, in prep).


\subsection{Non-dimensional parameters}
\label{sec:non-dimensional}
In this subsection, we evaluate several non-dimensional parameters for comparisons with the previous studies. Since we do not use any explicit diffusivities (viscosity $\nu$, magnetic diffusivity $\eta$, and thermal conductivity $\kappa$), the effective diffusivities, $\nu_\mathrm{eff}$, $\eta_\mathrm{eff}$, and $\kappa_\mathrm{eff}$ need to be evaluated. The evaluation procedure is shown in Appendix \ref{sec:evaluation_viscosity}. We evaluate the effective viscosity from the kinetic energy spectra. Since we use the same numerical scheme for the magnetic field and the entropy as the velocity, the Prandtl number $\mathrm{Pr}=\nu_\mathrm{eff}/\kappa_\mathrm{eff}$, and the magnetic Prandtl number $\mathrm{Pm}=\nu_\mathrm{eff}/\eta_\mathrm{eff}$ are assumed to be unity. The mean RMS velocity $\overline{v}_\mathrm{RMS}$ is defined as
\begin{align}
  \overline{v}_\mathrm{RMS} = \int_{r_\mathrm{min}}^{r_\mathrm{max}} v_\mathrm{(RMS)} r^2dr/
  \int_{r_\mathrm{min}}^{r_\mathrm{max}} r^2 dr
\end{align}
The non-dimensional numbers are defined as
\begin{itemize}
  \item Reynolds number $(\mathrm{Re})$ \citep{Featherstone_2015ApJ...804...67F}
  \begin{align}
    \mathrm{Re} = \frac{\overline{v}_\mathrm{RMS}d}{\nu_\mathrm{eff}}
  \end{align}
  where $d=r_\mathrm{max}-r_\mathrm{min}$ is the radial extent of the computational domain.
  \item Rayleigh number $\mathrm{(Ra)}$ \citep{gastine_2014MNRAS.438L..76G}
  \begin{align}
    \mathrm{Ra} = \frac{g_0d^3\Delta s}{c_\mathrm{p0}\nu_\mathrm{eff}\kappa_\mathrm{eff}}
  \end{align}  
  where $g_0$ and $c_{p0}$ are the gravitational acceleration and the heat capacity at constant pressure at the outer boundary $r=r_\mathrm{max}$. $\Delta s = \mathrm{max}(\tilde{s}_1) - \mathrm{min}(\tilde{s}_1)$, where $\tilde{s}_1$ is the horizontally averaged entropy perturbation.
    \item flux Rayleigh number ($\mathrm{Ra_F}$) \citep{Featherstone_2016ApJ...818...32F}.
  \begin{align}
    \mathrm{Ra_F} = \frac{g_0 F_0 d^4}{c_\mathrm{p0}\rho_\mathrm{t}T_\mathrm{t}\nu_\mathrm{eff}\kappa_\mathrm{eff}}
  \end{align}
  where $F_0$, $\rho_t$, and $T_\mathrm{t}$ are the energy flux density, the background density $\rho_0$, and the background temperature $T_0$ at $r=r_\mathrm{max}$, respectively.
    \item Ekman number $\mathrm{(Ek)}$ \citep{gastine_2014MNRAS.438L..76G}
  \begin{align}
    \mathrm{Ek} = \frac{\nu_\mathrm{eff}}{\Omega_0 d^2}
  \end{align}
  \item Convective Rossby number $\mathrm{Ro_c}$ \citep{gastine_2014MNRAS.438L..76G}
    \begin{align}
    \mathrm{Ro_c} = \sqrt{\frac{\mathrm{RaEk}^2}{\mathrm{Pr}}}
  \end{align}    
    \item Rossby number $\mathrm{(Ro)}$ \citep{Featherstone_2015ApJ...804...67F}
  \begin{align}
    \mathrm{Ro} = \frac{\overline{v}_\mathrm{RMS}}{2\Omega_0 d}
  \end{align}
  \item local Rossby number $\mathrm{(Ro_\ell)}$ \citep{christensen_2006GeoJI.166...97C}
  \begin{align}
    \mathrm{Ro_\ell} = \frac{\overline{\ell}_\mathrm{u}\overline{v}_\mathrm{RMS}}{\pi\Omega_0 d},
  \end{align}
  where $\overline{\ell}_\mathrm{u}$ is the mean spherical harmonic degree defined as
  \begin{align}
    \overline{\ell}_\mathrm{u} = \int_0^{\ell_\mathrm{max}} \ell \widehat{E}_\mathrm{h}(\ell) d\ell / \int_0^{\ell_\mathrm{max}} \widehat{E}_\mathrm{h}(\ell) d\ell.
  \end{align}
  $\widehat{E}_\mathrm{h}$ is the kinetic energy spectra of the horizontal velocity. We evaluate it at $r=0.83R_\odot$.
\end{itemize}
All these values are summarized in Table \ref{summary}. Thanks to a large number of grid points and the slope-limited artificial viscosity, the effective diffusivities are significantly reduced, and the Reynolds and Rayleigh numbers reach huge values. Convective and ordinary Rossby numbers $\mathrm{Ro_c}$ and $\mathrm{Ro}$ are typical values for the solar simulations \citep{Featherstone_2015ApJ...804...67F,mabuchi_2015ApJ...806...10M}. These are values for the transition between the fast equator and the fast pole. In these values, the effect of small-scale turbulence realized by the high resolution is not considered. The local Rossby number using the kinetic energy spectra is a way to consider the small-scale turbulence \citep{christensen_2006GeoJI.166...97C}. \cite{gastine_2014MNRAS.438L..76G} suggests that the transition between the fast equator and the fast pole occurs around $\mathrm{Ro}_\ell\sim1$. Our local Rossby numbers are much larger than the critical value due to the small-scale turbulence. This result indicates that only with the angular momentum transport by the Reynolds stress, i.e., the turbulence, we cannot maintain the solar-like differential rotation. This issue is again discussed in Section \ref{sec:summary}.

\subsection{Magnetic field generation}
\label{sec:magnetic_field_generation}

\begin{figure*}[htpb]
  \begin{center}
    \includegraphics[width=\textwidth]{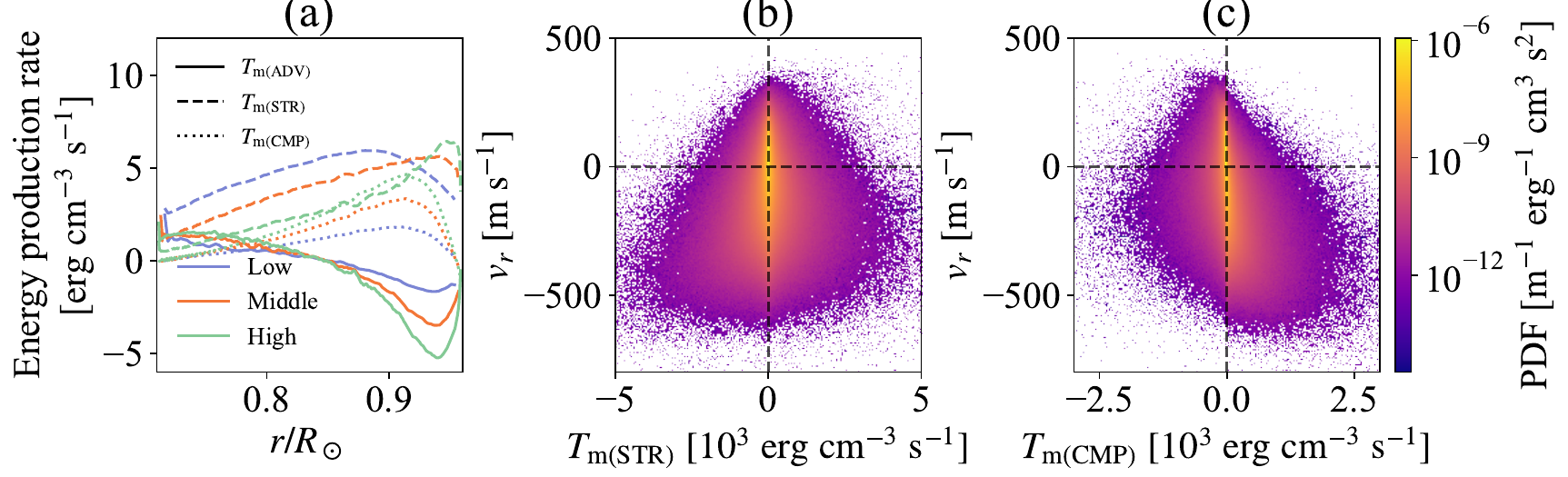}
    \caption{Magnetic field generation process is discussed. (a) Horizontally (spherically) averaged magnetic energy production rate is shown. The solid, dashed, and dotted lines indicate the magnetic energy production by the advection $T_\mathrm{m(ADV)}$, stretching $T_\mathrm{m(STR)}$, and compression $T_\mathrm{m(CMP)}$, respectively. The definition of each term is shown in eq.~(\ref{eq:energy_transfer}). Panels b and c show PDFs for $v_r$ vs. $T_{\mathrm{m(STR)}}$ and $v_r$ and $T_\mathrm{m(CMP)}$ at $r=0.9R_\odot$, respectively. The results for panels b and c are obtained from the High case.
    Panel a shows that the magnetic energy production by the stretching $T_\mathrm{m(STR)}$ is mostly reduced with increasing the resolution, while the compression $T_\mathrm{m(CMP)}$ increases. Large fraction of the stretching $T_\mathrm{m(STR)}$ has negative in the down flow region $v_r<0$ (panel b), while the compression is mostly positive there (pane c). 
    \label{magnetic_energy_transfer}}
  \end{center}
\end{figure*}

\begin{figure*}[htpb]
  \begin{center}
    \includegraphics[width=\textwidth]{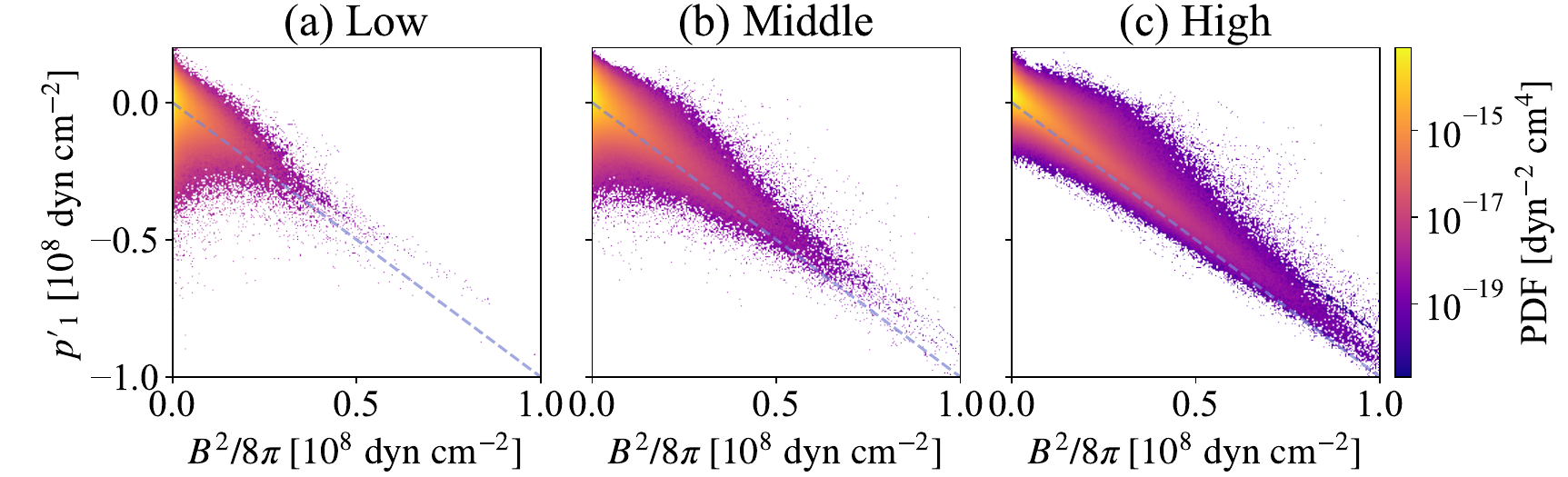}
    \caption{Probability density functions (PDFs) for $p\prime_1$ vs. $B^2/8\pi$ at $r=0.9R_\odot$ are shown. Panels a, b, and c are results from the Low, Middle, and High cases, respectively. The black dashed lines indicate $p\prime_1=-B^2/8\pi$.
    The strong magnetic field achieved in the High cases is primarily located on the dashed line. This result indicates that the strong magnetic field is maintained by the gas pressure.
    \label{PDF_prpm}}
  \end{center}
\end{figure*}

\begin{figure}[htpb]
  \begin{center}
    \includegraphics[width=0.45\textwidth]{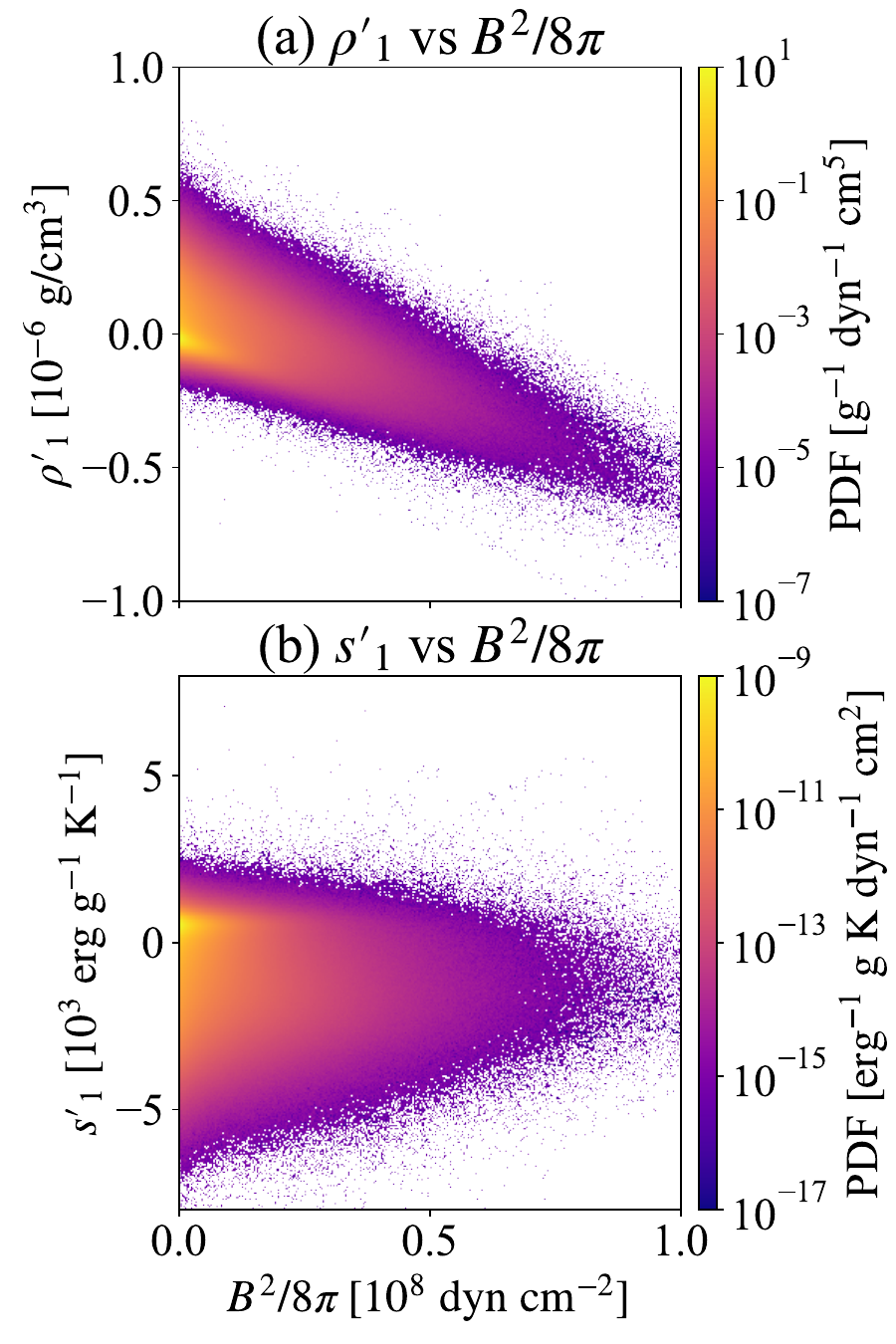}
    \caption{PDFs for $\rho\prime_1$ vs. $B^2/8\pi$ (panel a) and $s\prime_1$ vs. $B^2/8\pi$ (panel b) at $r=0.9R_\odot$ from the High case, respectively.
      The density fluctuation well correlates with the magnetic field strength (panel a), while the entropy fluctuation does not (panel b).
    \label{PDF_ropm_sepm}}
  \end{center}
\end{figure}

\begin{figure*}[htpb]
  \begin{center}
    \includegraphics[width=\textwidth]{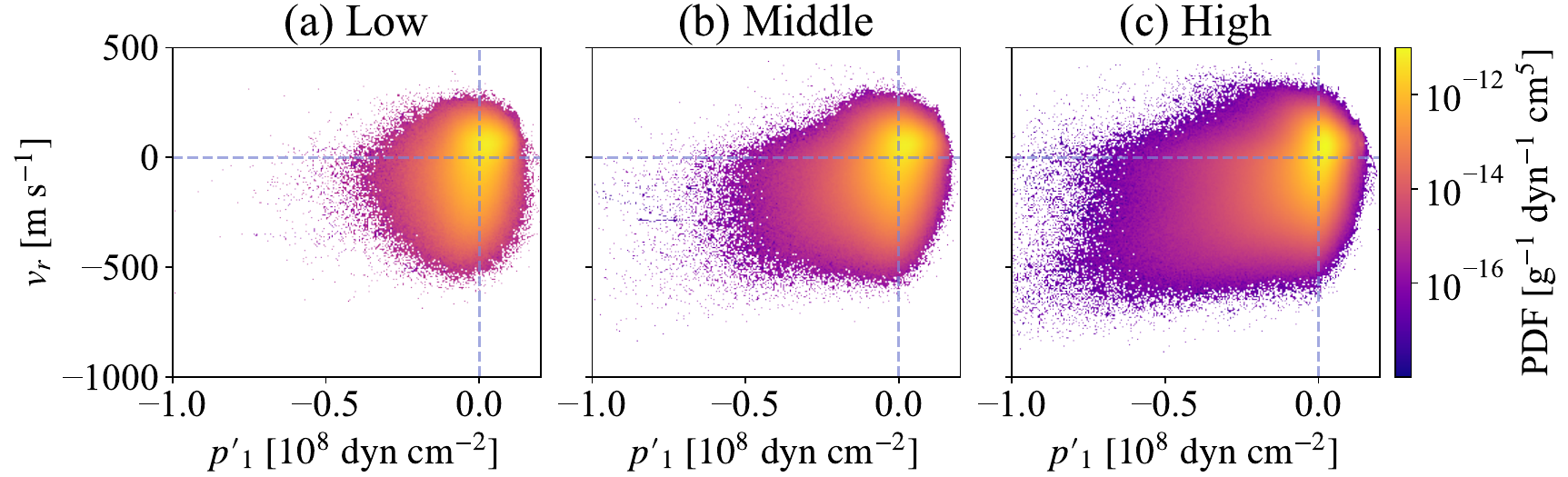}
    \caption{PDFs for $v_r$ vs. $p'_1$ at $r=0.9R_\odot$. Panels a, b, and c show the results from the Low, Middle, and High cases.
    The large gas pressure perturbation $p'_1$ in the High case is observed in downflow regions $v_r<0$.
    \label{PDF_vxpr}}
  \end{center}
\end{figure*}

\begin{figure}[htpb]
  \begin{center}
    \includegraphics[width=0.5\textwidth]{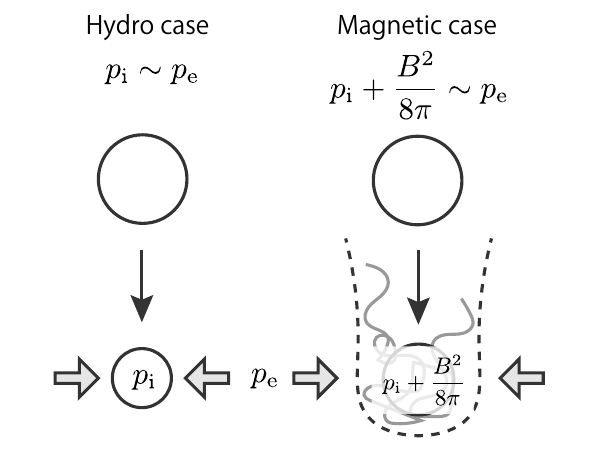}
    \caption{Explanation of compression mechanism to generate the strong magnetic field in thermal convection. The left panel shows the process in a hydrodynamic case without the magnetic field. The right panel shows a case with the magnetic field. The circle indicates the fluid parcel. The gray line in the right panel is a magnetic field line. The gray arrow indicates the gas pressure from an external fluid.
    When the magnetic field is absent, the gas pressure inside the downflow fluid parcel $p_\mathrm{i}$ is finally balanced with the external gas pressure $p_\mathrm{e}$. When the magnetic field exists, the external gas pressure is balanced with the internal gas pressure and the magnetic pressure, i.e., $p_\mathrm{i}+B^2/8\pi=p_\mathrm{e}$.
    \label{compression_exp}}
  \end{center}  
\end{figure}

\begin{figure}[htpb]
  \begin{center}
    \includegraphics[width=0.5\textwidth]{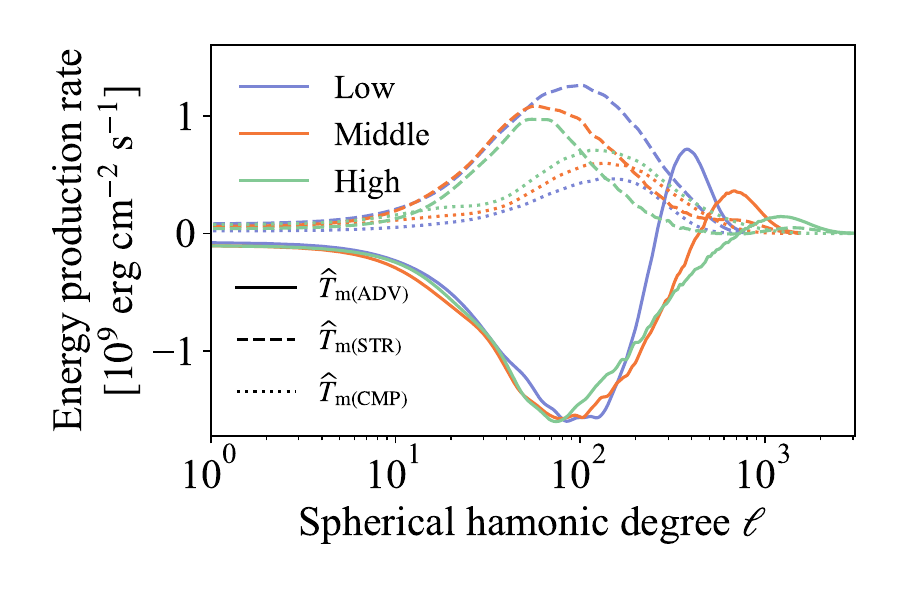}
    \caption{Spectra of the magnetic energy production rate at $r=0.9R_\odot$ are shown.
    The stretching $\widehat{T}_\mathrm{m(STR)}$ in the small scale is reduced in High case, while the compression $\widehat{T}_\mathrm{m(CMP)}$ increases.
    \label{magnetic_production_transfer}}
  \end{center}
\end{figure}

In this subsection, we discuss the generation mechanism of the magnetic field, especially the superequiparition magnetic field achieved in the High case (Fig.~\ref{rms_velocity}), i.e., the magnetic energy is larger than the kinetic energy.
We analyze the magnetic energy equation to investigate the mechanism. The equation is as follows.
\begin{align}
  \frac{\partial}{\partial t}\left(\frac{B^2}{8\pi}\right)
  =&
  \underbrace{
  -\frac{\bm{B}}{4\pi}\cdot\left[\left(\bm{v}\cdot\nabla\right)\bm{B}\right]
  }_{T_\mathrm{m(ADV)}} \nonumber\\
  &\underbrace{
  +\frac{\bm{B}}{4\pi}\cdot\left[\left(\bm{B}\cdot\nabla\right)\bm{v}\right]
  }_{T_\mathrm{m(STR)}} \nonumber\\
  &\underbrace{
  -\frac{B^2}{8\pi}\left(\nabla\cdot\bm{v}\right)
  }_{T_\mathrm{m(CMP)}} \label{eq:energy_transfer}
\end{align}
There are three contributions to change the magnetic energy, which are advection $T_\mathrm{m(ADV)}$, stretching $T_\mathrm{m(STR)}$, and compression $T_\mathrm{m(CMP)}$. Fig.~\ref{magnetic_energy_transfer}a shows the spherically averaged terms in eq.~(\ref{eq:energy_transfer}). The solid lines indicate the contribution from the advection $T_\mathrm{m(ADV)}$. Around the top boundary, the strong magnetic field is concentrated in the downflow region, and the magnetic energy is transported downward. As a result, the advection contribution is negative and positive in the near-surface layer and the deep convection zone, respectively. Because the higher resolution shows a stronger magnetic field, this effect increases with increased resolution. Next, the dashed lines are the contribution by the stretching term $T_\mathrm{m(STR)}$. In most of the convection zone, the amplitude decreases in the higher resolutions. Because the magnetic field increases, the Lorentz feedback is amplified, and the production rate of the magnetic field decreases. Fig.~\ref{magnetic_energy_transfer}b shows the PDF between the radial velocity $v_r$ and the stretching term $T_\mathrm{m(STR)}$. The result indicates that the main contribution of the stretching occurs at the downflows ($v_r<0$). While the net contribution of the stretching is positive, we can see a significant negative contribution (energy transfer from magnetic to kinetic energies) in the downflow region. 
The dependence of the stretching $T_\mathrm{m(STR)}$ on the resolution (Fig. \ref{magnetic_energy_transfer}a) indicates that the stretching is not responsible for the superequipartition magnetic fields in the High case. Finally, we discuss the compression term $T_\mathrm{m(CMP)}$ shown with the dotted lines in Fig.~\ref{magnetic_energy_transfer}a. The amplitude of the compression term monotonically increases with increased resolution. Also, Fig.~\ref{magnetic_energy_transfer}c shows the PDF between the radial velocity $v_r$ and the compression term $T_\mathrm{m(CMP)}$. Similar to the stretching, the important compression occurs at the downflow region, but the negative contribution of $T_\mathrm{m(CMP)}$ is not significant. The reason why the fluid can overcome and compress the strong magnetic field to amplify the field strength is shown in Fig.~\ref{PDF_prpm}. PDFs between the magnetic pressure (energy) and perturbation gas pressure are shown. The perturbation gas pressure is defined as
\begin{align}
  p\prime_1 = p_1 - \langle p_1\rangle,
\end{align}
and is the deviation from the longitudinal average. 
When we define the gas pressure inside and outside the strong magnetic field as $p_\mathrm{i}$ and $p_\mathrm{e}$, respectively. The perturbation gas pressure defined here is approximated as $p\prime_1 \sim p_\mathrm{i}-p_\mathrm{e}$.
The black dashed line in Fig.~\ref{PDF_prpm} indicates $p\prime_1 = -B^2/8\pi$, i.e., the magnetic pressure is balanced with the gas pressure. In the Low case (Fig.~\ref{PDF_prpm}a), the PDF distributes rather uniformly. Even the small magnetic energy ($\sim 10^8~\mathrm{dyn~cm^{-2}}$) has large perturbation gas pressure ($<-4\times10^7~\mathrm{dyn~cm^{-2}}$). In the High case (Fig.~\ref{PDF_prpm}), the magnetic field strength is amplified, and most of the strong magnetic field distributes on the $p\prime_1 = -B^2/8\pi$ line. Also, the region with a weak magnetic field and the low gas pressure disappears. These results indicate that the gas pressure maintains the superequipartition magnetic field realized in the High case. Because the solar interior is in a low Mach number situation, the internal energy is huge compared with the kinetic and magnetic energies. 
If the magnetic field is balanced with the gas pressure (internal energy), the magnetic field strength can be superequipartition to the kinetic energy.
The region with the weak magnetic field and the low gas pressure disappears in the High case, indicating that the dynamo is efficient enough to amplify all the small-scale fields once the magnetic field enters the low gas pressure region.\par
This process where the internal energy amplifies the magnetic field is similar to the explosion process \citep{Moreno-Insertis_1995ApJ...452..894M,Rempel_2001ApJ...552L.171R,Hotta_2012ApJ...759L..24H}. In the explosion process, the rising motion in the superadiabatic stratification leads to an entropy difference between the inside and outside of the flux tube. Fig.~\ref{PDF_ropm_sepm} shows the PDFs between (a) perturbation density and magnetic pressure and (b) perturbation entropy and the magnetic pressure. While the density well correlates with the magnetic pressure, the entropy does not. This result indicates that the entropy does not contribute to the amplification and that the process achieved in this study is different from the explosion process. \par
Also, as shown in \cite{Hotta_2015ApJ...803...42H}, the absolute amplitude of entropy perturbation increases with increased resolution. This increase also occurs in this study (see Subsection \ref{sec:convection_driving}), and this tends to lower the gas pressure in the downflow region ($s_1<0$) because the linearized equation of state is expressed as
\begin{align}
  \frac{p_1}{p_0} = \gamma\frac{\rho_1}{\rho_0} + \frac{s_1}{c_\mathrm{v}}.
\end{align}
Again, Fig.~\ref{PDF_ropm_sepm}b shows that the correlation between the entropy perturbation and the magnetic pressure is not good, and this fact indicates that the increase of the entropy perturbation does not contribute to amplifying the magnetic field.\par
We also investigate the location of the strong magnetic field amplification. Fig.~\ref{PDF_vxpr} shows the PDF between the gas pressure perturbation $p\prime_1$ and the radial velocity $v_r$. When we compare the Low (panel a) and High (panel c) cases, the low gas pressure region appears, especially at the downflow region. Considering the result shown, we can draw an overall picture of the amplification process of the superequipartition magnetic field. A schematic picture is shown in Fig.~\ref{compression_exp}. In a hydrodynamic case without the magnetic field (left panel), when a fluid parcel in the upper layer descends, the parcel has low pressure compared with the external fluid in the lower layer because of the stratification. This pressure imbalance is instantaneously relaxed by the sound wave. In a magnetic case, especially with an efficient small-scale dynamo like the High case, the situation changes. When a fluid parcel goes down to the lower layer, the small-scale magnetic field is involved. 
The low gas pressure inside the magnetic field $p_\mathrm{i}$ and the magnetic pressure $B^2/8\pi$ are balanced with the external gas pressure $p_\mathrm{e}$, i.e.,  $p_\mathrm{i} + B^2/8\pi = p_\mathrm{e}$.
Thus, the magnetic energy is amplified by the compression, i.e., maintained by the internal energy.
\par
We also discuss the spatial scale of magnetic field amplification. The spectral magnetic energy is expressed by 
\begin{align}
  \widehat{E}_\mathrm{mag}(\ell) = \frac{1}{8\pi}\widehat{\bm{B}}(\ell)\cdot\widehat{\bm{B}}^* (\ell),
\end{align}
where $\widehat{}$ and $^*$ denote the spherical harmonic transform and the complex conjugate, respectively. Then, the time evolution of $\widehat{E}_\mathrm{mag}(\ell)$ can be written as \citep[see details in][]{Pietarila_Graham_2010ApJ...714.1606P,Rempel_2014ApJ...789..132R},

\begin{align}
  \frac{\partial}{\partial t} \widehat{E}_\mathrm{mag}(\ell)
  = \widehat{T}_\mathrm{m(STR)} + \widehat{T}_\mathrm{m(ADV)} + \widehat{T}_\mathrm{m(CMP)}, 
\end{align}
where 
\begin{align}
  \widehat{T}_\mathrm{m(STR)} &= \frac{1}{8\pi}\widehat{\bm{B}}\cdot
  \widehat{\left(\bm{B}\cdot\nabla\right)\bm{v}}^* + c.c., \\
  \widehat{T}_\mathrm{m(ADV)} &= - \frac{1}{8\pi}\widehat{\bm{B}}\cdot
  \widehat{\left(\bm{v}\cdot\nabla\right)\bm{B}}^* + c.c.,\\
  \widehat{T}_\mathrm{m(CMP)} &= - \frac{1}{8\pi}\widehat{\bm{B}}\cdot
  \widehat{\left(\bm{B}\nabla\cdot\bm{v}\right)}^* + c.c.,
\end{align}
where $c.c.$ indicates the complex conjugate expression. Each term in the spectral magnetic energy evolution at $r=0.9R_\odot$ is shown in Fig.~\ref{magnetic_production_transfer}. The magnetic energy transfer by the advection $\widehat{T}_\mathrm{m(ADV)}$ does not depend on the resolution in a middle ($\ell\sim10^2$) to large scale ($\ell\sim1$). The advection term contribution $\widehat{T}_\mathrm{m(ADV)}$ is typically negative because the downward magnetic energy transport is dominant at this height. Around the smallest scale in each resolution, $\widehat{T}_\mathrm{m(ADV)}$ is positive. The dominant magnetic energy production source is the stretching $\widehat{T}_\mathrm{m(STR)}$, but the production rate decreases with increased resolution, especially at the small-scale because the magnetic field strength and the resulting Lorentz feedback increase. Meanwhile, the compression contribution $\widehat{T}_\mathrm{m(CMP)}$ increases with the resolution at middle scale ($\ell\sim 100$). The peak scale of the compression does not depend on the resolution. This result also supports our presented explanation of the amplification mechanism of the strong magnetic field. A complex small-scale magnetic field is concentrated at the downflow region. The field is strong enough to suppress the turbulent stretching, but the compression can still work. 

\subsection{Convection driving}
\label{sec:convection_driving}

\begin{figure}[htpb]
  \begin{center}
    \includegraphics[width=0.5\textwidth]{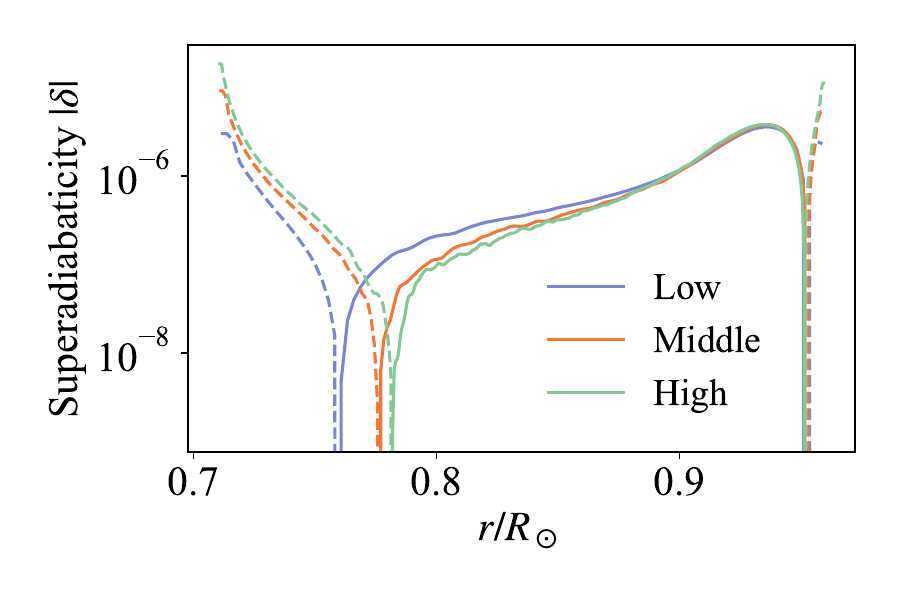}
    \caption{Superadiabaticity $|\delta|$ is shown. The solid and dashed lines indicate the positive and negative values of $\delta$, respectively. 
      Subadiabatic layer $(\delta <0)$ is extended with increasing the resolution.
    \label{superadiabaticity}}
  \end{center}
\end{figure}

\begin{figure}[htpb]
  \begin{center}
    \includegraphics[width=0.5\textwidth]{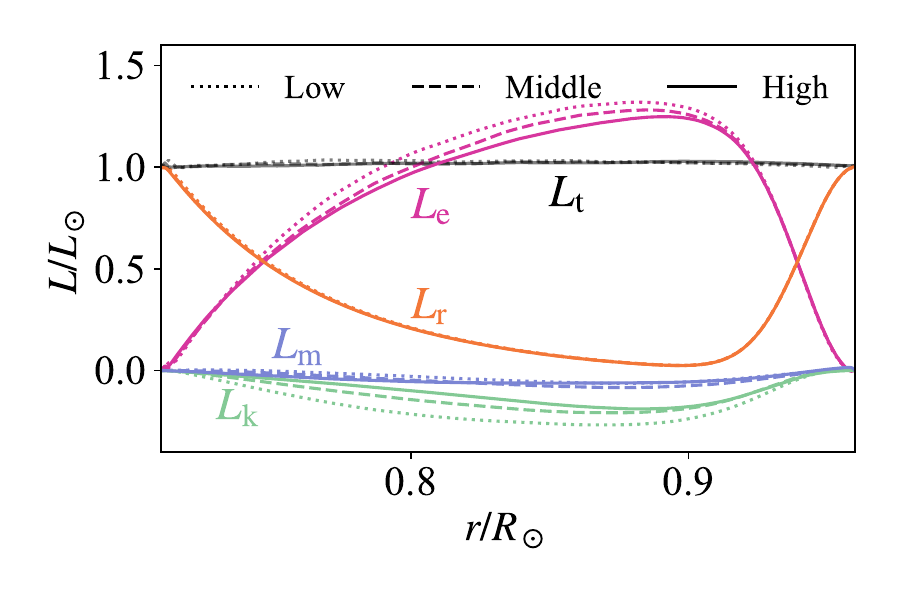}
    \caption{The enthalpy (magenta), radiative (orange), kinetic (green), and Poynting (blue) fluxes are shown. The solid, dashed, and dotted lines are the results from High, Middle, and Low cases, respectively.\label{1D_flux}}
  \end{center}
\end{figure}

\begin{figure}[htpb]
  \begin{center}
    \includegraphics[width=0.4\textwidth]{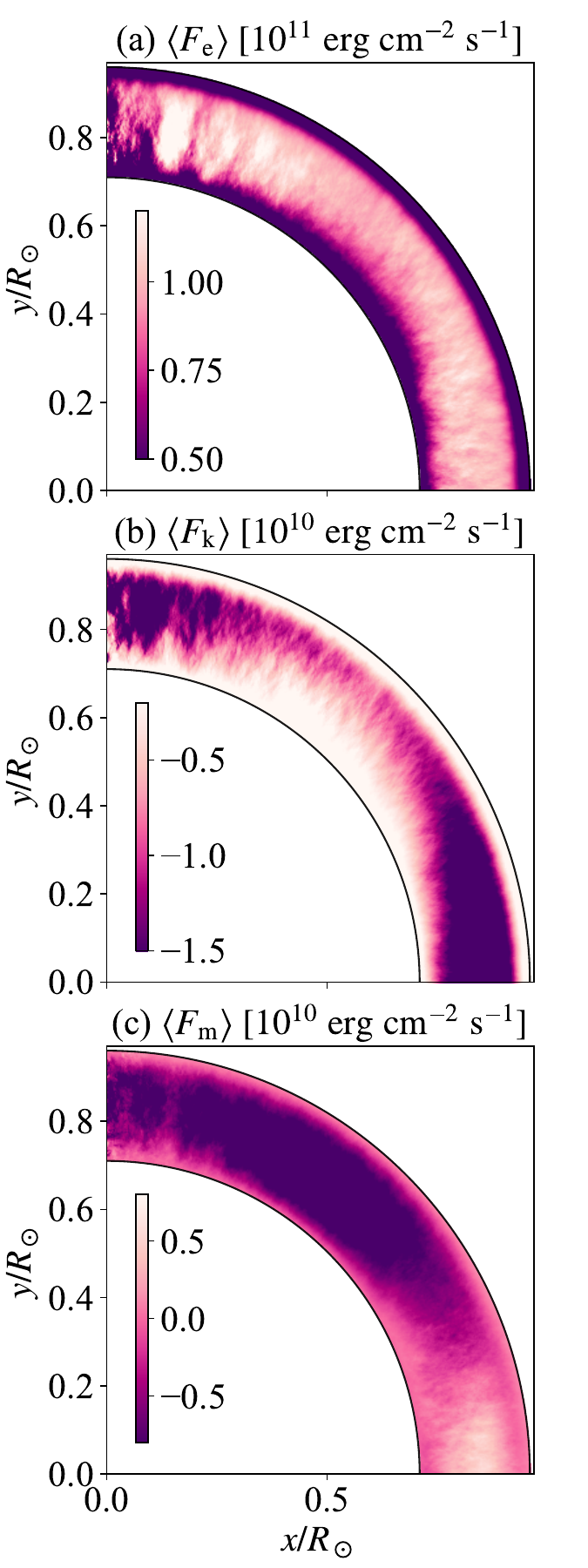}
    \caption{
      Latitudinal dependence of the energy fluxes in the High case are shown. Note that we adjust the color bar to emphasize the latitudinal dependence. Panels a, b and c show the range of $5\times10^{10}<\langle F_\mathrm{e}\rangle <1.5\times10^{11}~\mathrm{erg~cm^{-2}~s^{-1}}$, $-1.5\times10^{10}<\langle F_\mathrm{k}\rangle <-2\times10^{9}~\mathrm{erg~cm^{-2}~s^{-1}}$, and $-8\times10^{9}<\langle F_\mathrm{m}\rangle <8\times10^{9}~\mathrm{erg~cm^{-2}~s^{-1}}$, respectively.
    Panels a, b, and c show the enthalpy, kinetic, and Poynting fluxes, respectively.\label{2D_flux}}
  \end{center}
\end{figure}

\begin{figure*}[htpb]
  \begin{center}
    \includegraphics[width=0.8\textwidth]{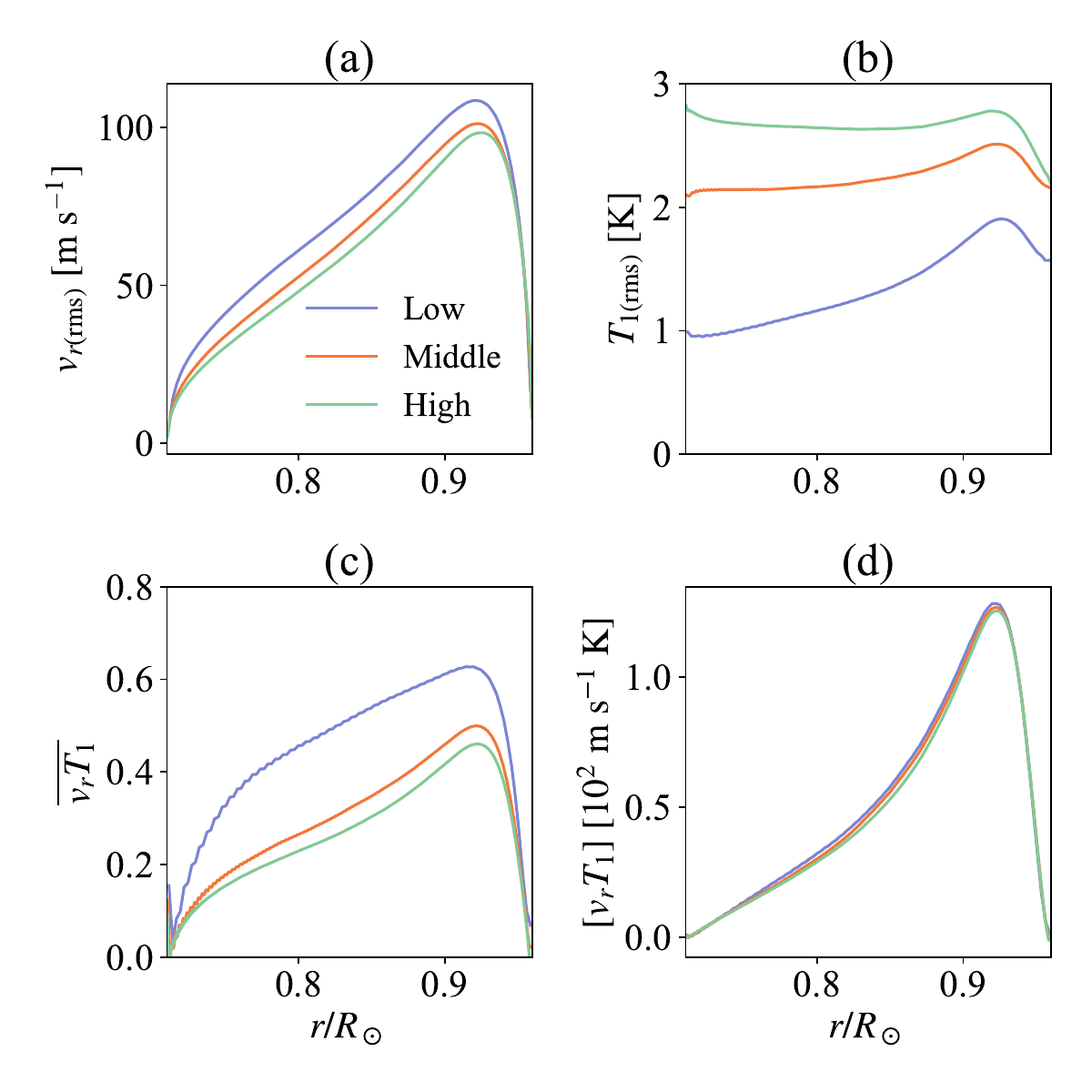}
    \caption{Each panel shows (a) RMS radial velocity, (b) RMS temperature perturbation, (c) correlation between $v_r$ and $T_1$ (see eq.~(\ref{eq:spherical_correlation})), and (d) normalized correlation between $v_r$ and $T_1$ (see eq.~(\ref{eq:spherical_normalized_correlation})).
      The convection velocity and the $v_r$ vs $T_1$ normalized correlation decreases, while the temperature perturbation increases. This balance maintains almost the same enthalpy flux between cases.
    \label{flux_ana}}
  \end{center}
\end{figure*}

\begin{figure*}[htpb]
  \begin{center}
    \includegraphics[width=0.8\textwidth]{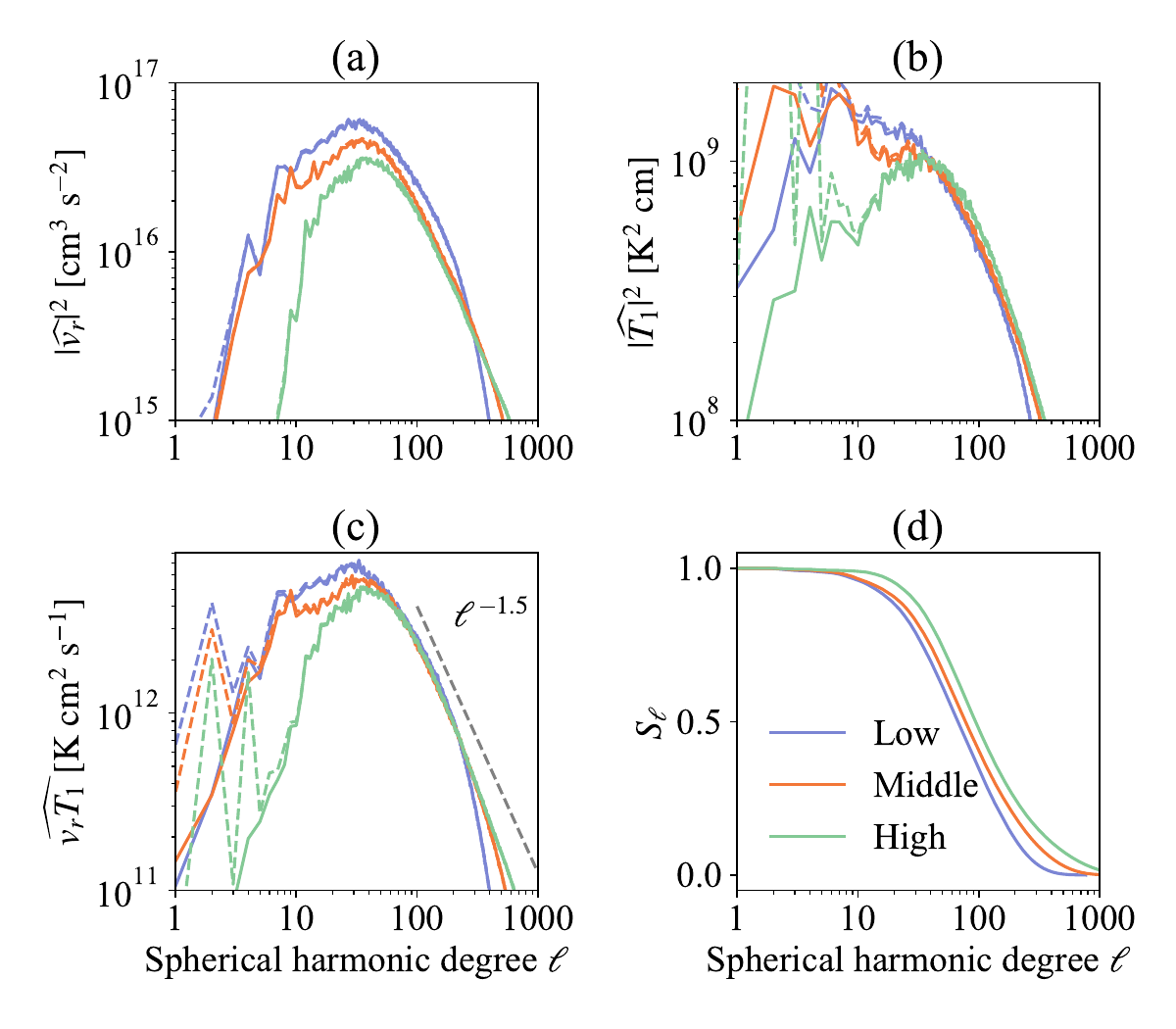}
    \caption{Panels show the spectra of  (a) radial velocity, (b) temperature perturbation, (c) correlation between radial velocity $v_r$ and temperature $T_1$. Panel d shows normalized summed correlation $S_\mathrm{\ell}$ defined at eq.~(\ref{eq:normalized_summed_correlation}). All the data are calculated at $r=0.9R_\odot$. The dashed lines indicate spectra including $m=0$ mode.
    Large fraction of the energy is transported in the small scale in the High case.
    \label{energy_flux_spectra}}
  \end{center}
\end{figure*}

\begin{figure}[htpb]
  \begin{center}
    \includegraphics[width=0.5\textwidth]{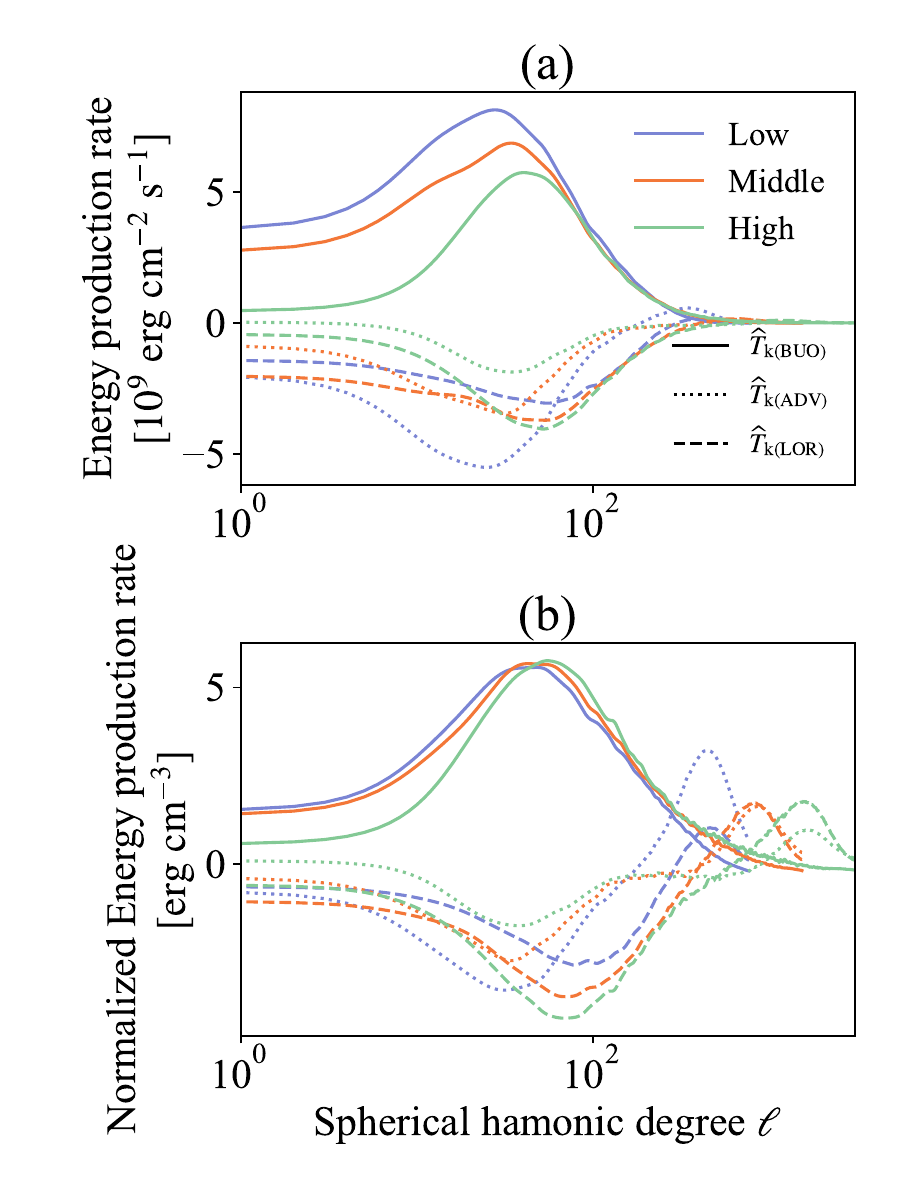}
    \caption{Spectra of the kinetic energy production rate at $r=0.9R_\odot$ are shown. Panel b shows the values shown in panel a normalized with
      ${\sqrt{\widehat{v_r}\widehat{v_r}^*/r}}$.
      The large-scale buoyancy $\widehat{T}_\mathrm{k(BUO)}$ and the Lorentz force $\widehat{T}_\mathrm{k(LOR)}$ are reduced in the High case. This result indicates that the suppression of the kinetic energy on the large scale in the High case is not directly caused by the Lorentz force.
    \label{kinetic_energy_transfer}}
  \end{center}
\end{figure}

In this subsection, we discuss the driving mechanism of the thermal convection. In particular, the mechanism in which the large-scale convection is suppressed in the High case is discussed.\par
For the discussion in this subsection, we additionally define statistical values, spherical average $\tilde{Q}$, spherical RMS $Q_\mathrm{(rms)}$, spherical correlation $[Q_1Q_2]$, and normalized spherical correlation $\overline{Q_1Q_1}$ as follows.
\begin{align}
  \widetilde{Q}(r) &= \frac{1}{4\pi}\int_S Q dS \label{eq:spharical_average}\\
  Q_\mathrm{(rms)}(r) &=\sqrt{ \frac{1}{4\pi}\int_S \left(Q-\widetilde{Q}\right)^2dS } \label{eq:sphrical_rms}\\
  \left[Q_1Q_2\right](r) &= \frac{1}{4\pi}\int_S Q_1Q_2 dS \label{eq:spherical_correlation}\\
  \overline{{Q_1Q_2}}(r) &= \frac{\left[Q_1Q_2\right]}{Q_{1\mathrm{(rms)}}Q_{2\mathrm{(rms)}}}
  \label{eq:spherical_normalized_correlation}
\end{align}
We note that the spherical RMS $Q_\mathrm{(rms)}$ defined in eq.~(\ref{eq:sphrical_rms}) is different from the longitudinal RMS $Q_\mathrm{(RMS)}$ defined in eq.~(\ref{eq:longitudinal_rms}).
Fig.~\ref{superadiabaticity} shows the superadiabaticity $\delta$ in different cases. The superadiabaticity is defined such as:
\begin{align}
  \delta = - \frac{H_p}{c_p}\frac{d\tilde{s}}{dr}.
\end{align}

We observe a thermal convectively stable region ($\delta<0$) in all cases. This layer is common in an effectively high Prandtl number convection \citep{Hotta_2017ApJ...843...52H,Bekki_2017ApJ...851...74B,Kapyla_2019A&A...631A.122K}.
The effective high Prandtl number is achieved with the strong small-scale magnetic field. In a high Prandtl number regime, the thermal structure does not diffuse, and low entropy material is accumulated at the base of the convection zone.
\cite{brandenburg_2016ApJ...832....6B} also shows that a non-local convection can cause this type of subadiabatic layer in his analytical model.
This process results in the convectively stable region ($\delta<0$). \cite{Bekki_2017ApJ...851...74B} shows that when the stable region is achieved around the base of the convection zone, the large-scale flow is suppressed because the convection driving scale in the deeper layer is larger because of the large pressure/density scale height. 
Because the stable region expands and the absolute value of superadiabaticity $|\delta|$ increases with the resolution, this effect should contribute to suppressing the large-scale convection. The difference of the superadiabaticity, however, between the Low and Middle cases is larger than that between the Middle and High cases, while the large-scale flow is significantly suppressed only in the High case. This indicates that the main reason for the large-scale suppression is not the change of the superadiabaticity.\par
The basic value that determines the convection velocity is the energy flux. In the solar convection zone, the energy flux is fixed by the efficiency of the nuclear fusion. Fig.~\ref{1D_flux} shows different types of fluxes. Definitions of the enthalpy $F_\mathrm{e}$, kinetic $F_\mathrm{k}$, Poynting $F_\mathrm{m}$, radiative $F_\mathrm{r}$, and total $F_\mathrm{\mathrm{t}}$ flux densities are \citep[see][]{Hotta_2014ApJ...786...24H}
\begin{align}
  F_\mathrm{e} =& \left(e_1 + \frac{p_1}{\rho_0} - \frac{p_0}{\rho_0^2}\rho_1 \right) \rho v_r, \\
  F_\mathrm{k} =& \frac{1}{2}\rho v^2 v_r, \\
  F_\mathrm{m} =& \  \frac{1}{4\pi}\left[\left(B_\theta^2 + B_\phi^2\right)v_r
   -\left(v_\theta B_\theta + v_\phi B_\phi\right)B_r \right],\\
  F_\mathrm{r} =& F_\mathrm{rad} + F_\mathrm{art},\\
  F_\mathrm{t} =& F_\mathrm{e} + F_\mathrm{k} + F_\mathrm{m} + F_\mathrm{r},
\end{align}
where $e$ is the internal energy
calculated with the OPAL repository.
$F_\mathrm{rad}$ and $F_\mathrm{art}$ are defined at eqs.~(\ref{eq:frad}) and (\ref{eq:fart}), respectively. For convenience, the sum of the physics-based radiation flux density $F_\mathrm{rad}$ and an artificial energy flux density $F_\mathrm{art}$ is called the radiative flux density $F_\mathrm{r}$ in this study. \par
In Fig.~\ref{1D_flux}, we integrate the flux densities over the full sphere and evaluate each corresponding luminosity (flux). The enthalpy flux $L_\mathrm{e}$ (magenta) slightly decreases with increased resolution. The decrease is more moderate than expected from the convection velocity suppression (Fig.~\ref{rms_velocity}). In the mixing length theory, the enthalpy flux scales as $L_\mathrm{e}\propto v^3_\mathrm{c}$, and the suppression of the convection velocity $v_\mathrm{c}$ should have a strong influence on the enthalpy flux. This deviation from the mixing length theory is essential to investigate the suppression mechanism of the convection velocity. The slight decrease of the enthalpy flux can be compensated for by the kinetic flux. As is usual, the kinetic flux is negative because the downflow has larger kinetic energy. This is reduced because of the convection velocity suppression. The Poynting flux has minor contributions, but the flux has a negative value. This downward Poynting flux is also caused because the downflow region has larger magnetic energy.
Fig.~\ref{2D_flux} shows two-dimensional energy flux density distribution in the High case 
The enthalpy flux density does not show a significant dependence on the latitude (Fig. \ref{2D_flux}a). 
$\langle F_\mathrm{k}\rangle$ is always negative in all latitudes. The inward kinetic energy flux is most effective at the equator and the poles (Fig. \ref{2D_flux}b).
Latitudinal variation is most prominent in the Poynting flux (Fig. \ref{2D_flux}c). $\langle F_\mathrm{m}\rangle$ is positive and negative at the low and high latitudes, respectively.
\par
In this paragraph, we discuss why the energy flux (especially the enthalpy flux) is maintained even with the suppressed convection velocity. With the equation of state for the perfect gas, the enthalpy flux density can be expressed as
\begin{align}
  F_\mathrm{e} \sim \rho_0 c_p \left[v_r T_1\right].
\end{align}
Because the background density, $\rho_0$, and heat capacity at constant pressure, $c_p$, do not change in a low Mach number situation, the correlation between the radial velocity, $v_r$, and the temperature perturbation, $T_1$, determines the enthalpy flux. Fig.~\ref{flux_ana} shows analysis to this end. Fig.~\ref{flux_ana}a shows the spherical RMS for the radial velocity $v_{r\mathrm{(rms)}}$. As discussed, the convection velocity is suppressed. Fig.~\ref{flux_ana}b shows the spherical RMS for the temperature perturbation $T_{1\mathrm{(rms)}}$. $T_{1\mathrm{(rms)}}$ increases with increased resolution. The magnetic field is amplified in higher-resolution simulations that suppress the mixing between the up and downflow. This process increases the temperature perturbation \citep[see also][]{Hotta_2015ApJ...803...42H}. In addition, the latitudinal temperature difference increases because of the presented process (see Subsection \ref{sec:meridional_force_balance}). The increased latitudinal temperature difference also contributes to increasing the spherical RMS for the temperature $T_{1(\mathrm{rms})}$. Fig.~\ref{flux_ana}c shows the normalized spherical correlation between the radial velocity $v_r$ and the temperature perturbation $T_1$. The correlation decreases with the increase in the resolution. This correlation should be good when the flow obeys the thermal convection. In the high-resolution simulations, small-scale turbulence, which does not behave as the thermal convection, increases, and the correlation decreases. As a result, the dimensional correlation $[v_r T_1]$, which directly determines the energy flux, stays the same among different resolutions (Fig.~\ref{flux_ana}). As a summary, the suppressed convection velocity and the worse normalized correlation are compensated by the increase in temperature perturbation to maintain the energy flux.\par
We also discuss the energy flux from the viewpoint of the spatial scale. Figs.~\ref{energy_flux_spectra}a, b, and c show the spectra of the radial velocity, the temperature perturbation, and these correlations, respectively. As discussed already, the radial velocity is suppressed in all the scales (Fig.~\ref{energy_flux_spectra}a). The increase of the temperature perturbation in the High case is mainly seen in the small scales ($\ell>40$: Fig.~\ref{energy_flux_spectra}b). These results support our interpretation of the increase of the temperature perturbation in the High case. We expect the suppression of the mixing by the magnetic field to increase the temperature perturbation effectively. This process is most effective on a small scale. The combination of the decrease of the radial velocity and the increase of the temperature perturbation in the small scales leads to a situation where the correlation $\widehat{v_r T_1}$ in the small scale ($\ell>40$) stays the same (Fig.~\ref{energy_flux_spectra}c). In addition, the higher-resolution calculation has a long tail of the correlation on a smaller scale. This result indicates that a significant fraction of the energy is transported by the small-scale turbulence in the High case. To evaluate the importance of the small-scale in energy transport, we calculate a value $S_\ell$ defined as follows.

\begin{align}
  S_\ell = \frac{\displaystyle\sum_{\ell'=\ell}^{\ell_\mathrm{max}}\widehat{v_r T_1}(\ell')}
  {\displaystyle\sum_{\ell'=0}^{\ell_\mathrm{max}}\widehat{v_r T_1}(\ell')}
  \label{eq:normalized_summed_correlation}
\end{align}

$S_\ell$ shows the fraction of the correlation from $\ell$ to $\ell_\mathrm{max}$ to the total correlation. Fig.~\ref{energy_flux_spectra}d shows the dependence of $S_\ell$ on the resolution. $S_\ell$ reaches unity around $\ell\sim5$ in the Low and Middle cases, while $\ell\sim20$ is enough for $S_\ell$ to reach unity in the High case. This indicates that in the High case, a significant fraction of the energy is transported by the middle to small scales ($\ell>20$), and the large-scale cannot transport the energy. We conclude that this is the main reason why the large-scale convection is suppressed in the High case.\par
We also investigate the convection driving mechanism in the viewpoint of the scale. In the analyses, we assume the background density is constant in time. Similar to the spectral magnetic energy $\widehat{E}_\mathrm{mag}$ discussed in Subsection \ref{sec:magnetic_field_generation}, the spectral kinetic energy $\widehat{E}_\mathrm{kin}$ evolution equation can be written such as:
\begin{align}
  \frac{\partial}{\partial t}\widehat{E}_\mathrm{kin} = 
  \widehat{T}_\mathrm{k(ADV)} + \widehat{T}_\mathrm{k(BUO)} +  \widehat{T}_\mathrm{k(LOR)},
\end{align}
where
\begin{align}
  \widehat{T}_\mathrm{k(ADV)} &= -\frac{1}{2}\rho_0 \widehat{\bm{v}}\cdot
  \widehat{\bm{v}\cdot\nabla\bm{v}}^* + c.c., \\
  \widehat{T}_\mathrm{k(BUO)} &= - \frac{1}{2}\rho_0 \widehat{\bm{v}}\cdot
  \widehat{-\rho_1\bm{g}+\nabla p}^* + c.c., \\
  \widehat{T}_\mathrm{k(LOR)} &= \frac{1}{8\pi}\widehat{\bm{v}}\cdot
  \widehat{\left(\nabla\times\bm{B}\right)\times\bm{B}}^* + c.c..
\end{align}
The result at $r=0.9R_\odot$ is shown in Fig.~\ref{kinetic_energy_transfer}. We note that while the Coriolis force should affect the spectral analysis, the amplitude is 1--2 orders of magnitude smaller than the other values, and we do not include it in our discussion. The general tendency is that the buoyancy drives the thermal convection and the advection, and the Lorentz force reduces the kinetic energy in almost all the scales. Also, both the energy production and the suppression on the large-scale is small in the High case. For all the contributions to the kinetic energy transfer, the velocity is multiplied. In the High case, the kinetic energy in the large-scale is reduced, and the reduction of the energy transfer seems an obvious result. To investigate the effective importance of the large-scale suppression, we normalize the kinetic energy transfer by $\sqrt{\widehat{v}\widehat{v}^*/r}$ (Fig.~\ref{kinetic_energy_transfer}b). The normalized kinetic energy transfer by the buoyancy $\widehat{T}_\mathrm{k(BUO)}$ is reduced only in the High case. We also observe the suppression of the Lorentz force contribution. These results indicate that the suppression of the large-scale kinetic energy is caused by the suppression of the buoyancy.
The magnetic field on a large scale does not directly contribute to the large-scale suppression.


\subsection{Meridional force balance}
\label{sec:meridional_force_balance}

\begin{figure*}[htpb]
  \begin{center}
\includegraphics[width=0.8\textwidth]{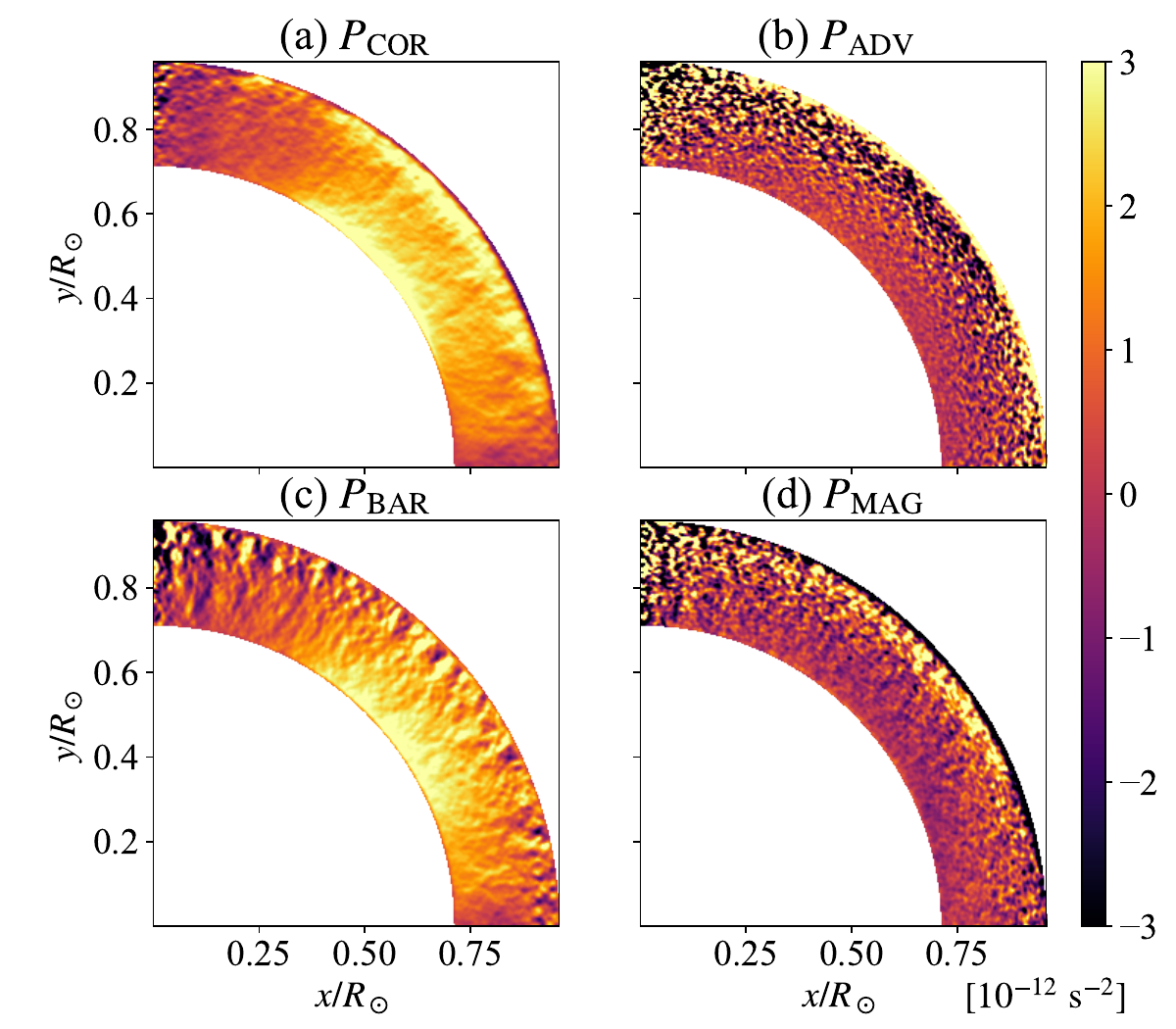}
\caption{Each term in the vorticity equation is shown.
  The non-Taylor-Proudman state $\partial \Omega/\partial z\neq0$ is mainly maintained by the baroclinic term $P_\mathrm{BAR}$ in the deep convection zone.
\label{vorticity_equation_2D}}
\end{center}
\end{figure*}

\begin{figure}[htpb]
  \begin{center}
\includegraphics[width=0.45\textwidth]{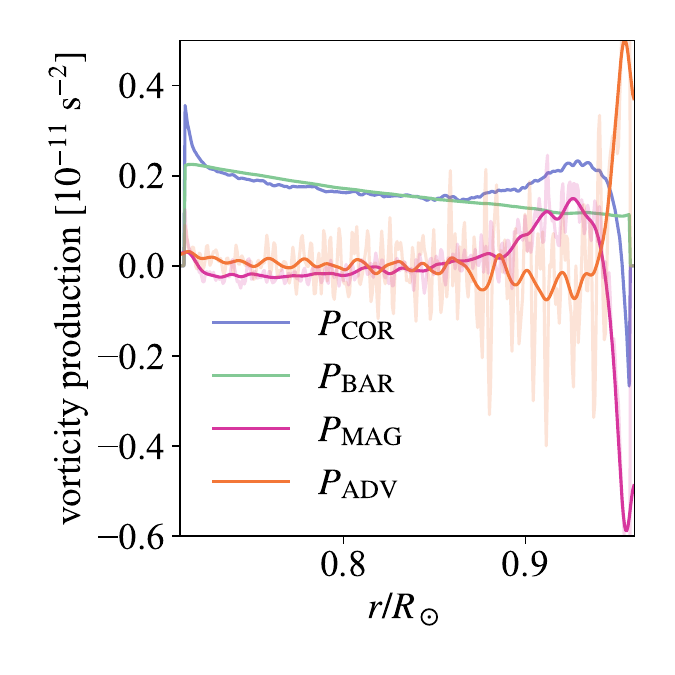}
\caption{The blue ($P_\mathrm{COR}$), green ($P_\mathrm{BAR}$), magenta ($P_\mathrm{MAG}$), and orange ($P_\mathrm{ADV}$) lines show the spherically averaged terms in the vorticity equation. The transparent lines indicate the raw data without radial filtering. \label{vorticity_equation_1D}}
\end{center}
\end{figure}

\begin{figure}[htpb]
  \begin{center}
  \includegraphics[width=0.45\textwidth]{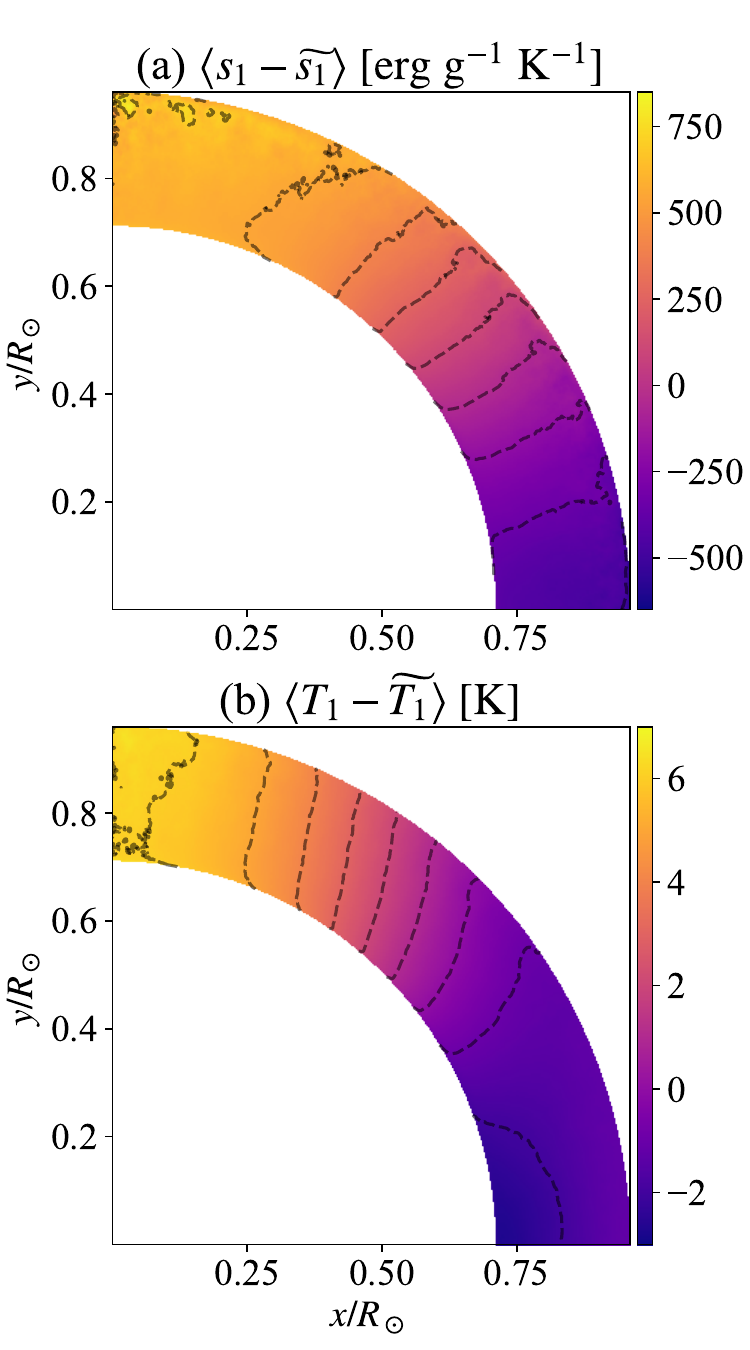}
  \caption{(a) $\langle s_1-\widetilde{s_1}\rangle$ and (b) $\langle T_1-\widetilde{T_1}\rangle$ in the High case are shown. $\langle\rangle$ and $\widetilde{~}$ indicate the longitudinal average and the spherical average, respectively (see eqs.~(\ref{eq:longitudinal_average}) and (\ref{eq:spharical_average})). \label{entropy_temperature_gradient_2D}}
  \end{center}
  \end{figure}

\begin{figure}[htpb]
\begin{center}
    \includegraphics[width=0.45\textwidth]{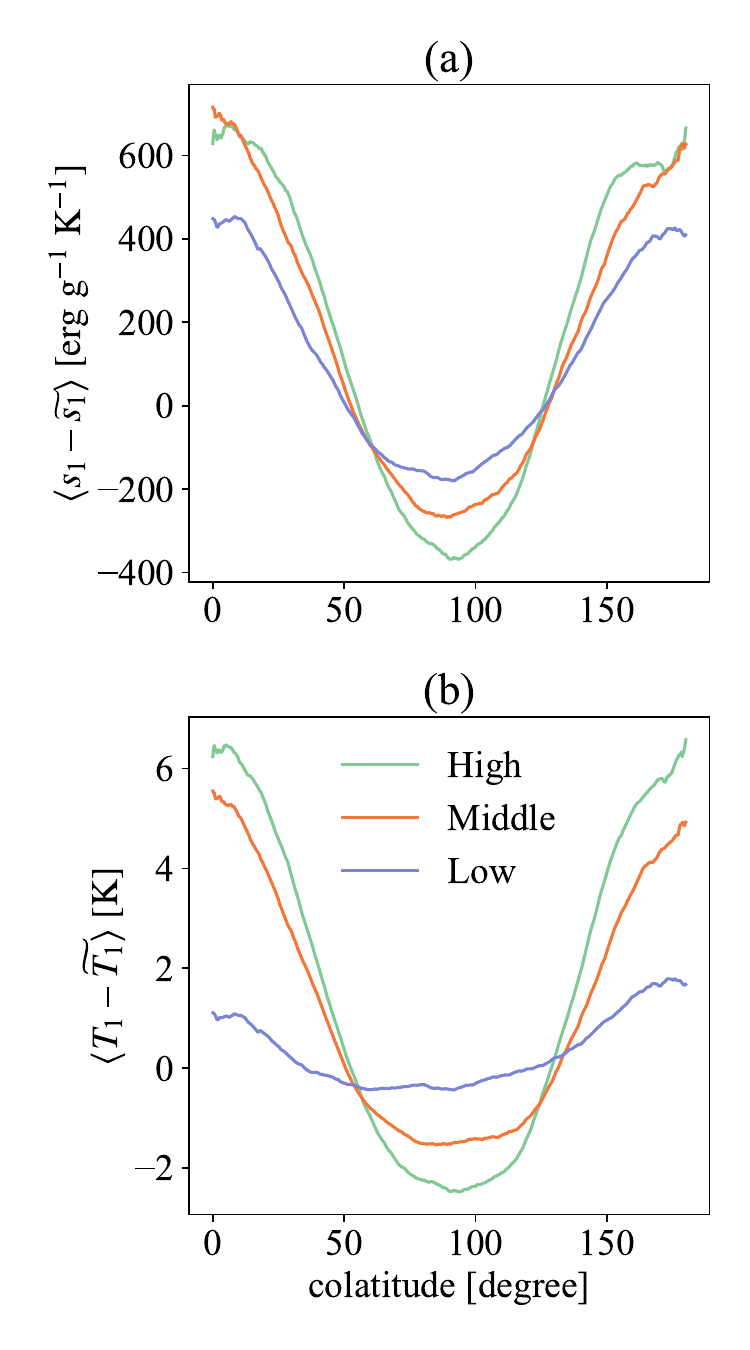}
\caption{Latitudinal dependence of (a) entropy and (b) temperature. The deviation from the spherical averaged longitudinally. \label{entropy_temperature_gradient_1D}}
\end{center}
\end{figure}

In this subsection, we discuss the force balance on the meridional plane, especially about the Taylor--Proudman constraint. The differential rotation in the High case does not obey the Taylor--Proudman constraints, i.e., $\partial \Omega/\partial z\ne 0$, where $z$ indicates the direction of the rotational axis. To address this aspect, we need to analyze the vorticity equation \citep[e.g., ][]{Miesch_2011ApJ...743...79M}. The longitudinal component of the vorticity equation in a steady-state $\partial/\partial t = 0$ is written as 
\citep[see also][for more detailed discussions about the balance]{balbus_2009MNRAS.400..176B}
\begin{align}
  \underbrace{-2r\sin\theta\Omega_0\frac{\partial\langle\Omega_1\rangle}{\partial z}}_{P_\mathrm{COR}}
  &=
  \underbrace{\left\langle\nabla\times\left(\bm{v}\times\bm{\omega}\right)\right\rangle_\phi}_{P_\mathrm{ADV}} \nonumber\\
  &+
  \underbrace{\frac{g}{\rho_0 r} \left(\frac{\partial\rho}{\partial s}\right)_p\frac{\partial \langle s_1\rangle }{\partial \theta}}_{P_\mathrm{BAR}}\nonumber \\
  &+
  \underbrace{\left\langle\nabla\times\left[\frac{1}{4\pi\rho_0}\left(\nabla\times\bm{B}\right)\times\bm{B}\right]\right\rangle_\phi}_{P_\mathrm{MAG}}
\end{align}

We show each term in the equation in Fig.~\ref{vorticity_equation_2D}. To suppress realization noise, especially in $P_\mathrm{ADV}$ ad $P_\mathrm{MAG}$, we use a Gaussian filter with a width of 5$\times$5 grid points. We also show the spherically averaged 1D profile of each term in Fig.~\ref{vorticity_equation_1D}. We use a Gaussian filter with a width of five grid points also for the 1D profile. The raw data are shown with transparent lines. The results clearly show that the deviation from the Taylor--Proudman theorem ($P_\mathrm{COR}$) is mainly balanced by the baroclinic term ($P_\mathrm{BAR}$). We see a significant deviation from the Taylor--Proudman theorem around the top boundary. This is maintained both by the advection ($P_\mathrm{ADV}$) and the magnetic field ($P_\mathrm{MAG}$). While this tendency is important to discuss the near-surface shear layer, we leave this for our future publication for the near-surface layer (Hotta, Kusano, \& Sekii in prep). In this paper, we focus on the discussion about the Taylor--Proudman theorem in the middle of the convection zone. The result shows that the Coriolis force is balanced with the baroclinic term, i.e., the latitudinal entropy gradient. \cite{Hotta_2018ApJ...860L..24H} argues that the efficient small-scale dynamo and generated magnetic field help construct the entropy gradient. As shown in Subsection \ref{sec:convection_driving}, the temperature perturbation increases with increased resolution. In addition, the convection velocity is reduced in the higher resolutions (Fig.~\ref{rms_velocity}).
The Coriolis force bends a warm upflow (cold downflow) poleward (equatorward). Both the high-resolution effects (increasing the temperature perturbation and reducing the convection velocity) enhance this process. Fig.~\ref{entropy_temperature_gradient_2D} shows the entropy and the temperature distributions in the High case. We succeed in reproducing the negative entropy and temperature gradient in the whole convection zone. \cite{Miesch_2006ApJ...641..618M} enforce the entropy gradient at the bottom boundary to avoid the Taylor--Proudman constraint \citep[see also][]{Miesch_2008ApJ...673..557M,Fan_2014ApJ...789...35F}. Also, \cite{brun_2011ApJ...742...79B} maintains the entropy gradient by a dynamical coupling of the convection and radiation zones \citep[see also][]{Rempel_2005ApJ...622.1320R}. In their studies, maintaining the negative entropy gradient in the near-surface equator is difficult, and the differential rotation tends to be the Taylor--Proudman type topology in the near-surface equator region \citep[For example, see Figs.~10 and 13 of ][]{brun_2011ApJ...742...79B}. In our simulations, the entropy gradient is generated by the turbulent process throughout the convection zone, and the differential rotation can avoid the Taylor--Proudman constraint, which is consistent with the observations \cite[e.g.,][]{Schou_1998ApJ...505..390S}. Fig.~\ref{entropy_temperature_gradient_1D} shows the resolution dependence of the entropy and the temperature gradient. It is clearly shown that the entropy and the temperature gradient increase with resolution. This result also indicates that the magnetic field maintains the entropy and temperature gradients because the magnetic strength increases with the resolution. The temperature difference between the equator and the pole at the base of the convection zone is 8 K in the High case. This value corresponds to the AB3 case in \cite{Miesch_2006ApJ...641..618M} with which they argue their most solar-like profile.
\par
Recently \cite{matilsky_2020ApJ...898..111M} show that the fixed flux boundary condition, which is similar to that in this study, is easier to generate the non-Taylor-Proudman differential rotation than the fixed entropy boundary condition. They show that the entropy gradient is generated by the anisotropic enthalpy flux caused by the Busse column in the low latitude. Since we use the fixed flux boundary condition, our calculation should also be benefited from the numerical setting.
      

\subsection{Angular momentum transport}
\label{sec:angular_momentum_transport}

\begin{figure*}[htbp]
  \begin{center}
    \includegraphics[width=\textwidth]{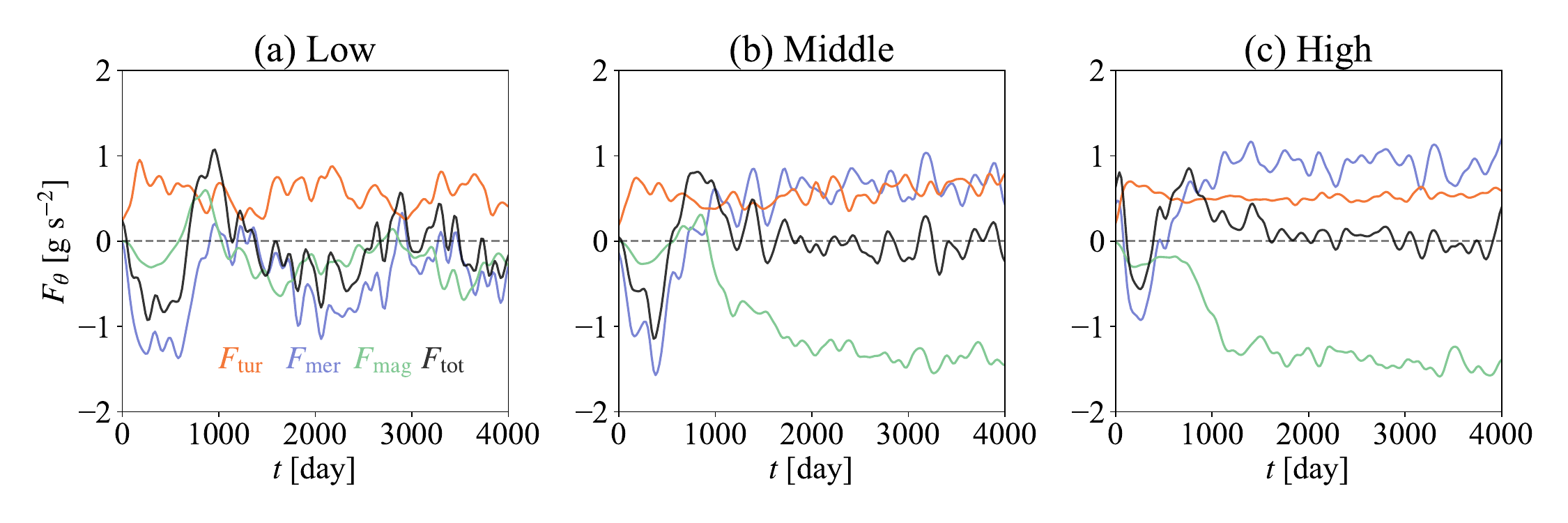}
    \caption{Radially averaged latitudinal angular momentum flux at $\theta=\pi/4$ is shown. Panels a, b, and c show the results from Low, Middle, and High cases, respectively. The orange, blue, green, and black lines show the turbulent $F_\mathrm{tur}$, meridional flow $F_\mathrm{mer}$, magnetic $F_\mathrm{mag}$, and total $F_\mathrm{tot}$ angular momentum fluxes, respectively. For the definition of the angular momentum of transport, see eqs.~(\ref{eq:ang_transport_turbulence}) to (\ref{eq:ang_transport_total}). We use a Gaussian filter with 60-day width to reduce the realization noise.
      The result shows that the sign of the transport by the meridional flow $F_\mathrm{mer}$ changed the sign from Low to Middle cases. This is the main reason for the fast equator.
    \label{ang_transport}}
  \end{center}
\end{figure*}

\begin{figure*}[htpb]
  \begin{center}
\includegraphics[width=\textwidth]{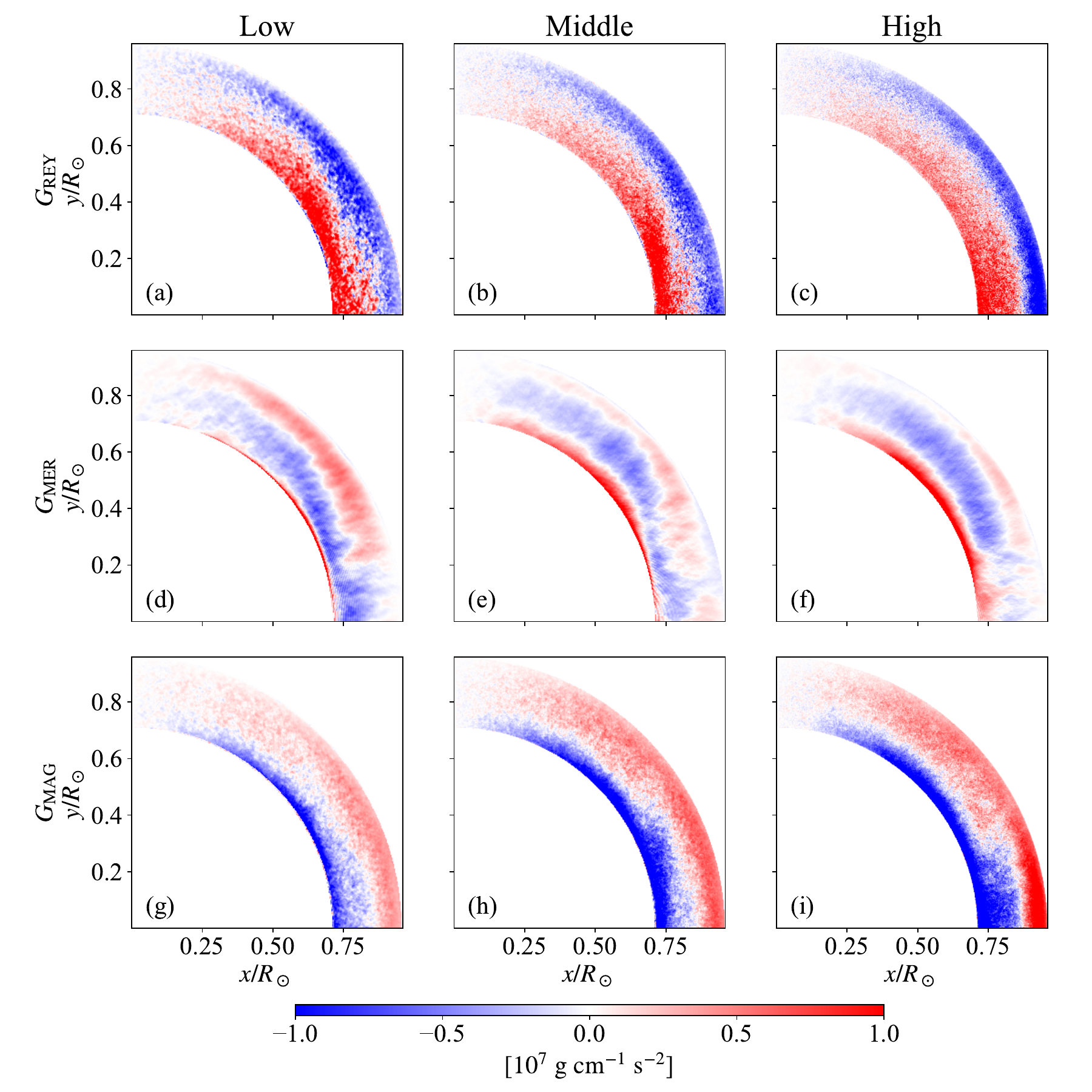}
\caption{Each term in gyroscopic pumping is shown. The left, middle, and right columns show the results from Low, Middle, and High cases, respectively. The top, middle, and low rows show $G_\mathrm{REY}$, $G_\mathrm{MER}$, and $G_\mathrm{MAG}$, respectively.\label{gyroscopic_pumping}}
\end{center}
\end{figure*}

\begin{figure*}[htpb]
  \begin{center}
\includegraphics[width=\textwidth]{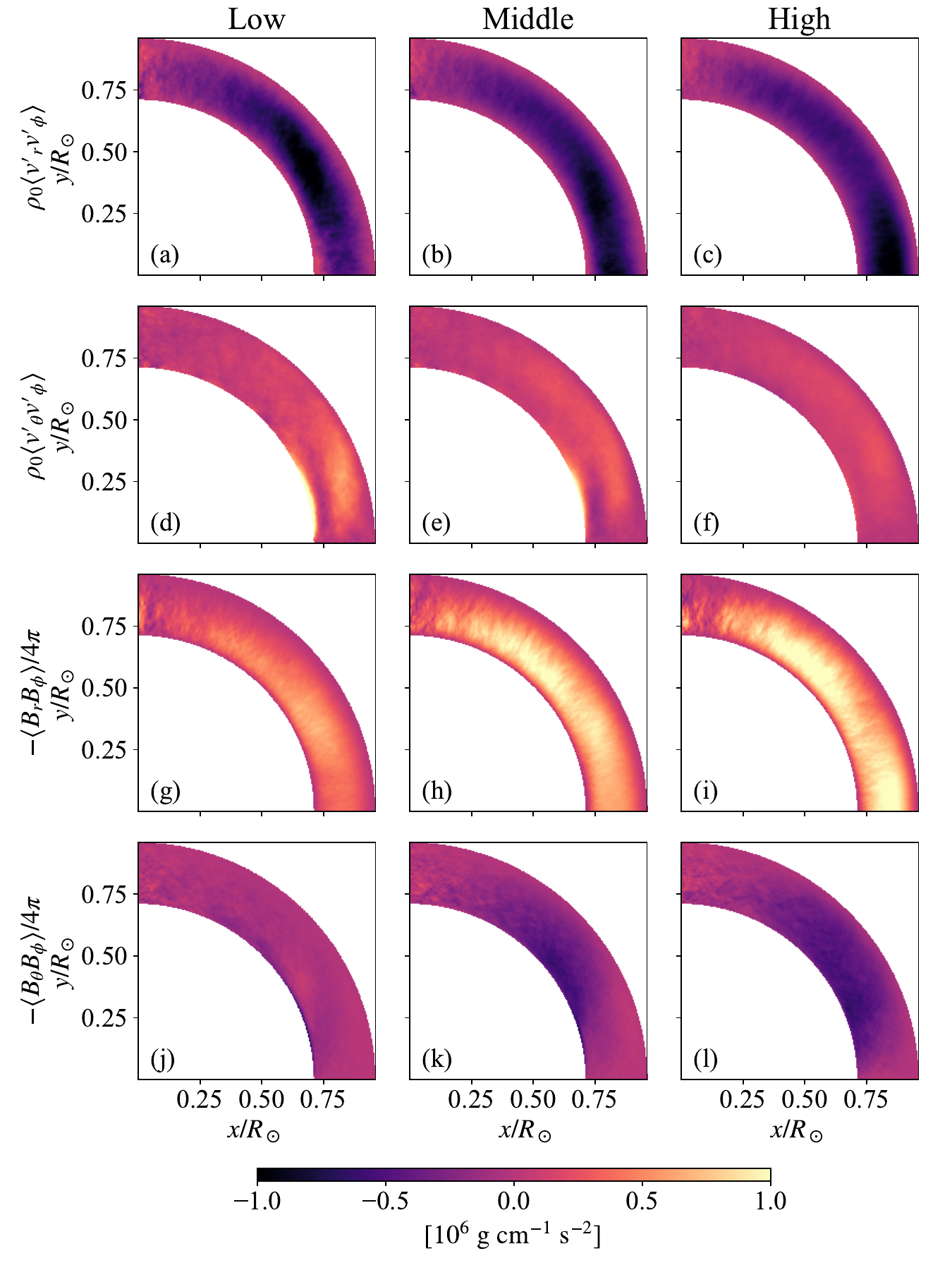}
\caption{Correlations for the angular momentum transport are shown. The left, middle, and right columns show the results from Low, Middle, and High cases, respectively. The first, second, third, and fourth rows show
$\rho_0\langle v'_r v'_\phi\rangle$, 
$\rho_0\langle v'_\theta v'_\phi\rangle$, 
-$\langle B_r B_\phi\rangle/4\pi$, and
-$\langle B_\theta B_\phi\rangle/4\pi$, respectively.
\label{angular_momentum_correlation}}
\end{center}
\end{figure*}

This subsection discusses the angular momentum transport, and we explain why the equator is rotating faster than the polar region. To discuss the angular momentum transport, we should start from the angular momentum conservation law that is approximated as
\begin{align}
  \frac{\partial}{\partial t} \left(\rho_0\langle\mathcal{L}\rangle\right)
  = G_\mathrm{REY} + G_\mathrm{MER} + G_\mathrm{MAG}, \label{eq:ang_conserve}
\end{align}
where 
\begin{align}
  G_\mathrm{REY} &= -\nabla\cdot
  \left(\rho_0 \lambda\langle \bm{v}'_\mathrm{m}v'_\phi\rangle\right), \label{eq:G_REY}\\
  G_\mathrm{MER} &= -\nabla\cdot
  \left(\rho_0
  \langle\bm{v}_\mathrm{m}\rangle
  \langle \mathcal{L}\rangle
  \right), \\
  G_\mathrm{MAG} &= -\nabla\cdot
  \left(-\lambda \frac{\langle\bm{B}_\mathrm{m}B_\phi\rangle}{4\pi}\right),\label{eq:G_MAG}
\end{align}
and $\lambda = r\sin\theta$. $\mathcal{L}=\lambda u_\phi=\lambda v_\phi + \lambda^2 \Omega_0$ is the specific angular momentum. $G_\mathrm{REY}$, $G_\mathrm{MER}$, and $G_\mathrm{MAG}$ are the angular momentum by the turbulence, mean flow (meridional flow), and magnetic field. We define $\bm{B}_\mathrm{m} = B_r \bm{e}_r + B_\theta \bm{e}_\theta$. Because the large-scale magnetic field $\langle\bm{B}\rangle$ is weak in this study (see Table \ref{summary}), we do not divide the magnetic contribution $G_\mathrm{MAG}$ to turbulent component $\bm{B}'$ and large-scale component $\langle \bm{B}\rangle$.\par
At first, we discuss how to transport the angular momentum equatorward in the High (and Middle) cases. To this end, we evaluate the temporal evolution of the latitudinal angular momentum flux density at $\theta=\pi/4$. The latitudinal angular momentum flux densities are:
\begin{align}
  F_\mathrm{tur} &= \rho_0 \lambda \langle v'_\theta v'_\phi\rangle,
  \label{eq:ang_transport_turbulence}
  \\
  F_\mathrm{mer} &= \rho_0  \langle v_\theta \rangle \langle \mathcal{L}\rangle,
  \label{eq:ang_transport_meridional_flow}
  \\
  F_\mathrm{mag}  &= -\lambda \frac{\langle B_\theta B_\phi\rangle}{4\pi}, 
  \label{eq:ang_transport_magnetic}
  \\
  F_\mathrm{tot} &= F_\mathrm{tur} + F_\mathrm{mer} + F_\mathrm{mag}.
  \label{eq:ang_transport_total}
\end{align}
These fluxes are radially averaged at $\theta=\pi/4$ in Fig.~\ref{ang_transport}. Although the turbulent angular momentum transport (orange line: $F_\mathrm{tur}$) is positive (equatorward) in all cases, the resulting differential rotation is different in all cases. This indicates that the equatorward angular momentum transport cannot be the reason why we have the fast equator in the High case. A prominent difference can be seen in the angular momentum transport by the meridional flow (blue line: $F_\mathrm{mer}$). $F_\mathrm{mer}$ is negative (poleward) in the initial phase ($<600~\mathrm{day}$) in all cases. This poleward angular momentum transport leads to the fast pole in the initial phase (see Fig.~\ref{mean_flow_initial} in Appendix \ref{sec:mean_flow_initial}). While $F_\mathrm{mer}$ stays almost negative in the Low case, the other cases clearly show positive $F_\mathrm{mer}$ in the latter phase. This is the reason why we have a fast equator in the Middle and High cases. Because of the low Mach number situation, $\nabla\cdot\left(\rho_0 \bm{v}_\mathrm{m}\right)=0$ is approximately satisfied. This leads to $\int \rho_0 v_\theta rdr\sim 0$ at an arbitrary latitude with the closed boundary condition for the radial velocity. Because the specific angular momentum is
\begin{align}
  \langle\mathcal{L}\rangle =& r^2\sin^2\theta\left(\langle\Omega_1\rangle + \Omega_0\right),
\end{align}
deeper layers (small $r$) tend to have smaller angular momentum than near-surface layers (large $r$). 
The equatorward meridional flow in the middle of the convection zone is the direct reason for accelerating the equator.
Due to the mass conservation, the fast meridional flow, which overcomes the poleward angular momentum transport near the surface, requires the poleward meridional flow around the base of the convection zone.
The poleward meridional flow around the base of the convection zone is the primary key to why we have the fast equator in the High case.
\par
Gyroscopic pumping is useful in understanding the maintenance mechanism of the meridional flow. Gyroscopic pumping is the angular momentum conservation law in a steady-state.
\begin{align}
  \rho_0 \langle\bm{v}_\mathrm{m}\rangle\cdot\nabla \langle\mathcal{L}\rangle \sim
  -\nabla\cdot\left(
    \rho_0\lambda\langle \bm{v}'_\mathrm{m}v'_\mathrm{\phi}\rangle
    - \lambda\frac{\langle \bm{B}_\mathrm{m}B_\phi\rangle}{4\pi}
    \right)
    \label{eq:gyroscopic_pumping}
\end{align}
In this study and the solar case, the differential rotation is weak, i.e., $\Omega_1/\Omega_0\sim0.1$ and the angular momentum $\langle\mathcal{L}\rangle$ does not change significantly even after the differential rotation is constructed. Thus, the gyroscopic pumping indicates that the angular momentum transport by the Reynolds stress and the magnetic field determines the topology of the meridional flow.
Fig.~\ref{gyroscopic_pumping} shows each term in eq.~(\ref{eq:ang_conserve}). From this figure, we can discuss two topics: one is the generation mechanism of the poleward meridional flow around the base of the convection zone, and the other is the acceleration mechanism of the near-surface equator. \par
At first, we discuss the generation mechanism of the meridional flow. Because of the poleward meridional flow around the base of the convection zone, the angular momentum increases (Figs.~\ref{gyroscopic_pumping}e and f). This is compensated for by the magnetic angular momentum transport ($G_\mathrm{MAG}$: Figs.~\ref{gyroscopic_pumping}h and i). In other words, the poleward meridional flow around the base of the convection zone is maintained by the magnetic angular momentum transport. The magnetic angular momentum transport decreases the angular momentum around the base of the convection zone, and the poleward meridional flow increases it as compensation. While the turbulent angular momentum transport tends to increase the angular momentum around the base of the convection zone (Figs.~\ref{gyroscopic_pumping}a, b, and c), this is not enough to compensate for the decrease by the magnetic angular momentum transport (see also Fig.~\ref{gyroscopic_pumping_appendix} for the sum of $G_\mathrm{REY}$ and $G_\mathrm{MER}$).\par
As for the increase of the angular velocity in the near-surface equator, the magnetic angular momentum has the main contribution. The major difference in the differential rotation between the Middle and High cases is the angular velocity in the near-surface equator (Fig.~\ref{mean_flow}). The High case has a large angular velocity there, which is more consistent with the solar observation. It is apparent that this increase in the angular velocity in the High case is caused by the magnetic angular momentum transport ($G_\mathrm{MAG}$: Fig.~\ref{gyroscopic_pumping}i). In all cases, the near-surface equator is accelerated by $G_\mathrm{MAG}$, but the amplitude of $G_\mathrm{MAG}$ increases in the High case because the magnetic field strength increases with the resolution (see Subsection \ref{sec:magnetic_field_generation}).\par
In summary, the equatorward latitudinal angular momentum transport is done by the meridional flow constructed by the magnetic angular momentum transport. To have the large angular velocity in the near-surface equator, we need additional contributions by the magnetic field, which is stronger in the higher resolutions. For the angular momentum transport, both the strength and correlation are important. We analyze the result in this regard in the following paragraph.\par
Fig.~\ref{angular_momentum_correlation} shows the correlation between the velocities and the magnetic fields.
For both the Reynolds stress $\langle v'_i v'_j\rangle$ and the Maxwell stress $\langle B_i B_j\rangle$, the main contribution is the radial transport. The distributions of $G_\mathrm{REY}$ and $G_\mathrm{MAG}$ are roughly explained by the radial angular momentum transport. The Reynolds stress transports the angular momentum radially inward, while the Maxwell stress transports it in the opposite direction. The radially inward angular momentum transport is a usual result with a high Rossby number (weak rotational influence) situation 
\citep[see][]{Gastine_2013Icar..225..156G,Featherstone_2015ApJ...804...67F,Hotta_2015ApJ...798...51H,karak_2015A&A...576A..26K}.
While the Rossby number $\mathrm{(Ro)}$ decreases from the Low to High cases (see Table \ref{summary}), the radially inward angular momentum transport does not change much or even increase (see also normalized correlation in Fig. \ref{normalized_correlation}). This result indicates the weak influence of rotation on the small-scale flow in all the cases. As the local Rossby number $\mathrm{(Ro_\ell)}$, which increases from the Low to High cases, measures the rotational influence is not strong enough to maintain the solar-like differential rotation by the Reynolds stress.
As explained in the previous paragraph, the essential reason for the poleward meridional flow around the base of the convection zone and the large angular velocity around the near-surface equator is the magnetic angular momentum transport. Fig.~\ref{angular_momentum_correlation} shows that the negative correlation $\langle B_r B_\phi\rangle$, i.e., the radially outward magnetic angular momentum transport, is responsible for both of these. The radially outward transport decreases and increases the angular momentum at the base and the top of the convection zone, respectively. 
\begin{figure*}[htpb]
  \begin{center}
\includegraphics[width=\textwidth]{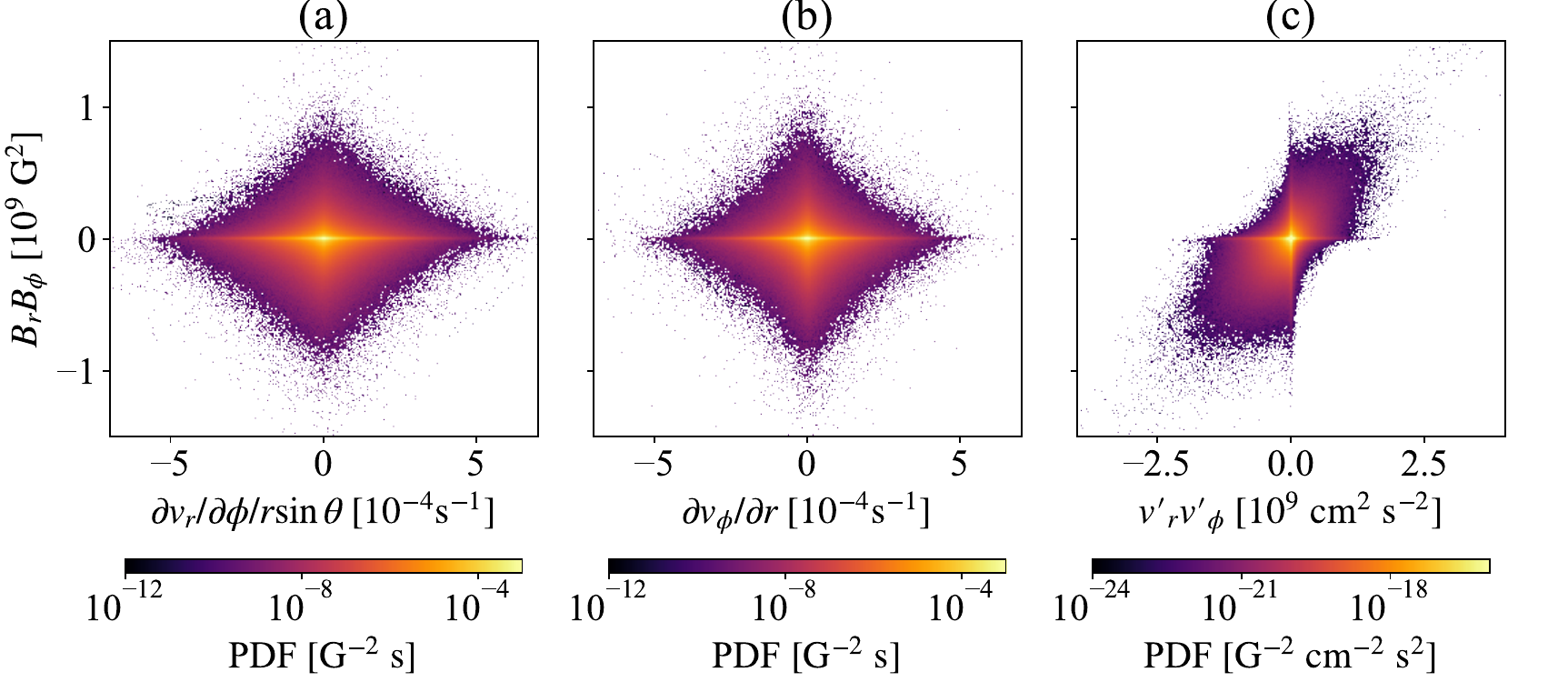}
\caption{2D PDFs of (a) $B_rB_\phi$ vs. $\partial v_r/\partial \phi/r\sin\theta$, (b) $B_rB_\phi$ vs. $\partial v_\theta/\partial r$, and (c) $B_rB_\phi$ vs. $v'_rv'_\phi$ at $r=0.9R_\odot$ in the High case, are shown.
  $B_rB_\phi$ is well correlates with $v'_rv'_\phi$, while the others do not. This indicates that $B_rB_\phi$ correlation is possibly originated from $v'_rv'_\phi$.
\label{cor_bxbzvxvz1}}
\end{center}
\end{figure*}

\begin{figure*}[htpb]
  \begin{center}    
\includegraphics[width=\textwidth]{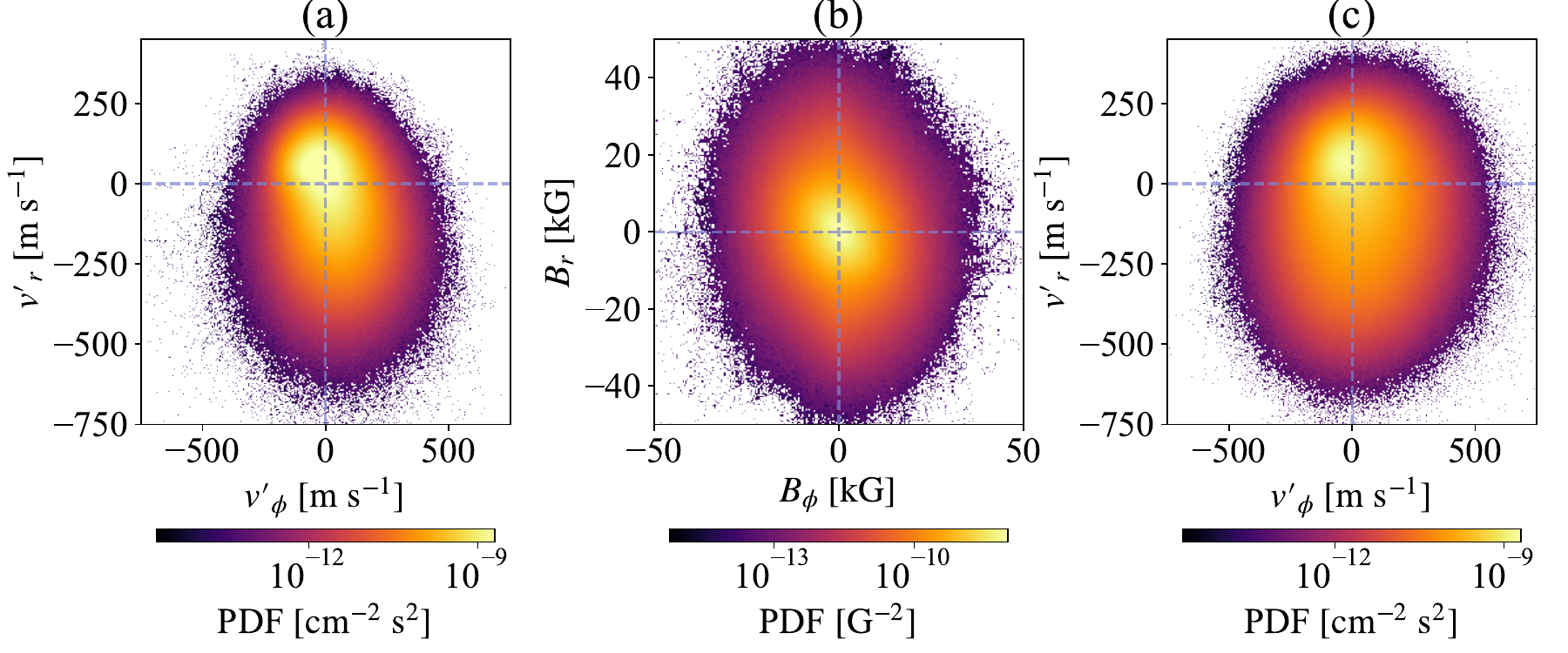}
\caption{Correlations of (a) $v'_r$ vs. $v'_\phi$, (b) $B_r$ and $B_\phi$ from the High case are shown. Panel c shows the correlation of $v'_r$ vs. $v'_\phi$ from the High-HD case. All the data are at $r=0.9R_\odot$.
\label{cor_bxbzvxvz2}}
\end{center}
\end{figure*}

The main possible reasons for the magnetic field correlation are the shear and the alignment to the flow. The shear term of the induction equation is written as
\begin{align}
  \frac{\partial B_r}{\partial t} =& \frac{B_\phi}{r\sin\theta}\frac{\partial v_r}{\partial \phi} + [...], \\
  \frac{\partial B_\phi}{\partial t} =& B_r\frac{\partial v_\phi}{\partial r} + [...].
\end{align}
Thus, the shear of the flow can correlate the magnetic field components. In addition, the magnetic induction equation in high conductivity limit is
\begin{align}
  \frac{\partial \bm{B}}{\partial t} = 
  \nabla\times\left(\bm{v}\times\bm{B}\right).
\end{align}
This means that when the magnetic field is parallel to the velocity, $\bm{v}\times\bm{B}=0$, 
the magnetic field does not evolve more. Conversely, the magnetic field tends to be parallel to the velocity. To understand the origin of the negative correlation of $\langle B_r B_\phi\rangle$, Fig.~\ref{cor_bxbzvxvz1} shows the PDF of (a) $B_rB_\phi$ vs. $\partial v_r/\partial \phi/r\sin\theta$, (b)$B_rB_\phi$ vs. $\partial v_\theta/\partial r$, and (c) $B_rB_\phi$ vs. $v'_rv'_\phi$ at $r=0.9R_\odot$ in the High case. While we do not see clear correlation between $B_rB_\phi$ and shears (Figs.~\ref{cor_bxbzvxvz1}a and b), $B_rB_\phi$ and $v'_rv'_\phi$ correlate well. This indicates that the origin of the negative $\langle B_r B_\phi\rangle$ is not the flow shear but the negative correlation of velocities $\langle v'_r v'_\phi\rangle$. Figs.~\ref{angular_momentum_correlation}a, b, and c show that $\langle v'_rv'_\phi\rangle$ is negative at all latitudes. This is caused by the Coriolis force. The Coriolis force in the longitudinal equation of motion is
\begin{align}
  \frac{\partial v_\phi}{\partial t} = [...] - 2\Omega_0\left(v_r \sin\theta + v_\theta\cos\theta\right).
\end{align}
Thus, the radial velocity, which is the source of the thermal convection, is bent by the Coriolis force and the negative $\langle v'_r v'_\phi\rangle$ is caused. Because the magnetic induction equation only suggests that the magnetic field tends to be parallel to the velocity, it is possible that $\langle B_r B_\phi\rangle$ is the origin of the $\langle v'_r v'_\phi\rangle$.  To confirm the origin of $\langle v'_r v'_\phi\rangle$, we compare the hydro case (High-HD) and the magnetic case (High) in Fig.~\ref{cor_bxbzvxvz2} with PDFs. Fig.~\ref{cor_bxbzvxvz2} shows PDFs of (a) $v'_r$ vs. $v'_\phi$, (b) $B_r$ and $B_\phi$ from the High case are shown.  Fig.~\ref{cor_bxbzvxvz2}c shows the correlation of $v'_r$ vs. $v'_\phi$ from the High-HD case. Even in the hydro case, we see a similar correlation between $v'_r$ and $v'_\phi$ (Fig.~\ref{cor_bxbzvxvz2}c) to the magnetic case (Fig.~\ref{cor_bxbzvxvz2}a). This result shows that the magnetic field is not the main origin of $\langle v'_r v'_\phi\rangle$, but the velocity is the origin of $\langle B_r B_\phi\rangle$.
\par
We decompose the Reynolds stress to radial $r$ and colatitudinal $\theta$ components. We note that decomposition of parallel and perpendicular to the rotational axis directions, i.e., $z$ and $\lambda$ directions, are also useful. Since the constant surface of the specific angular momentum is parallel to the cylindrical surface (constant $\lambda$) in the leading order, it is difficult for the meridional flow to transport the angular momentum through the constant $\lambda$ surface, i.e., $\int \rho_0\langle v_\lambda\rangle \langle\mathcal{L}\rangle dz\sim 0$ due to the anelastic approximation $\int \rho_0 \langle v_\lambda \rangle dz\sim0$. Thus the meridional flow mainly transports the angular momentum in the $z$ direction. This tendency indicates that $\langle v'_z v'_\phi\rangle$ and $\langle v'_\lambda v_\phi\rangle$ are responsible for the generation of the meridional flow and the differential rotation, respectively.
\par
In this study, we analyze the Reynolds stress just as the velocity correlation $\langle v'_i v'_j\rangle$. The stress includes diffusive part, i.e., so-called turbulent viscosity and non-diffusive part, so-called $\Lambda$ effect \citep{rudiger_1980GApFD..16..239R}. If we can distinguish these two from the Reynolds stress, we can directly evaluate the anisotropy. 

\section{Summary and Discussion}
\label{sec:summary}

\begin{figure*}[htpb]
  \begin{center}
    \includegraphics[width=0.8\textwidth]{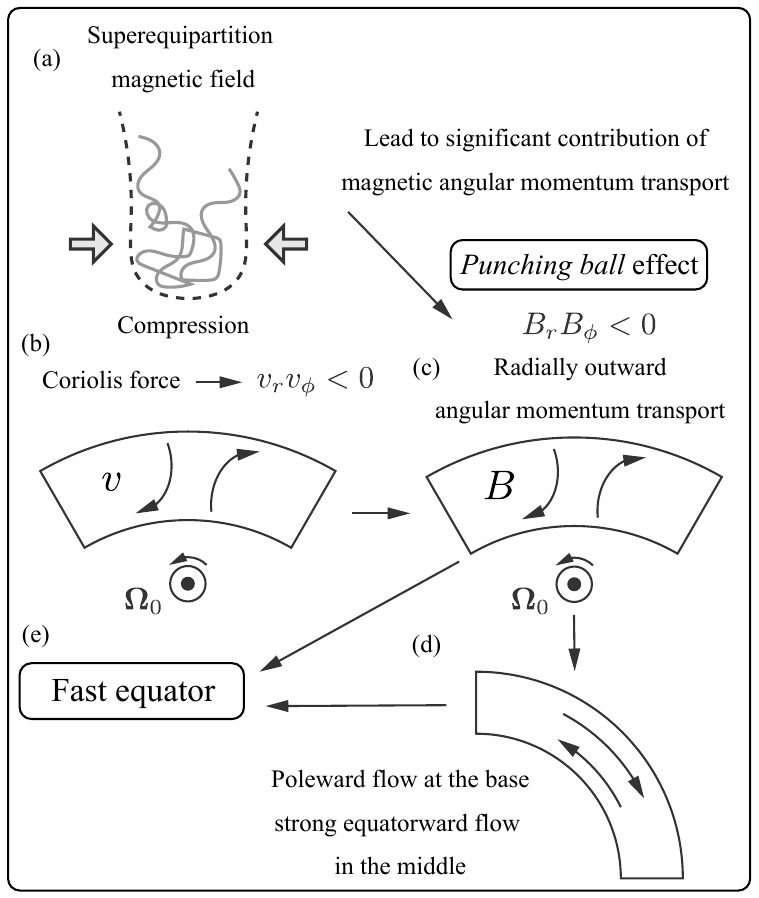}
    \caption{Summary explanation of the process for the fast equator.\label{summary_exp}}
  \end{center}  
\end{figure*}

We analyze the simulation data of \cite{hotta_2021NatAs...5.1100H} in which the solar-like differential rotation, i.e., the fast equator and the slow pole, is presented. Fig.~\ref{summary_exp} summarizes our revealed processes for the fast equator. (a) High resolution suppresses numerical diffusion and enhances the amplification of the magnetic field. The compression is the main mechanism to generate the superequipartition magnetic field. Because the strong magnetic field is balanced with the gas pressure, the internal energy is available for amplification. (b) The Coriolis force causes the negative correlation of velocities $\langle v'_r v'_\phi\rangle$, which is typical in the large Rossby number regime. (c) The magnetic field tends to be parallel to the flow and also has a negative correlation $\langle B_r B_\phi\rangle<0$. This transports the angular momentum radially outward. We can simply think that the radially outward magnetic angular momentum transport is the back reaction to the Coriolis force. We call it the {\it Punching ball} effect because the magnetic field behaves as if it is being punched by the Coriolis force. The {\it Punching ball}, which is radially outward magnetic angular momentum transport,  is the essential process and our new finding for the fast equator. (d) Because of the radially outward angular momentum transport, the angular momentum around the base of the convection zone decreases. To compensate for the decrease, the meridional flow becomes poleward around the base of the convection zone. To satisfy the mass conservation, an equatorward meridional flow in the middle of the convection zone is caused. Because the specific angular momentum is larger in the middle of the convection zone than at the base with the same latitude, the equatorward flow leads to the net equatorward angular momentum transport by the meridional flow. (e) Both the Maxwell stress (panel c) and the meridional flow (panel d) are essential to the fast equator in the near-surface layer. A prominent difference between the Middle and High cases is the angular velocity at the near-surface equator (see Figs.~\ref{mean_flow}b and c). This difference is caused by the magnetic field strength in these two cases. In conclusion, we suggest that the magnetic field has two roles in the construction of differential rotation. One is the maintenance of the meridional flow; the other is the angular momentum transport to maintain the fast near-surface equator.
\par
\cite{Brun_2004ApJ...614.1073B} have already shown that the Maxwell stress tends to be opposite to the Reynolds stress. In their study, the radial Reynolds stress is positive, i.e., the radially outward angular momentum transport and the radial Maxwell stress is negative. This indicates that the differential rotation is maintained by the Reynolds stress in \cite{Brun_2004ApJ...614.1073B} and the Maxwell stress suppresses it. In this study, we find that the radial Maxwell stress is positive and is the main driver through the {\it Punching ball} effect. This is qualitatively different from the previous models.
\par
Many studies have already suggested that the magnetic field can relax the criterion of the rotation rate for the fast equator \citep{Fan_2014ApJ...789...35F,gastine_2014MNRAS.438L..76G,mabuchi_2015ApJ...806...10M,karak_2015A&A...576A..26K} since the magnetic field suppresses the convection velocity. The new role found in this study is qualitatively different from these studies. While the convection velocity is also suppressed in this study, it looks that only the suppression is not enough for the fast equator since increasing the resolution leads to a larger inward angular momentum transport by turbulence (Fig. \ref{angular_momentum_correlation}) which is a negative factor for the fast equator (see also Fig. \ref{normalized_correlation} for normalized correlations). The magnetic angular momentum transport has an essential role in the fast equator. The negative correlation of $\langle B'_r B'_\phi \rangle$ is also found by \cite{karak_2015A&A...576A..26K}, but the amplitude is more than two orders of magnitude smaller than the Reynolds stress (see their Fig. 18). 
\par
In the following subsections, we discuss the remaining issues and our future perspective in several aspects.
\subsection{Magnetic field intensification}
In this study, compression is an important process to amplify the magnetic field. In the process, we can use the internal energy, which is about $10^6$ times larger than the kinetic energy in the deep convection zone. We have not reached numerical convergence, i.e., higher resolutions show stronger magnetic field (see Fig.~\ref{rms_velocity}b). 
\cite{Featherstone_2016ApJ...818...32F} show that the kinetic energy can converge to a specific value by increasing the Rayleigh number in their hydrodynamic run around $\mathrm{Ra}\sim10^5$. \cite{hindman_2020ApJ...898..120H} carry out a similar survey with the rotation and find the saturation around $\mathrm{Ra}\sim10^7$ with the solar rotation case. Our Rayleigh number is larger than these critical Rayleigh numbers. These results indicate that when we carry out a hydrodynamic run, the kinetic energy is expected to be converged. In addition, our result shows that the magnetic field generation requires further resolution to be numerically converged than hydrodynamic models.
Currently, we cannot conclude the magnetic field strength in the real Sun, but it is most probably stronger than our simulation result. The strength may be determined by a balance between the generation of compression and the destruction by the small-scale turbulence. At the same time, 
provided our suggested mechanism of the magnetic angular momentum transport is correct, we do not expect that the magnetic field strength in the real Sun is much stronger than our simulation because our differential rotation is similar to the observational results. In our simulation, the magnetic field directly determines the differential rotation topology. If our magnetic field strength is completely different from reality, the differential rotation is also away from reality. This is not the case in the simulation. Of course, there is a possibility that our suggested mechanism is incorrect.
We need to perform higher-resolution simulations to reach numerical convergence and to understand magnetic field strength in reality and the validity of the mechanism.
\par
We note that we use the RSST in our calculation. When the Alfven velocity exceeds the reduced speed of sound, the amplification efficiency should be decreased. In this study, the maximum magnetic pressure is $B^2/(8\pi)\sim1\times10^8~\mathrm{dyn~cm^{-2}}$ (see Fig. \ref{PDF_prpm}), and the effective gas pressure evaluated with the reduced speed of sound at $r=0.9R_\odot$ is $\rho_0c_\mathrm{s}^2/\xi^2\sim2.4\times10^9~\mathrm{dyn~cm^{-2}}$. These values indicate that we can ignore the influence of the RSST on the compression in this study. We also emphasize that even if the RSST influenced the result, it would weaken the magnetic field strength. Our conclusion, i.e., strong magnetic field constructs the differential rotation, should be robust.
\subsection{Convection suppression}
\begin{figure}[htbp]
  \centering
  \includegraphics[width=0.5\textwidth]{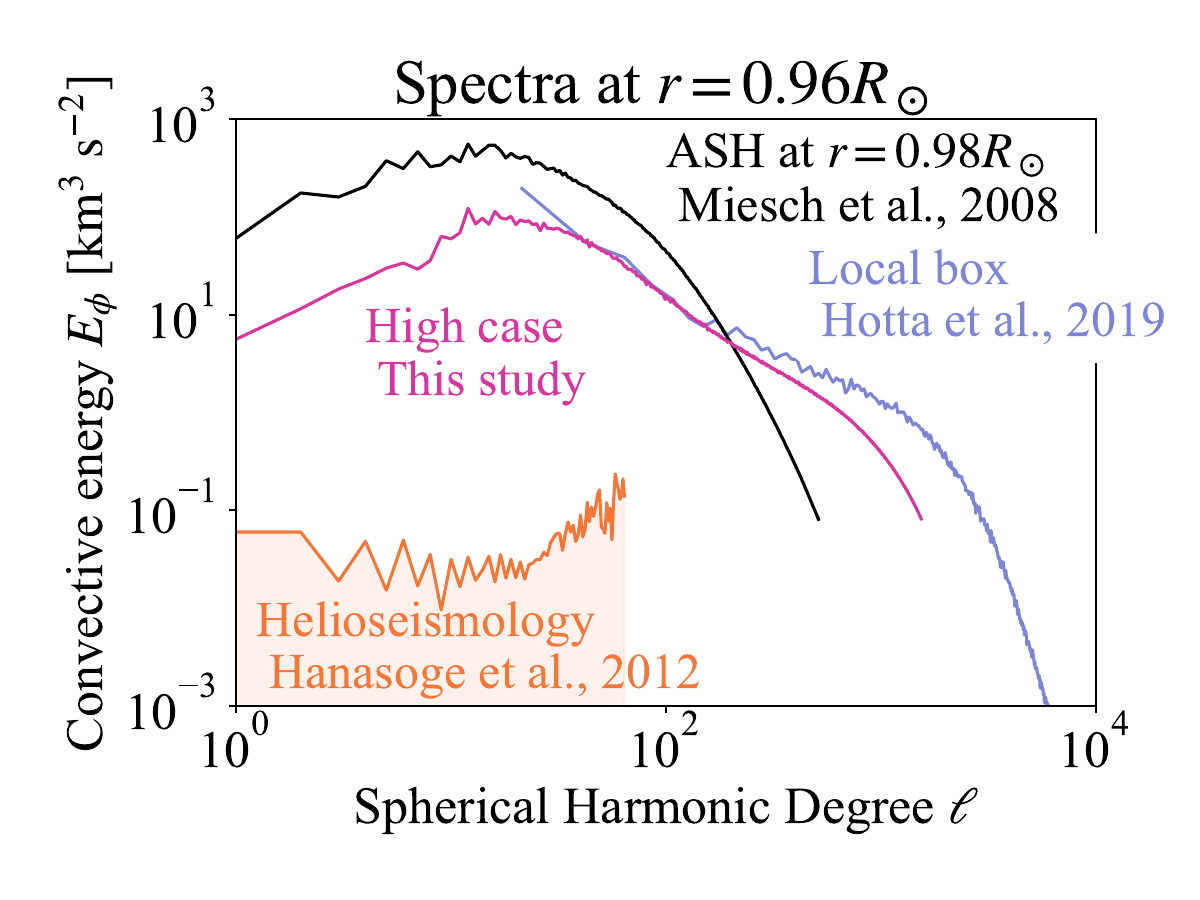}
  \caption{Comparison of kinetic energy spectra $E_\phi$ of longitudinal solar velocities in simulations and an observation.
  We show data from the local helioseismology \citep[orange]{Hanasoge_2012PNAS..10911928H}, ASH simulation as $r=0.98R_\odot$ \citep[black]{Miesch_2008ApJ...673..557M}, a local calculation \citep[blue]{Hotta_2019SciA....eaau2307}, and the High case in this study (magenta). Except for the ASH simulation, we show the energy spectra at $r=0.96R_\odot$. \label{comp_spectra}}
\end{figure}

Fig.~\ref{comp_spectra} shows a comparison of kinetic energy spectra of the longitudinal velocity $E_\phi$ between simulations and an observation. We follow the definition of the spectra of \cite{gizon_2012PNAS..10911896G}, where
\begin{align}
  \frac{\int_V v_\phi^2/2 dV}{\int_V dV} = \sum_{\ell>0} \frac{E_\phi}{r}. 
\end{align}
To exclude the contribution of the differential rotation, we exclude $m=0$ mode, where $m$ is the azimuthal wavenumber. The integration is carried out in the whole computational domain. The magenta line shows the result from the High case in this study. The blue line shows the result from \cite{Hotta_2019SciA....eaau2307}. In the calculation, the horizontal extent is restricted to 200 Mm, but it covers the whole convection zone vertically from the base to the photosphere. As suggested by \cite{Hotta_2019SciA....eaau2307}, the existence of the photosphere does not change the energy spectra in deeper layers, and the magenta and blue lines are consistent. The black line shows the result from another global calculation \citep{Miesch_2008ApJ...673..557M} at $r=0.98R_\odot$. The orange line indicates the upper limit suggested by the local helioseismology \citep{Hanasoge_2012PNAS..10911928H}. While we still have a large discrepancy between the simulation and the observation, the difference is relaxed. In our simulation, the large-scale convection is suppressed because the small-scale turbulence can efficiently transport the energy. Also, in this regard, the higher resolution possibly changes the result more. Meanwhile, recently, the helioseismology results have been revised \citep{proxauf_2021arXiv210607251P}. Our simulation results in the High case are highly consistent with the revised result of \cite{Greer_2015ApJ...803L..17G}. Currently, we cannot conclude if our convective velocity is correct or not. A more detailed comparison between simulations and observation is needed.\par
\cite{miesch_2012ApJ...757..128M} evaluate the lower limit of the convective velocity from the dynamical balance for the differential rotation. The evaluated value is not consistent with the local helioseismology \citep{Hanasoge_2012PNAS..10911928H}. In their study, they do not consider the magnetic contribution for the construction of the differential rotation. In this study, we find that the magnetic field is a dominant contribution. This means that the convection velocity has large freedom. The Rossby number does not solely determine the differential rotation.
One remaining restriction on the convection velocity is the energy flux. The solar luminosity $L_\odot$ is determined; thus, there should be a lower limit on the convection velocity to transport the required energy. Our simulation also shows that the temperature perturbation increases with the resolution. If the temperature perturbation increases, the lower limit on the convective velocity should be relaxed.  At the same time, significantly large temperature perturbation should be detected by the local helioseismology with mean travel time. Future observations for the convection velocity as well as the temperature perturbation will contribute to solving the problem.

\subsection{Meridional flow}

Currently, the local helioseismology for the meridional flow is still controversial. \cite{Zhao_2013ApJ...774L..29Z} indicate the double cell flow with the poleward meridional flow around the base of the convection zone. On the other hand, \cite{Gizon_2020Sci...368.1469G} show equatorward meridional flow around the base. In this regard, our result is more consistent with \cite{Zhao_2013ApJ...774L..29Z}'s result. 
This discrepancy is caused by the difference in observations. \cite{Zhao_2013ApJ...774L..29Z} adopt SDO data, and \cite{Gizon_2020Sci...368.1469G} use both SoHO (Solar and Heliospheric Observatory) and GONG (Global Oscillation Network Group) data. \cite{Gizon_2020Sci...368.1469G} find that SoHO and GONG data are consistent, but SDO data show a systematic difference from these two data.
We should also note that observations still have tiny sensitivity in the deep convection zone because it requires long enough separated two endpoints of $\Delta\sim45~\mathrm{degree}$ for evaluating the travel time \citep{giles_2000PhDT.........9G}. The observation has not accomplished enough precise observations for these separated two endpoints. For example, \cite{Gizon_2020Sci...368.1469G} show meridional flow results with and without the data with $\Delta> 30~\mathrm{degree}$, but the result does not change. This indicates that the data with $\Delta >30~\mathrm{degree}$ are not used for their inversion because of the large error, and the equatorward meridional flow is caused by the constraint of the mass conservation. This result indicates that we cannot conclude that our meridional flow is inconsistent with \cite{Gizon_2020Sci...368.1469G}'s result. 
In order to check whether our MC model is compatible with the helioseismic observations, it is needed to compute the seismic travel times based on this solution and to compare them with the observations.
Observations from different viewing angles, such as the Solar Orbiter \citep{muller_2013SoPh..285...25M} also enable us to understand the whole topology of the meridional flow, which should be of significant impact on the understanding of the convection and magnetic fields in the solar convection zone.

\begin{acknowledgments}
  We thank the anonymous referee for helpful comments, especially for the non-dimensional numbers. The authors thank L. Gizon, M. Rempel Y. Bekki and K. Mori for their comments on the manuscript. H.H. is supported by JSPS KAKENHI grants No. JP20K14510, JP21H04492, JP21H01124, JP21H04497, and MEXT as a Program for Promoting Researches on the Supercomputer Fugaku (Toward a unified view of the universe: from large-scale structures to planets, grant no. 20351188). The results were obtained using the Supercomputer Fugaku provided by the RIKEN Center for Computational Science.
  The authors are grateful to Rachel Howe for giving us the HMI inversion data for the solar differential rotation, S. Hanasoge, and M. Miesch for providing the spectral data.
\end{acknowledgments}

\software{R2D2 \citep{Hotta_2019SciA....eaau2307,Hotta_2020MNRAS.494.2523H,hotta_2021NatAs...5.1100H}}


\appendix
\section{Stream function}
\label{sec:stream_function}
In this appendix, we explain our method to calculate the stream function. Because a low Mach number situation is kept in our calculation, the meridional flow $\langle\bm{v}_\mathrm{m}\rangle=\langle v_r\rangle\bm{e}_r + \langle v_\theta\rangle \bm{e}_\theta$ should obey the anelastic approximation $\nabla\cdot\left(\rho_0 \langle\bm{v}_\mathrm{m}\rangle\right)\sim 0$. This indicates that the meridional flow can be written as a stream function $\Psi(r,\theta)$ as follows.

\begin{align}
  \rho_0\langle \bm{v}_\mathrm{m}\rangle = \nabla\times \left(\Psi\bm{e}_\phi\right)\label{stream1}
\end{align}

Taking the rotation of eq.~(\ref{stream1}) leads to
\begin{align}
  \nabla\times\left(\rho_0\langle \bm{v}_\mathrm{m}\rangle\right) = - \nabla^2 \left(\Psi\bm{e}_\phi\right),
\end{align}
because $\nabla\cdot \left(\Psi(r,\theta)\bm{e}_\phi\right)=0$. Thus, we need to solve the Poisson equation of
\begin{align}
  \left[\nabla\times\left(\rho_0\langle \bm{v}_\mathrm{m}\rangle\right)\right]_\phi = - \nabla^2 \Psi + 
  \frac{\Psi}{r^2\sin^2\theta}.
\end{align}
The solution of the Poission equation is a steady-state ($\partial/\partial t=0$), and the solution of the diffusion equation with a source term is as follows
\begin{align}
  \frac{\partial \Psi}{\partial t}
   = \nabla^2 \Psi - \frac{\Psi}{r^2 \sin^2\theta} - \left[\nabla\times\left(\rho_0\langle \bm{v}_\mathrm{m}\rangle\right)\right]_\phi.
   \label{diffusion_equation}
\end{align}
We simply integrate eq.~(\ref{stream1}) for the initial condition of eq.~(\ref{diffusion_equation}). Then, we evolve eq.~(\ref{diffusion_equation}) for several time steps, and the solution reaches a steady-state. We use the obtained value for the stream function used in Figs.~\ref{mean_flow} and \ref{mean_flow_initial}.

\section{Gyroscopic pumping}

\begin{figure}[htbp]
  \begin{center}
    \includegraphics[width=\textwidth]{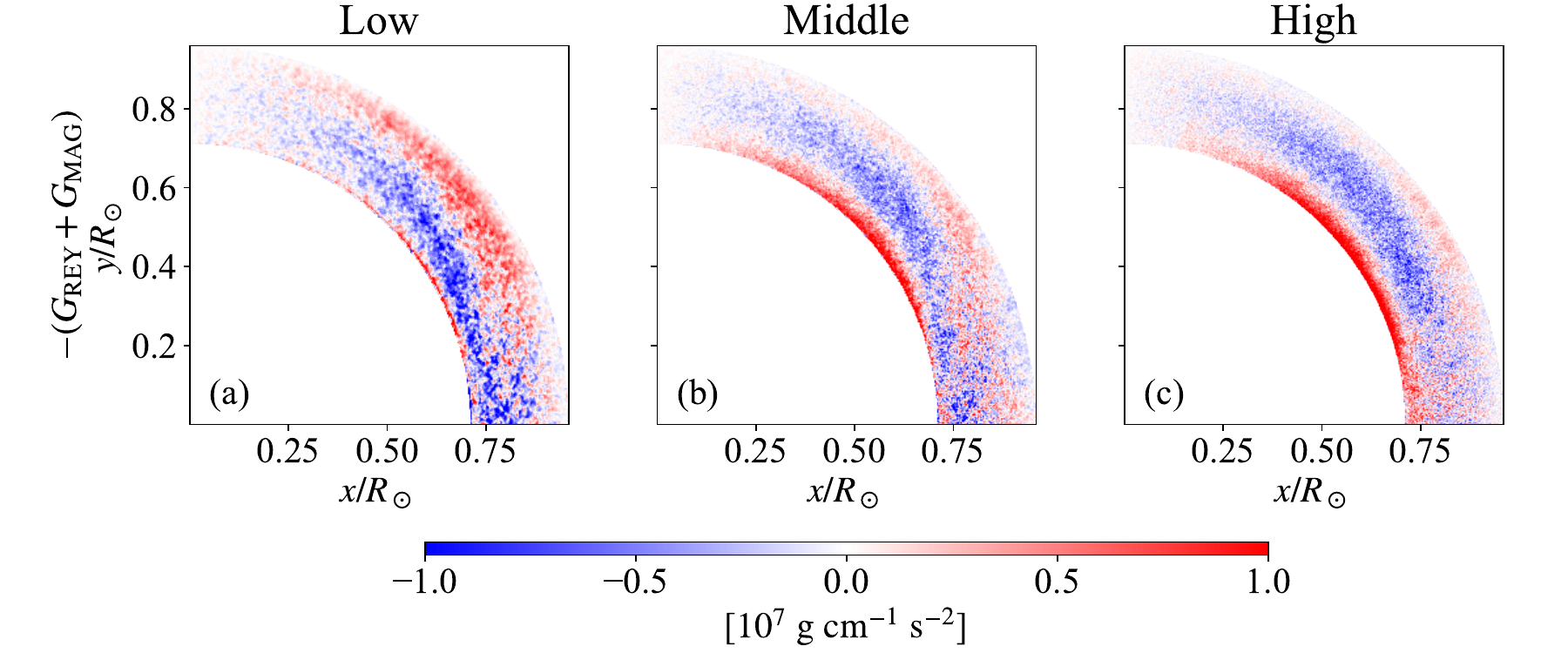}
    \caption{\label{gyroscopic_pumping_appendix}}
  \end{center}
\end{figure}

Fig.~\ref{gyroscopic_pumping_appendix} shows $-(G_\mathrm{REY} + G_\mathrm{MAG})$ (see eqs.~(\ref{eq:G_REY}) and (\ref{eq:G_MAG}) for the definitions). While these values fluctuate much because of the nature of turbulent flow and the magnetic field, we certainly confirm that $G_\mathrm{REY} + G_\mathrm{MAG}$ is balanced with $G_\mathrm{MER}$ (Figs.~\ref{gyroscopic_pumping}g, h, and i). Fig.~\ref{gyroscopic_pumping_appendix} indicates that the gyroscopic pumping including the magnetic field (eq.~(\ref{eq:gyroscopic_pumping})) is at least roughly accomplished in our analyzed period in all cases.

\section{Mean flows in an initial phase}
\label{sec:mean_flow_initial}

\begin{figure}[htbp]
  \begin{center}
    \includegraphics[width=\textwidth]{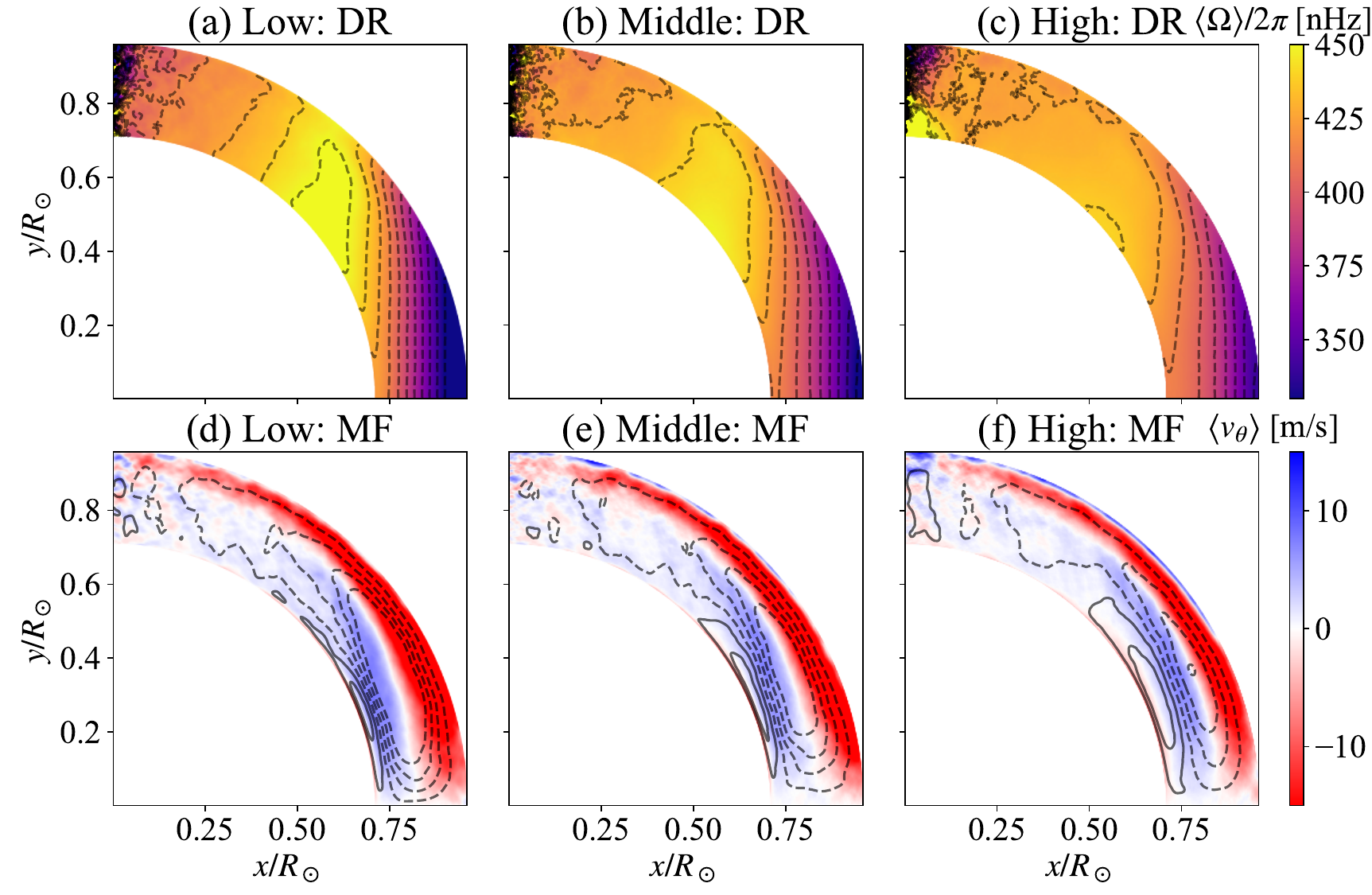}
    \caption{Format is the same as Fig.~\ref{mean_flow}, but the time average is between $t=200$ to 600 day.\label{mean_flow_initial}}
  \end{center}
\end{figure}

Fig.~\ref{mean_flow_initial} shows the differential rotation and the meridional flow in an initial phase (200--600 days). While we can reproduce the fast equator in the High case in the latter phase (Fig.~\ref{mean_flow}), all cases show the fast pole in the initial phase. During the long calculation, the magnetic field evolves and is amplified, and then the fast equator is constructed in the final steady phase in the High case.

\section{Normalized velocity correlations}
\begin{figure}[htbp]
  \begin{center}
    \includegraphics[width=\textwidth]{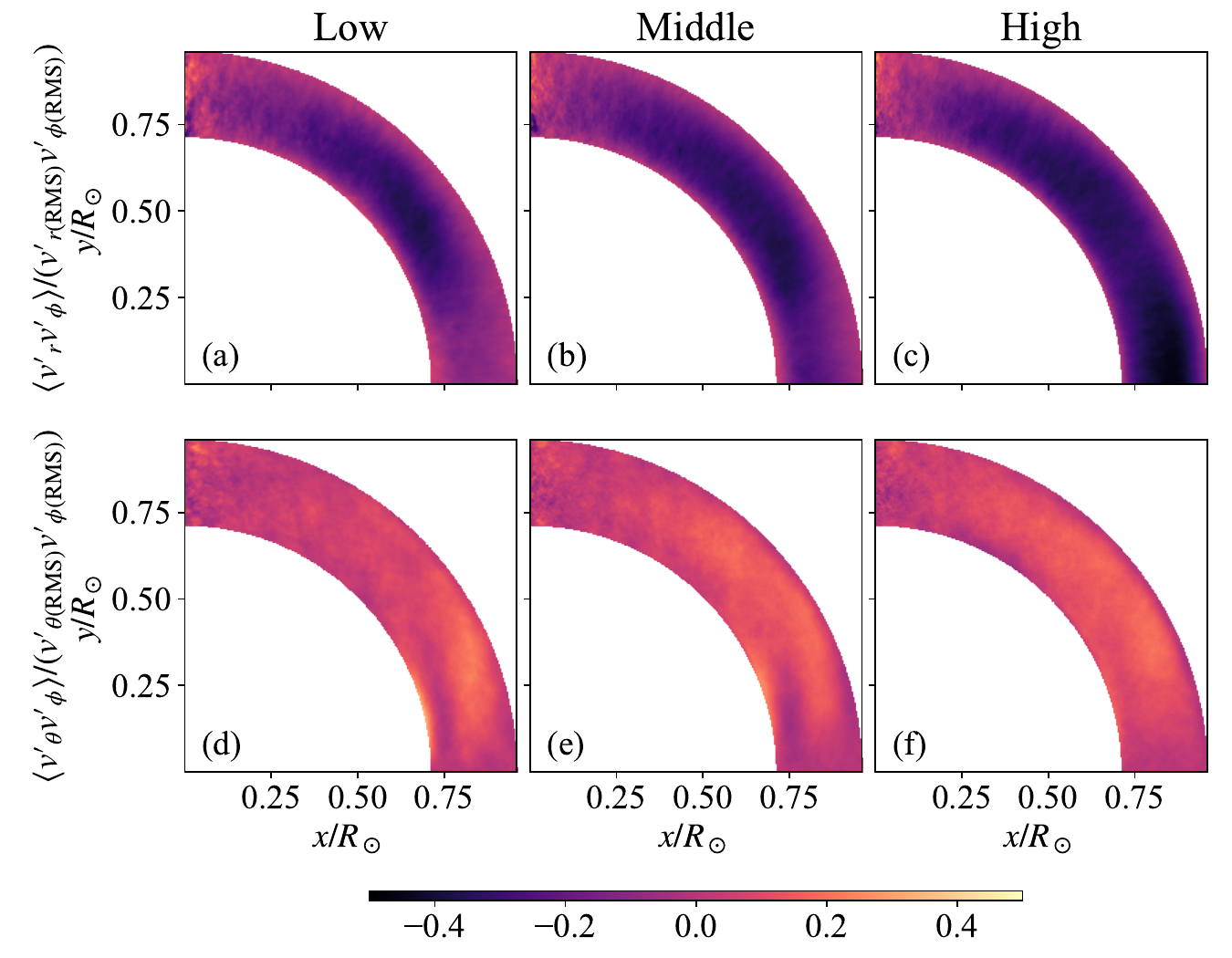}
    \caption{Normalized correlations between velocities $\langle v'_i v'_\phi\rangle/(v'_{i\mathrm{(RMS)}}v'_{\phi\mathrm{(RMS)}})$ are shown. The upper and lower panels show the radial and colatitudinal component, respectively. The left, middle, and right colums show the results from Low, Middle, and High cases, respectively. The radial component of the normalized correlation $\langle v'_r v'_\phi\rangle/(v'_{r\mathrm{(RMS)}}v'_{\phi\mathrm{(RMS)}})$ clearly increases with the negative sign with indicating the higher effective Rossby number in the High case.\label{normalized_correlation}}
  \end{center}
\end{figure}

Fig. \ref{angular_momentum_correlation} show the velocity correlations. Since the velocity amplitude changes in the cases, we cannot directly evaluate the anisotropy of the turbulence from Fig. \ref{angular_momentum_correlation}. In this appendix, we additionally show normalized correlation between velocities $\langle v'_i v'_\phi\rangle/(v'_{i\mathrm{(RMS)}}v'_{\phi\mathrm{(RMS)}})$, with which the variation of the velocity amplitude is removed. Fig. \ref{normalized_correlation} shows the result. It is clear that the anti-correlation between $v'_r$ and $v'_\phi$ increases with the resolution. This is a tendency of the high Rossby number regime \citep[e.g.,][]{karak_2015A&A...576A..26K}. The result supports our idea that the High case stays in the high Rossby number regime even though the fast equator is reproduced.

\section{Evaluation of viscosity}
\label{sec:evaluation_viscosity}
\begin{figure}[htbp]
  \begin{center}
    \includegraphics[width=0.5\textwidth]{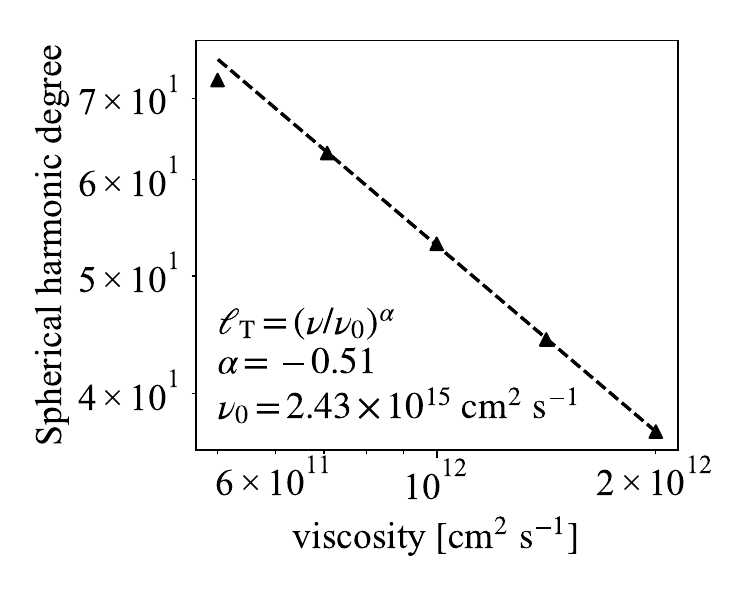}
    \caption{Dependence of the spherical harmonic degree for the Taylor microscale $\ell_\mathrm{T}$ on the explicit diffusivities are shown. The triangles show the raw data and the dashed line show the power-law fitting with the data except for $\nu=5\times10^{11}~\mathrm{cm^2~s^{-1}}$. \label{viscosity_evaluation}}
  \end{center}
\end{figure}

Since we use numerical diffusion for all the variables, evaluating effective diffusivity is not a simple task. At the same time, it is good to show a rough estimate of these for comparison purposes with previous and future research. To this end, we adopt a similar way to \cite{Hotta_2016Sci....351..1427}. The spherical harmonic degree for the Taylor microscale $\ell_\mathrm{T}$ is evaluated as \citep{Pietarila_Graham_2010ApJ...714.1606P},
\begin{align}
  \ell^2_\mathrm{T} = \int_{\ell_\mathrm{min}}^{\ell_\mathrm{max}} \ell^2 \widetilde{E}_\mathrm{h}(\ell) d\ell
  /\int_{\ell_\mathrm{min}}^{\ell_\mathrm{max}} \tilde{E}_\mathrm{h}(\ell) d\ell \label{eq:taylor}
\end{align}
\cite{batchelor_1953theory} and \cite{weygand_2007JGRA..11210201W} suggest that the effective Reynolds number is determined by the Taylor microscale $\lambda_\mathrm{T}=2\pi r/\ell_\mathrm{T}$ and the integral scale for the turbulent motions $L_0$ as
\begin{align}
  \mathrm{Re}_\mathrm{eff} \propto \left(\frac{L_0}{\lambda_\mathrm{T}}\right)^2
\end{align}
In this study, the large-scale convection is significantly influenced by the magnetic field and the situation makes it difficult to evaluate $L_0$ from the spectra data. Thus we assume the Taylor microscale is determined by the diffusivity (viscosity). In order to investigate the dependence of the Taylor microscale on the viscosity, we carry out additional five simulations. We adopt the explicit viscosity $\nu$, magnetic diffusivity $\eta$, and thermal conductivity on the entropy $\kappa$ as
\begin{align}
  \frac{\partial}{\partial t}\left(\rho \bm{v}\right) &= [...] -\nabla\cdot\bm{D} \\
  \frac{\partial \bm{B}}{\partial t} &= [...]-\nabla\times\left(\eta\nabla\times \bm{B}\right), \\
  \rho T\frac{\partial s_1}{\partial t } &= [...] + \nabla\cdot\left(\kappa \rho T \nabla s\right),
\end{align}
where the viscous stress tensor is 
\begin{align}
  D_{ij} = -2\rho \nu \left[e_{ij} - \frac{1}{3}\left(\nabla\cdot\bm{v}\right)\delta_{ij}\right],
\end{align}
and $e_{ij}$ and $\delta_{ij}$ are the deformation tensor and the Kronecker delta, respectively. We adopt the same grid poits as the Low case, $(N_r,N_\theta,N_\phi,N_\mathrm{YY})=(96,384,1152,2)$. We adopt five values of diffusivities as 
$\nu=\eta=\kappa=5\times10^{11},~7.1\times10^{11},~1\times10^{12}, 1.4\times10^{12}$, and $2\times10^{12}~\mathrm{cm^2~s^{-1}}$. These are constant in space.
The rotation is not included in the simulations. The other settings are identical to the simulations in the main text.
We evaluate the Taylor micro scale for these calculations at $r=0.83R_\odot$. We set $\ell_\mathrm{min}=10$ in eq. (\ref{eq:taylor}) to exclude the global scale convection. Fig. \ref{viscosity_evaluation} shows the dependence of $\ell_\mathrm{T}$ on the diffusivities.
The results show a power-law relation and the result with $\nu=5\times10^{11}~\mathrm{cm^2~s^{-1}}$ is slightly deviated from the relation. We carry out a fitting the data between 
$7.1\times10^{11}~\mathrm{cm^2~s^{-1}}\leq\nu\leq2\times10^{12}~\mathrm{cm^2~s^{-1}}$.
The result with $\nu=5\times10^{11}~\mathrm{cm^2~s^{-1}}$ seems affected by the numerical diffusivity and the data is excluded from the fitting. The fitting result is $\ell_\mathrm{T}=(\nu/\nu_0)^\alpha$ with $\nu_0=2.43\times10^{15}~\mathrm{cm^2~s^{-1}}$ and $\alpha=-0.51$. The result is consistent with the theoretical expectation $\ell_\mathrm{T}\propto\nu^{-1/2}$. We use the fitting result to evaluate the effective diffusivities for the simulation result in the main text. $\ell_\mathrm{T}$ for the Low, Middle, High, and High-HD cases are 110, 188, 340, and 346, respectively. These lead to the effective viscosity of $2.41\times10^{11}$, $8.44\times10^{10}$, $2.64\times10^{10}~\mathrm{cm^2~s^{-1}}$, and $2.55\times10^{10}~\mathrm{cm^2~s^{-1}}$ for the Low, Middle, High, and High-HD cases, respectively. Since the parameter runs in this appendix should have the same numerical diffusivity as the Low case, i.e., $2.4\times10^{11}~\mathrm{cm^2~s^{-1}}$, it is reasonable the run with $\nu=5\times10^{11}~\mathrm{cm^2~s^{-1}}$ is affected by the numerical diffusivity. Since we use the same numerical scheme for the velocity, the magnetic field and the entropy, we can assume that the effective diffusivities $\eta$, and $\kappa$ have the same values as $\nu$. We emphasize that the small scale features are significantly influenced by the magnetic field in the simulations. The Taylor microscale must be altered. Thus, the evaluated value is just a reference for comparisons with different calculations.

\end{document}